\documentclass[longauth]{aa}

\usepackage{graphicx}
\usepackage[dvipsnames]{xcolor}
\usepackage[hidelinks,colorlinks=true,linkcolor=blue,citecolor=blue]{hyperref}
\usepackage{txfonts}
\usepackage{gensymb}
\usepackage{graphicx}   
\usepackage{amsmath}    
\usepackage{amssymb}    
\usepackage{float}
\usepackage[caption = false]{subfig}
\usepackage{listings}
\definecolor{codegreen}{rgb}{0,0.6,0}
\definecolor{codegray}{rgb}{0.5,0.5,0.5}
\definecolor{codepurple}{rgb}{0.58,0,0.82}
\definecolor{backcolour}{rgb}{0.95,0.95,0.92}

\usepackage{natbib}
\bibpunct{(}{)}{;}{a}{}{,}
\newcommand{\hi}{H{\sc i}}

\begin{document} 

\title{The BINGO project V:\\ Further steps in component separation and bispectrum analysis}


\author{
Karin~S.~F.~Fornazier\inst{1,2}\thanks{karin.fornazier@gmail.com},
Filipe~B.~Abdalla\inst{1,2,3}\thanks{filipe.abdalla@gmail.com}, 
Mathieu~Remazeilles\inst{4}\thanks{mathieu.remazeilles@manchester.ac.uk},
Jordany~Vieira\inst{1}, 
Alessandro~Marins\inst{1}, 
Elcio~Abdalla\inst{1},
Larissa~Santos\inst{5}, 
Jacques~Delabrouille\inst{6,7,8}, 
Eduardo~Mericia \inst{3},
Ricardo~G.~Landim\inst{9}, 
Elisa~G.~M.~Ferreira\inst{1,10}, 
Luciano~Barosi\inst{11},
Francisco~A.~Brito\inst{11,18},
Amilcar~R.~Queiroz\inst{11}, 
Thyrso~Villela\inst{3,12,13}, 
Bin~Wang\inst{5,14}, 
Carlos~A.~Wuensche\inst{3},
Andre~A.~Costa\inst{5},
Vincenzo~Liccardo\inst{3},
Camila~Paiva~Novaes\inst{3},
Michael~W.~Peel\inst{15,16},
Marcelo~V.~dos~Santos\inst{11},
Jiajun~Zhang\inst{17}}

\institute{Instituto de F\'isica, Universidade de S\~ao Paulo -  05315-970, S\~ao Paulo, Brazil\\
 \email{bingotelescope@usp.br}
\and
Department of Physics and Astronomy, University College London, Gower Street, London, WC1E 6BT, UK 
\and
Instituto Nacional de Pesquisas Espaciais - INPE, Divis\~ao de Astrof\'isica,  Av. dos Astronautas, 1758, CEP: 12227-010 - S\~ao Jos\'e dos Campos, SP, Brazil
\and
Jodrell Bank Centre for Astrophysics, Department of Physics and Astronomy, The University of Manchester, Oxford Road, Manchester, M13 9PL, U.K. 
\and Center for Gravitation and Cosmology, College of Physical Science and Technology, Yangzhou University, Yangzhou 225009, China
\and
Laboratoire Astroparticule et Cosmologie (APC), CNRS/IN2P3, Universit\'e Paris Diderot, 75205 Paris Cedex 13, France 
\and IRFU, CEA, Universit\'e Paris Saclay, 91191 Gif-sur-Yvette, France 
\and
Department of Astronomy, School of Physical Sciences, University of Science and Technology of China, Hefei, Anhui 230026
\and
Technische Universit\"at M\"unchen, Physik-Department T70, James-Franck-Stra\text{$\beta$}e 1, 85748 Garching, Germany
\and
Max-Planck-Institut f{\"u}r Astrophysik, Karl-Schwarzschild Str. 1, 85741 Garching, Germany
\and
Unidade Acad\^emica de F\'{i}sica, Universidade Federal de Campina Grande, R. Apr\'{i}gio Veloso, 58429-900 - Bodocong\'o, Campina Grande - PB, Brazil 
\and 
Instituto de F\'{i}sica, Universidade de Bras\'{i}lia, Bras\'{i}lia, DF, Brazil
\and 
Centro de Gest\~ao e Estudos Estrat\'egicos - CGEE,
SCS Quadra 9, Lote C, Torre C S/N Salas 401 - 405, 70308-200 - Bras\'ilia, DF, Brazil
\and 
School of Aeronautics and Astronautics, Shanghai Jiao Tong University, Shanghai 200240, China   
\and 
Instituto de Astrof\'{i}sica de Canarias, 38200, La Laguna, Tenerife, Canary Islands, Spain 
\and 
Departamento de Astrof\'{i}sica, Universidad de La Laguna (ULL), 38206, La Laguna, Tenerife, Spain
\and 
Center for Theoretical Physics of the Universe, Institute for Basic Science (IBS), Daejeon 34126, Korea
\and 
Departamento de Física, Universidade Federal da Paraíba, Caixa Postal 5008, 58051-970 João Pessoa, Paraíba, Brazil
}
 \authorrunning{K.S.F. Fornazier  et al.}
   \date{Received; accepted }

 
  \abstract
{
Observing the neutral hydrogen (\hi\,) distribution across the Universe via redshifted 21cm line intensity mapping (IM) constitutes a powerful probe for cosmology. However, the redshifted 21cm signal
is obscured by the foreground emission from our Galaxy and other extragalactic foregrounds. This paper addresses the capabilities of the BINGO survey to separate such signals.
}
{
We show that the BINGO instrumental, optical, and simulations setup is suitable for component separation, and that we have the appropriate tools to understand and control foreground residuals. Specifically, this paper  looks in detail at the different residuals left over by foreground components, shows that a noise-corrected spectrum is unbiased, and  shows that we understand the remaining systematic residuals by analyzing nonzero contributions to the three-point function. 
}
{
We use the generalized needlet internal linear combination ({\tt GNILC}), which we apply to sky simulations of the BINGO experiment for each redshift bin of the survey. We use binned estimates of the bispectrum of the maps to assess foreground residuals left over after component separation in the final map.
}
{
We present our recovery of the redshifted 21cm signal from sky simulations of the BINGO experiment, including foreground components. We test the recovery of the 21cm signal through the angular power spectrum at different redshifts, as well as the recovery of its non-Gaussian distribution through a bispectrum analysis. 
We find that non-Gaussianities from the original foreground maps can be removed down to, at least, the noise limit of the BINGO survey with such techniques.
}
{
Our component separation methodology allows us to subtract the foreground contamination in the BINGO channels down to levels below the cosmological signal and the noise, and to reconstruct the 21cm power spectrum for different redshift bins without significant loss at multipoles $20 \lesssim \ell  \lesssim 500$. Our bispectrum analysis yields strong tests of the level of the residual foreground contamination in the recovered 21cm signal, thereby allowing us to both optimize and validate our component separation analysis.
}


   \maketitle
   \flushbottom
  { \hypersetup{linkcolor=black}
 }

\section{Introduction}

Over the last century dedicated experiments mapping different large-scale observables in the Universe have increased our understanding of the cosmic history, and established modern observational cosmology as a precise and quantitative science. However, despite great successes in building a cosmological concordance model with tightly constrained parameters, a number of questions regarding the constituents of the Universe are yet to be fully answered. The nature of dark energy, which leads to the observed accelerated expansion of the Universe, is one of the great mysteries in modern cosmology. The Baryon Acoustic Oscillations from Integrated Neutral Gas Observations (BINGO) telescope, which is designed to measure one of the most powerful observables used to  characterize dark energy, the Baryon Acoustic Oscillations (BAO), may enlighten this late evolution of the Universe \citep{2020_project,2020_instrument,2020_forecast}. 

BINGO will map the integrated sky emission of the neutral hydrogen (\hi\ signal) 21cm line transition within a redshift interval of $ 0.127 < z < 0.449$. The intensity mapping (IM) technique \citep{Peterson:2006bx,Battye:2013} thus allows the measurement of the entire \hi\ flux density of a wide patch of the sky at different redshift bins, producing \hi\ maps that can be used as input data to estimate cosmological parameters \citep{Abdalla:2004ah}.

However, a crucial intermediate stage comes before the production of \hi\ maps and the estimation of cosmological parameters. Mitigating the foreground emission in radio sky observations is critical for the reliable recovery of the 21cm signal, which is much fainter than the diffuse emission from the Galactic interstellar medium (ISM). Several methods have been proposed in the literature to separate the astrophysical foregrounds from the cosmological 21cm signal, with the aim of accurately reconstructing the power spectrum of the 21cm signal without biasing the estimation of the cosmological parameters.

Most component separation methods in the literature have been devised to primarily deal with foreground contamination in cosmic microwave background (CMB) data, for which one can rely on the known blackbody frequency spectrum of the CMB to disentangle the signal from the foregrounds through multi-frequency observations. Some of these methods rely on a parametric model for the foregrounds, such as {\tt Commander} \citep{Eriksen:2004ss, Eriksen:2007mx}, a joint CMB and foreground Bayesian fitting method based on Gibbs sampling. Other methods, the so-called blind (or semi-blind) methods, do not rely on any assumption about the frequency dependence of the foregrounds, but mostly exploit statistical correlations between frequency channels to mitigate the foreground contamination. Such blind methods include internal linear combinations (ILCs) such as the needlet ILC \citep[NILC;][]{delabrouille2009full,Basak:2011yt,basak2013needlet}, a constrained variance minimization implemented on spherical wavelets; the Spectral Matching Independent Component Analysis \citep[SMICA;][]{2003MNRAS.346.1089D,cardoso2008component}; and the Correlated Component Analysis  \citep[CCA;][]{bedini2005separation,Bonaldi:2006qw}, which use statistical decorrelation to disentangle independent components; the Spectral Estimation Via Expectation Maximization \citep[SEVEM;][]{FernandezCobos:2011bm}, which builds internal foreground templates from different maps between pairs of channels; the Generalized Morphological Component Analysis \citep[GMCA;][]{2013MNRAS.429..165C}, which exploits sparsity to separate CMB and foregrounds; the Independent Component Analysis (ICA), which maximizes some measure of  non-Gaussianity (NG)  to disentangle independent sources, such as {\tt FastICA} \citep{Maino:2001vz}; and Bayesian formulations of the ICA \citep{Vansyngel:2014dfa}.

In contrast to the CMB, the frequency dependence of the cosmological 21cm signal is nontrivial, and the emission law somewhat random or decorrelated across frequencies, hence making it more challenging to model 21cm emission. For this reason, component separation algorithms dedicated to 21cm data analysis in the literature mostly reduce to foreground subtraction techniques, with a known risk of partial loss of the 21cm signal during the subtraction. Typical component separation algorithms that have been applied to 21cm data include principal component analysis \citep[PCA;][]{2015MNRAS.447..400A,2019AJ....157...34Z}, independent component analysis \citep[ICA;][]{2012MNRAS.423.2518C,2014MNRAS.441.3271W}, and generalized morphological component analysis
\citep[GMCA;][]{2013MNRAS.429..165C,2020MNRAS.499..304C}. 
For the present analysis we use the generalized needlet ILC  \citep[{\tt GNILC};][]{Remazeilles:2011,Olivari2015}, an extension of the blind NILC  method that compensates for the lack of information on the frequency dependence of the cosmological signal by some prior information on its spatial statistics (power spectrum), and also exploits here the decorrelation between cosmological signals originating from different redshift bins.

Here we present the reconstructed maps and power spectra of the cosmological 21cm signal in the presence of various foregrounds and white noise. However, differently from the primordial fluctuations generated by inflationary models, the 21cm brightness temperature fluctuations depend on the densities, temperatures, and velocity gradients at late cosmic times, hence giving rise to 
NGs. Higher order statistics are thus necessary to completely characterize the intrinsic non-Gaussian 21cm field. However, residual foreground contamination can also leave a non-Gaussian imprint in the reconstructed 21cm field after the component separation step. We can hence use higher order statistics to probe the non-Gaussian features of our signal at different scales and discern them from foreground residuals. The bispectrum, which is the Fourier transform of the three-point correlation function, is well suited to capture NGs,
either intrinsic or left over by foreground residuals, in our recovered 21cm field.

The map  that describes the temperature of our Galaxy as a function of sky position does not behave at all like a Gaussian random field; therefore, it should have a very large nonzero bispectrum. We expect this bispectrum to be large compared to the intrinsic NGs
produced by the log-normal large-scale structure (LSS)
field stated above. 
Estimating the bispectrum should therefore be a very good test to detect any residual foregrounds in the recovered 21cm maps.

One of our goals here is to identify spurious non-Gaussian features in the recovered 21cm maps that would be larger than the intrinsic 
NG of the 21\,cm signal, and which is due to any residual foreground contamination after component separation or greater than the noise present in the simulated data set. If we  had such a scenario, it would be clear that residuals from foreground separation have been injected into our output maps, and we would then have a clear tool to identify if this is the case. 

Alternatively, it is possible that we might find non-Gaussian features in the bispectrum that are partially due to the 21\,cm signal from the above-mentioned nonlinear evolution of matter, but also due to any poorly subtracted foregrounds. In such scenarios, measurements of the bispectrum would have to model the contributions to determine if there are any residual non-Gaussian modes in the final maps that have been left over from the foreground subtraction method. We expect the bispectrum of the 21\,cm to be small compared to the non-Gaussian bispectrum signal from the Galaxy, hence this scenario is unlikely.

It is important to note here that  for the conditions we use to simulate the 21cm signal in this paper 
the cosmological non-Gaussian signal is small, and  is compatible with zero within our error bars. In these conditions, the bispectrum output is used in this paper as a double-check for the cleaning procedure and not as a tool to get cosmological information.


This paper is organized as follows. Section~\ref{sec:simulations} describes the simulation codes used on this work, both for foregrounds and the 21cm signal with non-Gaussian information. Section~\ref{sec:powerspectum} describes our component separation procedure and debiasing approach, including residual foreground contamination. 
In Section~\ref{sec:bispectrum} we present the bispectrum module as well as its abilities for pinpointing subtle foreground contamination. In Section~\ref{sec:conclusions} we present the conclusion of this work.

This  paper is the fifth in a series of papers presenting the BINGO project. Companion paper I is the project paper \citep{2020_project},  paper II describes the instrument \citep{2020_instrument}, paper III shows the optical design \citep{2020_optical_design}, paper IV describes the mission simulations \citep{2020_sky_simulation}, paper VI discusses the $21$cm catalog simulations \citep{2020_mock_simulations}, and paper VII presents  the cosmological forecasts for the BINGO telescope \citep{2020_forecast}.

\section{Simulations}\label{sec:simulations}

In this paper we implement the {\tt GNILC} method on simulated data sets for the BINGO telescope specifications, as given by \cite{2020_instrument}. Simulations are crucial to the analysis toolkit in order to test our entire pipeline: from observed maps to the estimation of the cosmological parameters. 
 We use a {\tt HEALPix} \citep{Gorski:2005} pixelization scheme at $N_{\rm side}=256$, a Gaussian beam with a full width at half maximum (FWHM) of $40'$, and celestial coordinates for our simulated maps.


Our simulations of the sky as observed by the BINGO telescope include the non-Gaussian 21cm signal, several foregrounds, and white noise. The 21cm power spectrum is generated using the Unified Cosmological Library for $C_\ell$s ({\tt UCLCL}) code \citep{mcleod2017joint,Loureiro:2018qva} that enters as an input to the Full-sky Lognormal Astro-fields Simulation Kit ({\tt FLASK}) code \citep{Xavier2016}, which in turn generates the non-Gaussian simulated 21cm maps. 

{\tt FLASK} can generate fast full-sky simulations of cosmological  LSS observables, such as multiple matter density tracers (galaxies, quasars, dark matter halos), CMB temperature anisotropies and weak lensing convergence, and shear fields.
{\tt UCLCL} is a library for computing two-point angular correlation function of various cosmological fields that are related to LSS surveys. It uses the formalism of angular power two-point correlations, and then derives the exact analytical equations for the angular power spectrum of cosmological observables. We describe below the simulated maps of 21\,cm emission obtained with {\tt FLASK}, of foreground components from the {\tt PSM} code \citep{Delabrouille:2012ye} and of instrumental noise.

\subsection{Cosmological signal}


We simulate our non-Gaussian 21cm map as a log-normal field 
for the 30 redshift bins observed by the BINGO telescope (see Table~\ref{tab:table1}).
%
The {\tt FLASK} code
produces non-Gaussian fields by applying a transform to an originally Gaussian field in such a way that the transformed field obeys the two-point function originally supplied to the code. This transformation also produces a modification to the one-point function (to produce a log-normal field) chosen by the user and an associated bispectrum that cannot be chosen explicitly by the user. This transformation produces a non-Gaussian signal that is meant to skew the Gaussian field into a log-normal field, especially enhancing the overdensities into higher peaks, which is what is expected by the gravitational evolution of density perturbations. A summary of {\tt FLASK} characteristics and specific choices for the BINGO series of papers can be found in Paper IV \citep{2020_sky_simulation}.

\begin{table}[tb]
\begin{centering}
\begin{tabular}{|r|l|r|l|}
\hline
\multicolumn{4}{|c|}{ Frequency (MHz)} \\
\hline
Channel & $\nu_{\rm{min}}$--$\nu_{\rm{max}}$ & Channel  & $\nu_{\rm{min}}$--$\nu_{\rm{max}}$\\
\hline
\phantom{0}0 & \phantom{0}960--\phantom{0}971   & \phantom{0}1 & \phantom{0}971--982\\
\phantom{0}2 &  \phantom{0}982--\phantom{0}993   & \phantom{0}3 & \phantom{0}993--1004 \\
\phantom{0}4 &  1004--1015 & \phantom{0}5 & 1015--1026 \\
\phantom{0}6 &  1026--1037 & \phantom{0}7 & 1037--1048 \\
\phantom{0}8 &  1048--1059 & \phantom{0}9 & 1059--1070 \\
10 & 1070--1081 & 11 & 1081--1092 \\
12 & 1092--1103 & 13 & 1103--1114 \\
14 & 1114--1125 & 15 & 1125--1136 \\
16 & 1136--1147 & 17 & 1147--1158 \\
18 & 1158--1169 & 19 & 1169--1180 \\
20 & 1180--1191 & 21 & 1191--1202 \\
22 & 1202--1213 & 23 & 1213--1224 \\
24 & 1224--1235 & 25 & 1235--1246 \\
26 & 1246--1257 & 27 & 1257--1268 \\
28 & 1268--1279 & 29 & 1279--1290  \\
\hline
\end{tabular}
\caption{BINGO frequency channels (in MHz).}
    \label{tab:table1}
\end{centering}
\end{table}

\begin{figure}[tb]
\includegraphics[trim=1cm 0.4cm 6cm 18cm, clip=true, width=0.5\textwidth]{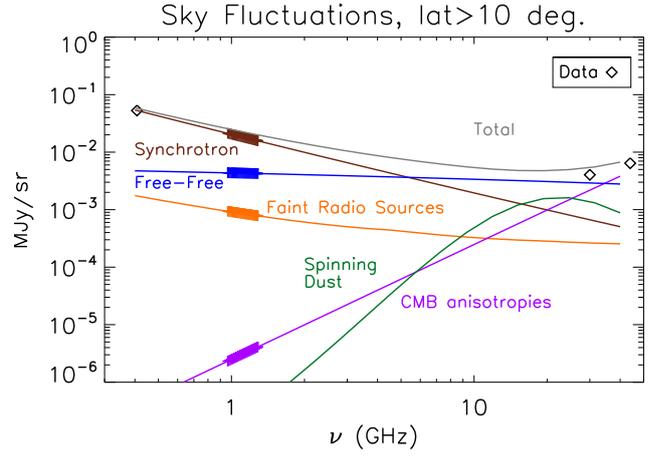}
 \caption{Frequency scaling of the main foreground components simulated with the PSM. The various curves display the standard deviation of the maps of the various components as a function of frequency, at an angular resolution of $40^\prime$, and at galactic latitudes greater than $10^\circ$. The general shapes of the curves illustrate the average frequency scaling, while the relative amplitudes in the BINGO frequency range show the relative importance of the various components for the detection of 21cm fluctuations by BINGO. The data points are from the 408\,MHz map of \cite{Remazeilles:2015}, and the 28.4\,GHz and 44.1\,GHz maps of the LFI instrument on board the Planck satellite.}\label{fig:cmb}
\end{figure}

\begin{figure*}
\centering
    \includegraphics[width=.45\textwidth]{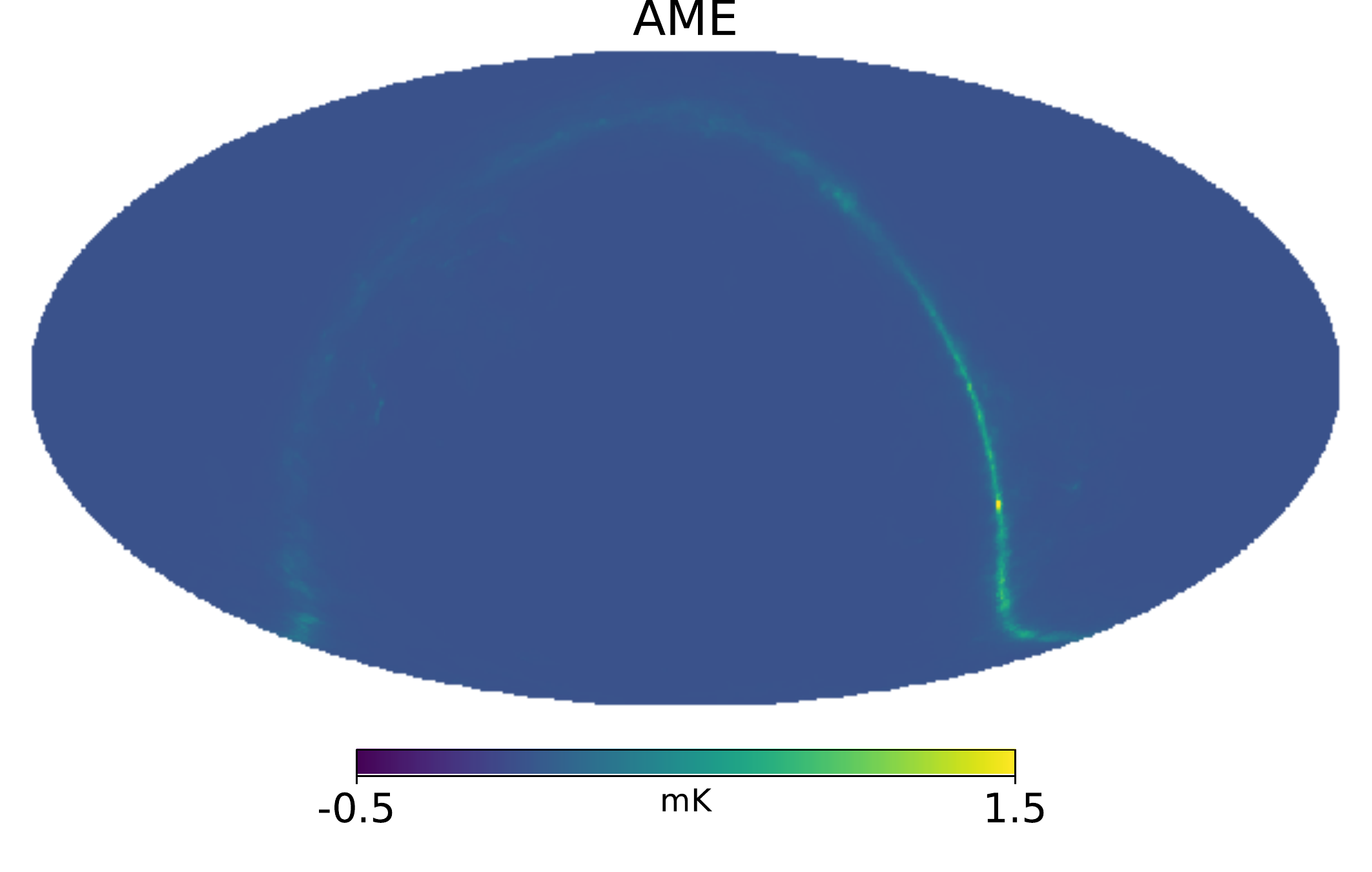}
    \includegraphics[width=.45\textwidth]{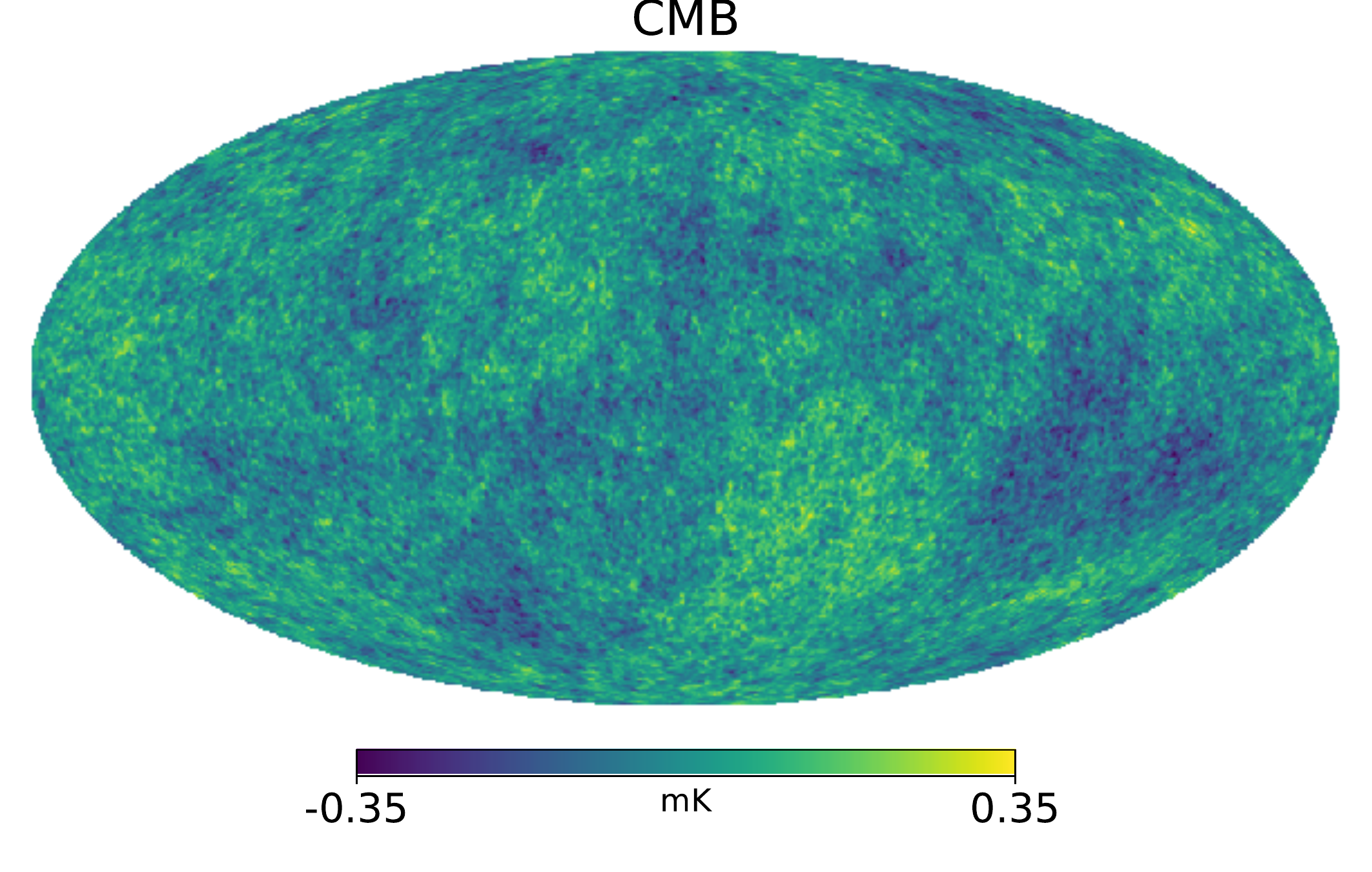}
    \\[\smallskipamount]
    \includegraphics[width=.45\textwidth]{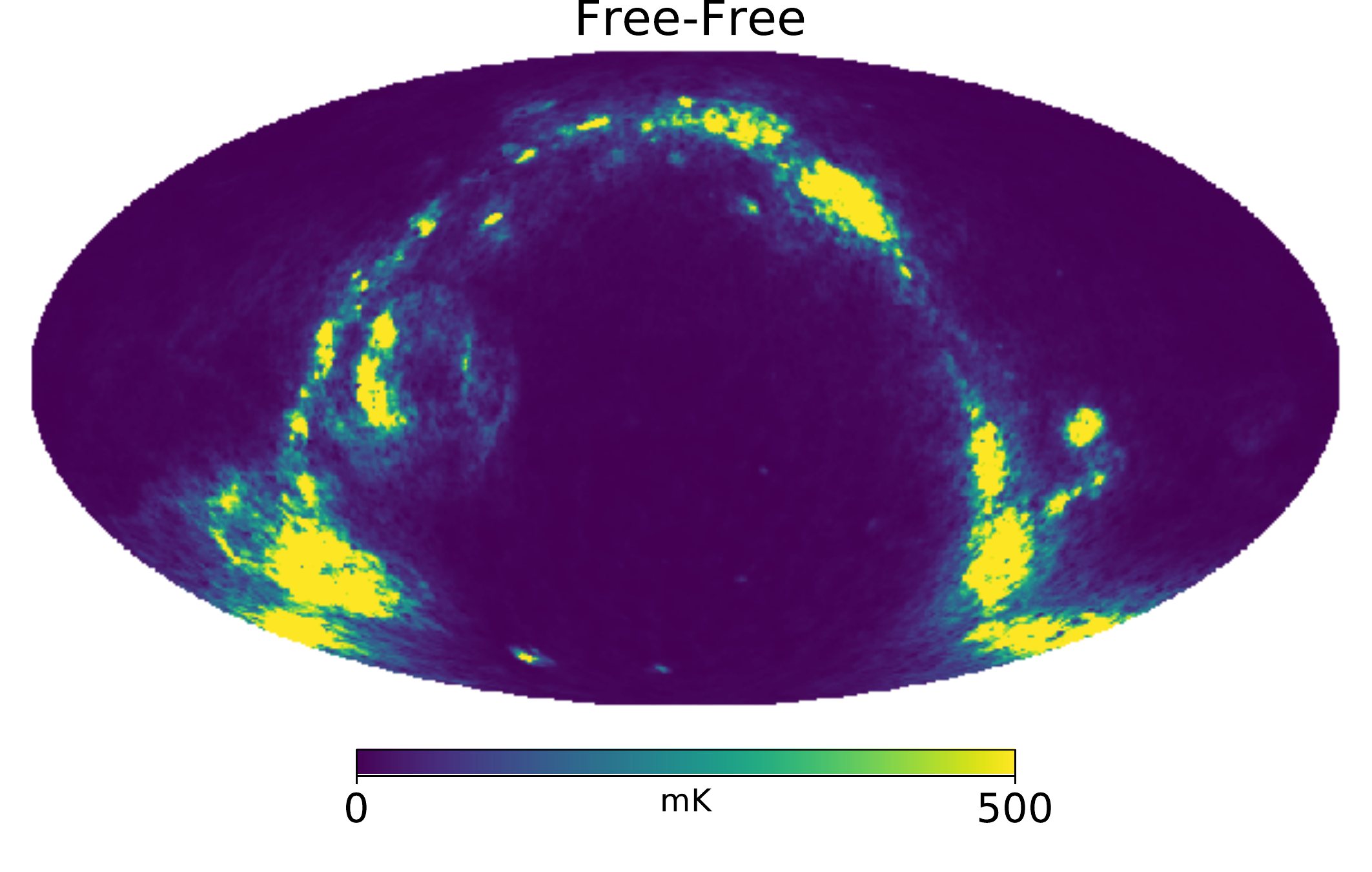}
    \includegraphics[width=.45\textwidth]{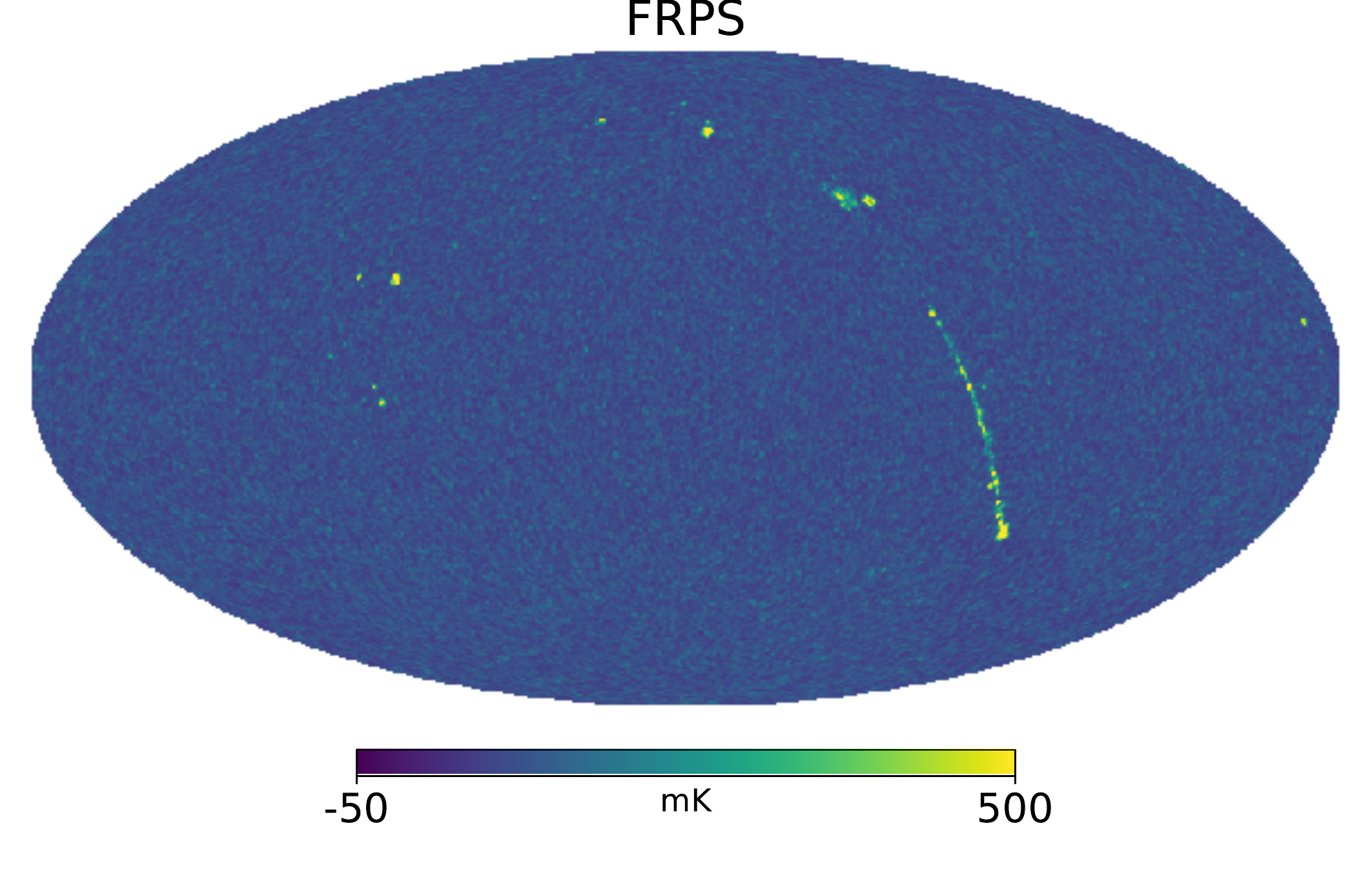}
    \\[\smallskipamount]
    \includegraphics[width=.45\textwidth]{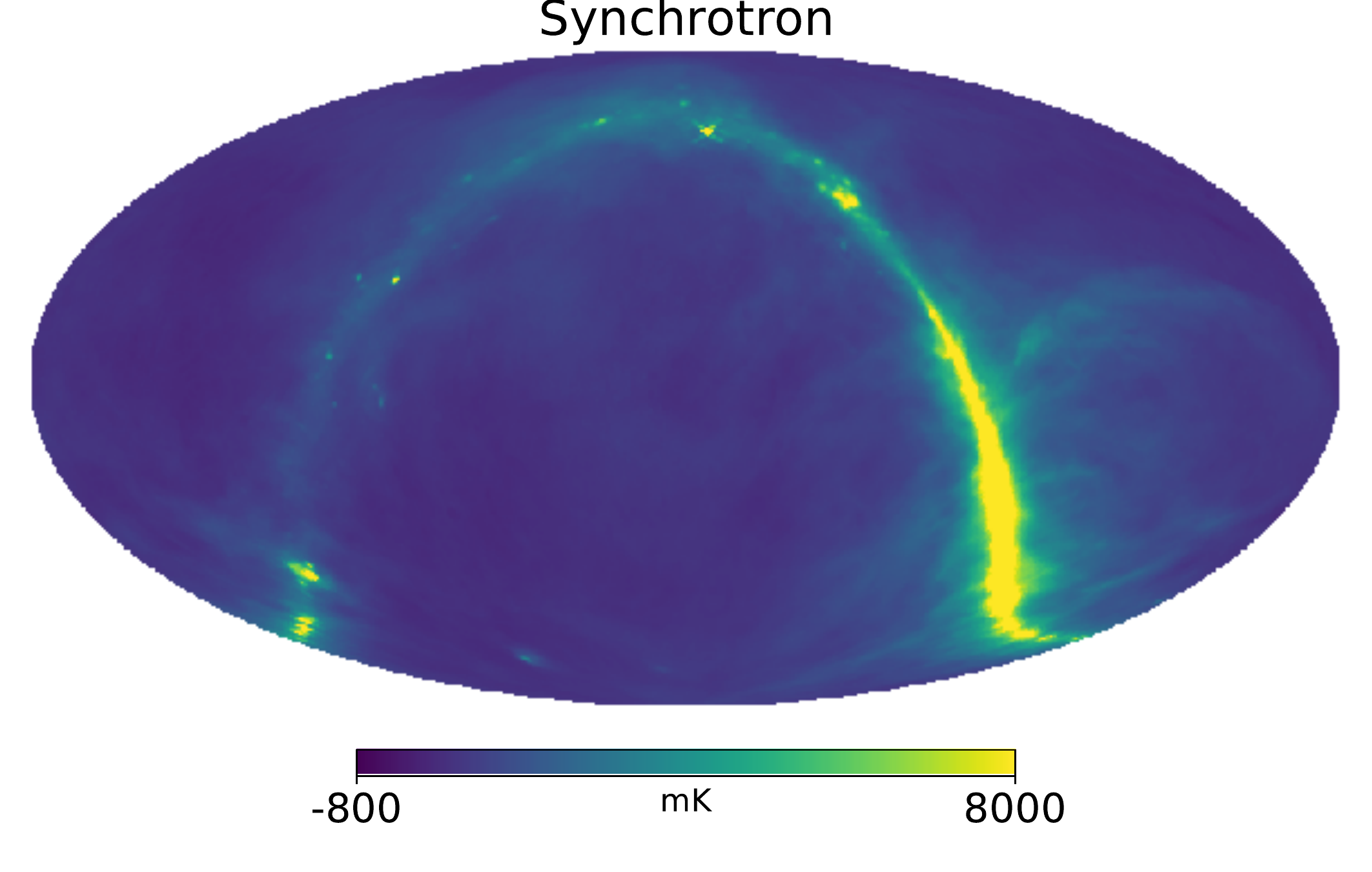}
    \includegraphics[width=.45\textwidth]{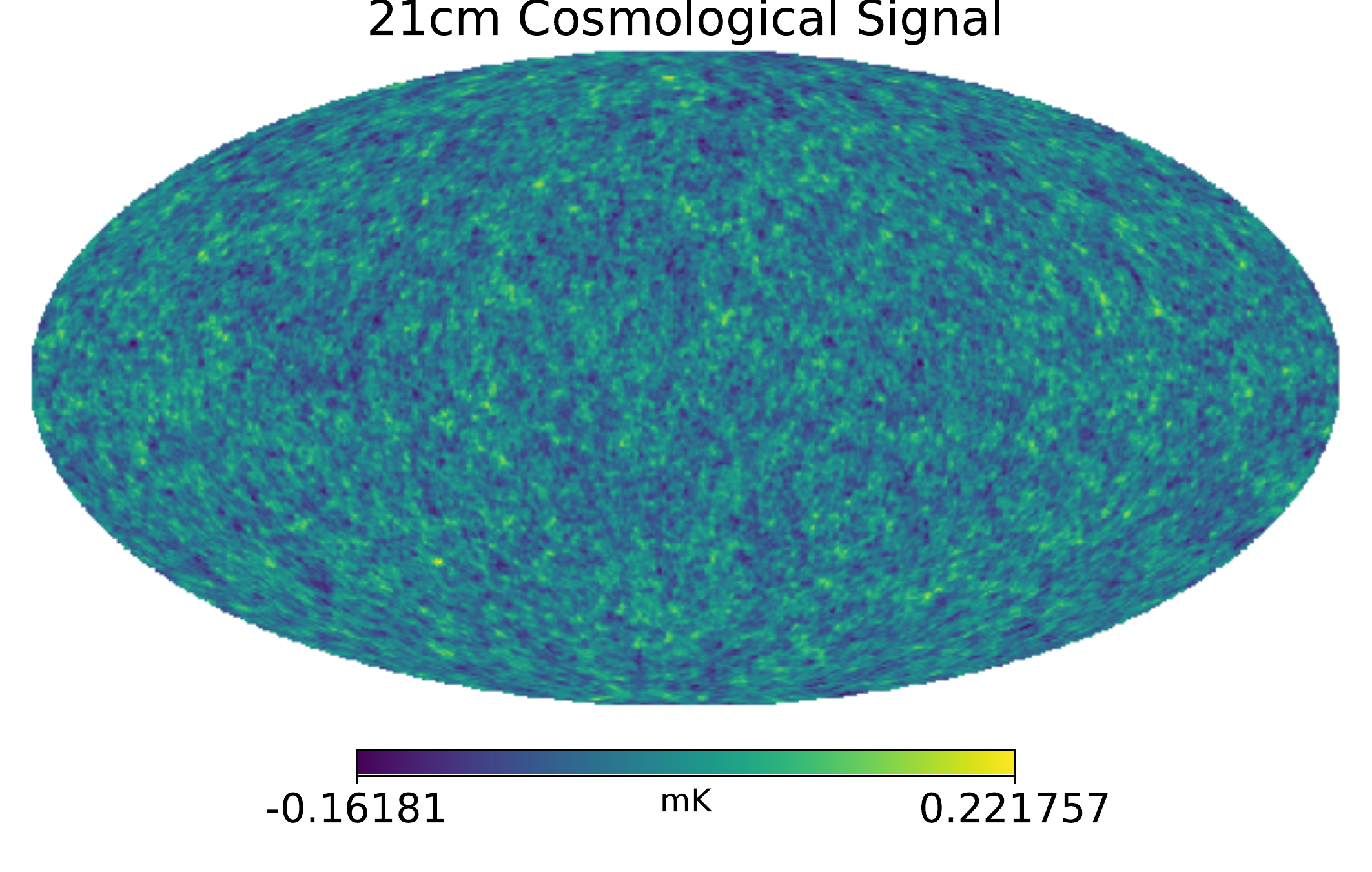}
    
     \caption{Simulated maps of the foreground components and the 21cm cosmological signal in mK units, as observed in the BINGO frequency band 16 (see Table~\ref{tab:table1}): AME (\textit{top left}), CMB (top right), free-free (\textit{middle left}), faint radio point sources (FRPS)  (\textit{middle right}), synchrotron (\textit{bottom left}), and the 21cm log-normal cosmological signal (\textit{bottom right}). All maps are shown in celestial coordinates, have a {\tt HEALPix} resolution of $N_{\rm{side}}=256$, and are convolved with a $40'$ beam.}\label{fig:allcomponents}
\end{figure*}

\subsection{Foregrounds}

We use the {\tt PSM} code \citep{Delabrouille:2012ye} to generate our simulated foreground maps (see Table \ref{tab:table1} for the frequency band distribution; we have simulated bands that start slightly before and extend slightly above the nominal BINGO filter, see \citealp{2020_instrument} for the full instrument description).  We consider three Galactic foreground emissions: synchrotron, free-free, and AME (assumed to be spinning dust), which we generate consistently for all the BINGO channels. The thermal dust emission is  subdominant in the BINGO frequency range and is neglected in our calculations. Extragalactic foregrounds due to CMB temperature anisotropies and a background of unresolved radio point sources are also a significant source of contamination in the BINGO frequency range, and are included. 
Figure \ref{fig:cmb} shows the relative amplitude of all main foreground components as a function of frequency, also illustrating  the average frequency scaling of those components of emission in the frequency range of interest for BINGO observations.

\begin{figure}
\centering
\includegraphics[width=0.45\textwidth]{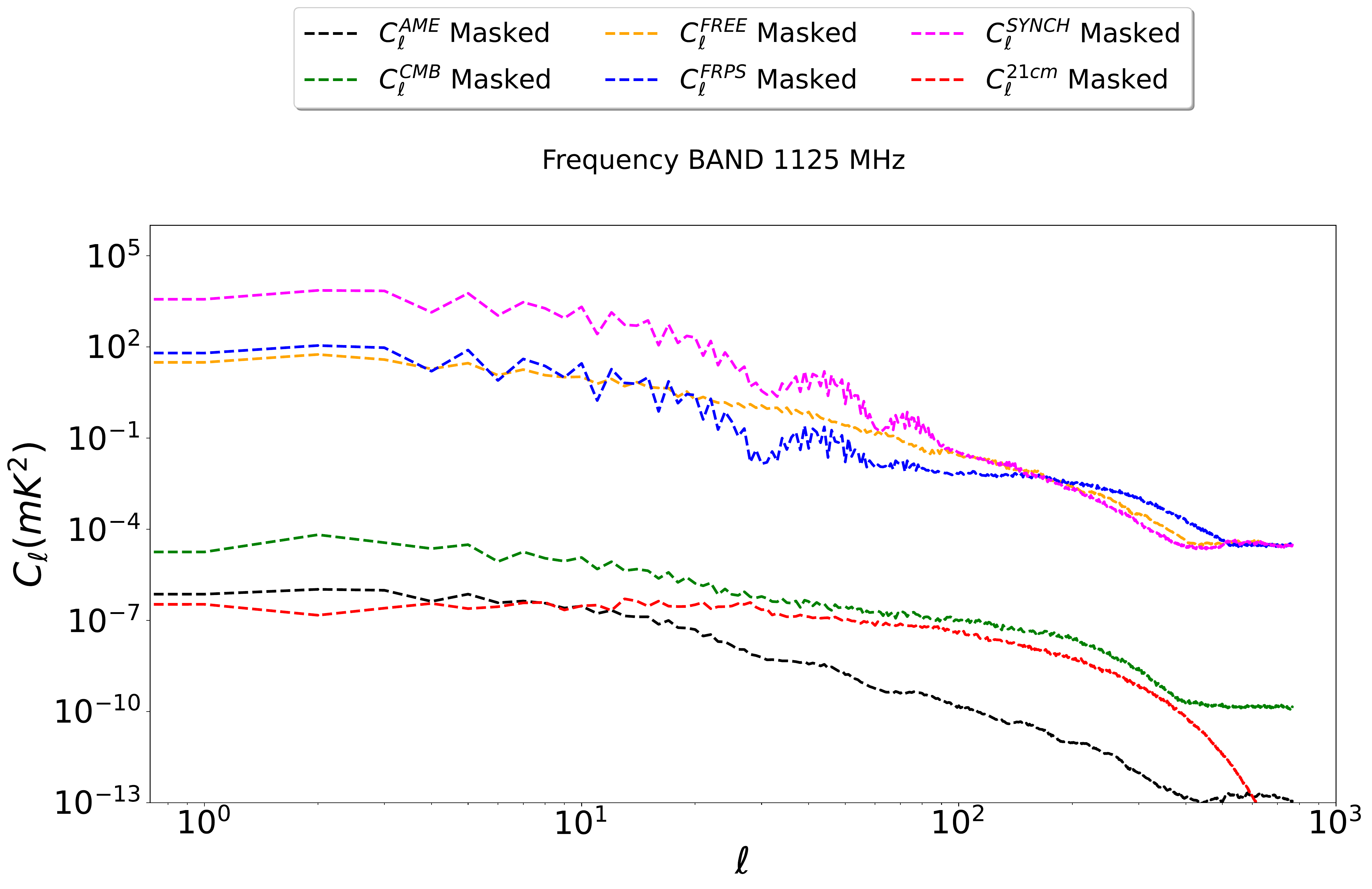}

\caption{Power spectra of the foregrounds and 21cm signal in the masked and beam-convolved sky (in the $\ell$ range of 0 to 300) in the BINGO frequency band 15 ($\sim 1125$\,MHz; see Table~\ref{tab:table1}). Before computing the power spectrum, all maps were convolved with a 40 arcmin beam, and masked according to the BINGO sky coverage. We also apply a galactic mask described in Sect.~\ref{sec:sky-coverage}. 
}
\label{fig:component_PS}
\end{figure}

Figures~\ref{fig:allcomponents} and \ref{fig:component_PS} show the maps of the components contributing to our simulated sky and their power spectra, respectively. As is evident from Fig.~\ref{fig:component_PS}, the large foreground contribution to the sky is a challenge for component separation and for the recovery of the 21cm signal. We need to suppress the foreground power down to the level of the expected 21\,cm spectrum, which implies a foreground rejection level better than one part in $10^4$ for the synchrotron, whose power  is about eight to nine orders of magnitude above that of the 21\,cm signal in the sky region observed by BINGO, after masking the brightest regions around the Galactic plane.

In the next subsections we briefly describe the five components mentioned above. For a more detailed description of the physics behind astrophysical foregrounds, see \cite{2020_project} and references therein.

\subsubsection{Galactic synchrotron}

Galactic synchrotron emission originates from interactions between the Galactic magnetic field and relativistic cosmic ray electrons. It is the dominant contaminant for the 21cm signal in the frequency range covered by BINGO. The synchrotron frequency dependence is modeled as a power law, in units of antenna (Rayleigh-Jeans) temperature, $T_{\rm sync}(\nu) \propto \nu^{-\beta_s}$. We use as a template the 408\,MHz all-sky map produced by \cite{Remazeilles:2015}, which is extrapolated to BINGO frequencies through a power law considering a non-uniform spectral index, $\beta_s$,  over the sky \citep{MivilleDeschenes:2008hn}.

In this simulation we do not include any effect due to polarized synchrotron emission. We  made this choice because BINGO will measure the intensity with horns that are specifically sensitive to the circular polarization that has a high attenuation to a linear polarization response. This attenuation is  on the order of 40dB, as  simulated in \cite{2020_instrument} and \cite{2020_optical_design}.

In other words, the BINGO collaboration chose to remove this contamination instrumentally, creating the horns described in \cite{2020_instrument} and  \cite{2020_optical_design} (mainly Figure 18), whose response is polarized, and using a cross-dragonian optical system that suppresses this type of polarization.

\subsubsection{Galactic free-free}

Free-free emission is produced by electron-ion interaction in the Galactic ISM. It is, together with synchrotron, the main source of contamination to the 21cm signal. The free-free emission can be traced by the H$\alpha$ emission line, since both depend on the emission measure $\text{EM} = \int n_{e}^{2}dl$. We simulated free-free emission following \cite{Dickinson:2003vp}, using their composite template of H$\alpha$ and a single spectral emission law, which is uniform over the sky, given by
\begin{equation}
\label{eq:ff-law}
    T_b(T_e,\nu) = 8.396 \times 10^3 \, a(T_e,\nu) \, T_4 ^{0.667} \times 10^{0.029/T_4} \times 1.08 \, \nu_{\rm GHz}^{-2.1},
\end{equation}
where $T_e$ is the electron temperature, $T_4=T_e/10^4$, and $a(T,\nu)$ is the Gaunt correction factor given by
\begin{eqnarray*}
    a(T_e,\nu) & = & 0.366 \, \nu_{\rm GHz}^{0.1} \, T_e^{-0.15} \\
    & & \times \left[ \ln{\left( 4.995 \times 10^{-2} \nu_{\rm GHz}^{-1} \right) } + 1.5 \ln{(T_e)}  \right].
\end{eqnarray*}
This free-free emission law is a slowly varying function of frequency, depending slightly on the electron temperature, which is assumed here to be constant ($T_e = 7000$\,K).

\subsubsection{Galactic anomalous microwave emission}

Dipole emission from spinning dust grains is currently the most commonly accepted explanation for the so-called dust-correlated anomalous microwave emission (AME). Radio emission is produced when the electric dipoles in small dust grains spin up. This component is mainly observed in the frequency range of 10--60\,GHz, and is subdominant in the BINGO frequency range. There is, however, significant uncertainty about the exact emission spectrum.



Here we use the {\tt PSM} prescription for the spinning dust, as described in \cite{Delabrouille:2012ye},  which is modeled using a spinning dust template map extrapolated in frequency using a single emission law. We use a 353\,GHz thermal dust template obtained using GNILC from the Planck satellite observations \citep{PlanckXLVIII}, which is scaled to the spinning dust emission at 22.8\,GHz on the basis of the scaling found in \cite{2016A&A...594A..25P}. To extrapolate to lower frequencies, the spinning dust emission law is parameterized following the model of \cite{Draine:1998gq}, using a mix of 96.2\% warm neutral medium and 3.8\% reflection nebulae. 
Our choice for the AME modeling differs somewhat from that in  companion Paper IV \citep{2020_sky_simulation}, which used the \textit{Planck} GNILC dust optical depth map $\tau_{353}$ \citep{PlanckXLVIII} instead of the GNILC dust intensity map at $353$\,GHz, and scaled it down from 22.8\,GHz to the BINGO frequencies of $\sim$ 1\,GHz using the publicly available {\tt spdust2} code instead of the Draine and Lazarian model as implemented in the PSM. 
The differences between the two are representative or current uncertainties. They do not matter much as for both models spinning dust emission is subdominant.
 
 \subsubsection{Cosmic microwave background}

Most of the electromagnetic radiation in the Universe is in the form of a near-isotropic background of thermal radiation originating from the time when free electrons and nuclei in the primordial plasma first combined to form neutral atoms, the cosmic microwave background (CMB). The spectrum of emission is very close to that of a blackbody at an average temperature of ${T_{\rm CMB} \simeq 2.726}$\,K, with small temperature fluctuations across the sky, at a level on the order of 100\,$\mu$K. In the BINGO frequency range, the corresponding brightness fluctuations are comparable in amplitude to those of the 21cm signal of interest (see Fig.~\ref{fig:allcomponents}). 

In our sky simulations, we generate random CMB temperature fluctuations with a harmonic power spectrum based on the best fit spectrum from \citet{Planck2018:cosmoparams}. We note, however, that CMB maps from the Planck satellite could also be used to subtract most of this component from the BINGO observations.



\subsubsection{Radio point sources}

In addition to emission from the diffuse interstellar medium in our own Galaxy, we must take into account emission from the background of distant compact radio sources, both Galactic and extragalactic. The most luminous of these objects can be identified as individual sources, while the rest contribute a diffuse background from the integrated emission of faint objects that cannot be detected individually. 

Although not resolved by BINGO, the brightest objects are gathered in a catalog that contains sources characterized by their parameters: type of the source, position on the sky, flux density, and polarization fraction as a function of frequency. In the model we adopt the objects in this catalog are described as a population of point sources. On the other hand, the diffuse background of sources that are not detected individually is gathered in the form of sky background inhomogeneities represented using frequency-dependent brightness fluctuation maps. Although individual sources are not kept in the format of a catalog at each frequency, maps are effectively produced by summing the contribution of a large population of sources. 

The population of radio sources is based on observations at 0.8\,GHz, 1.4\,GHz, and 4.85\,GHz. The emission of each source in the frequency range of interest for BINGO is modeled as a power law, with a distribution of spectral indices for steep and flat radio sources. Details of the radio source model can be found in the description of the Planck Sky Model \citep{Delabrouille:2012ye}.



Sources with flux densities above than the \textit{Planck} 5$\sigma$ detection limit in the 30, 70, 353, and 857\,GHz are considered  ``bright'' and are not included in the simulation as it is assumed that their contribution will be cut out or fitted out from the BINGO data.



\subsection{BINGO sky coverage}
\label{sec:sky-coverage}

\begin{figure}[tb]
\centering

\includegraphics[width=0.40\textwidth]{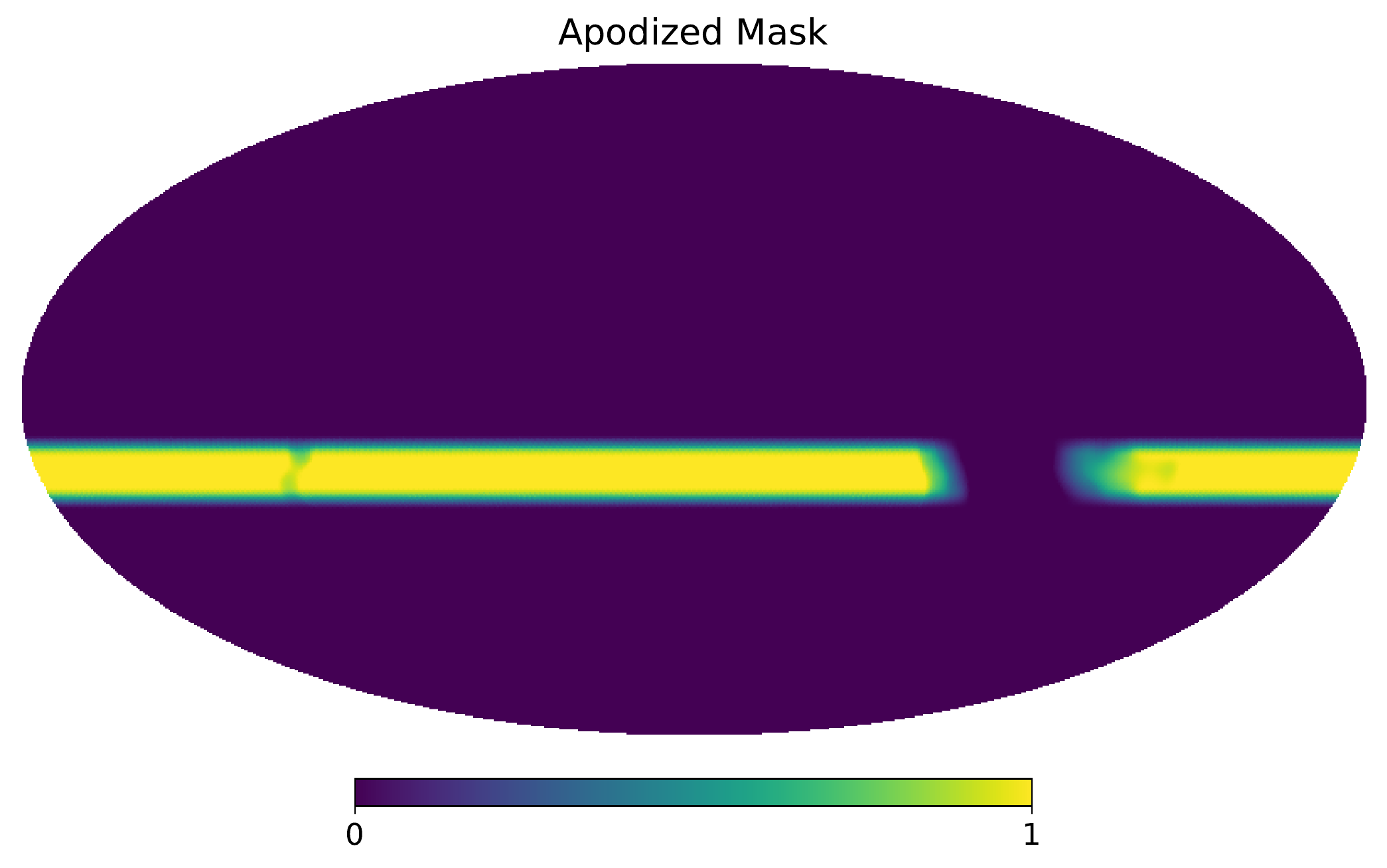}
\includegraphics[width=0.23\textwidth]{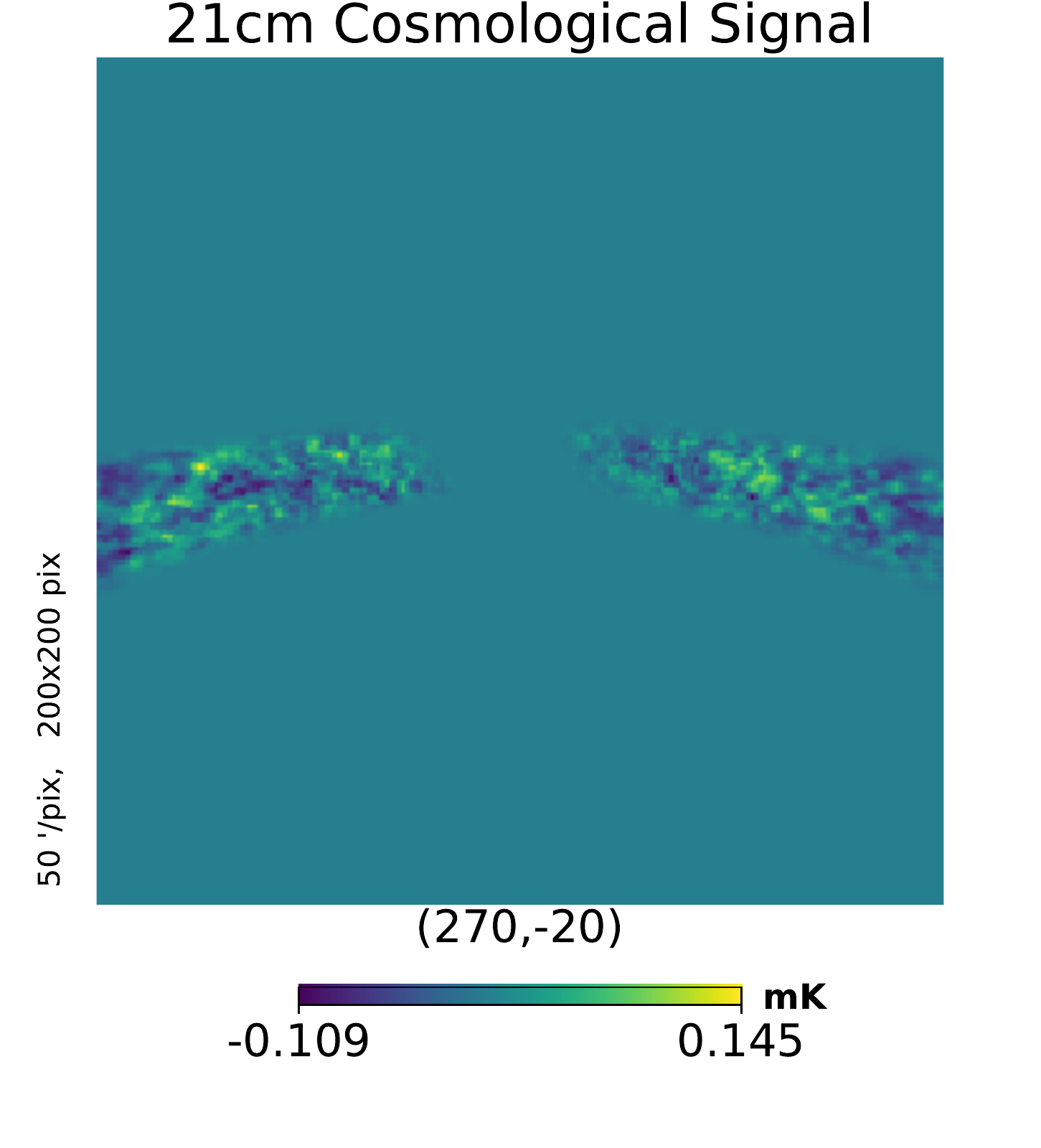}
\includegraphics[width=0.23\textwidth]{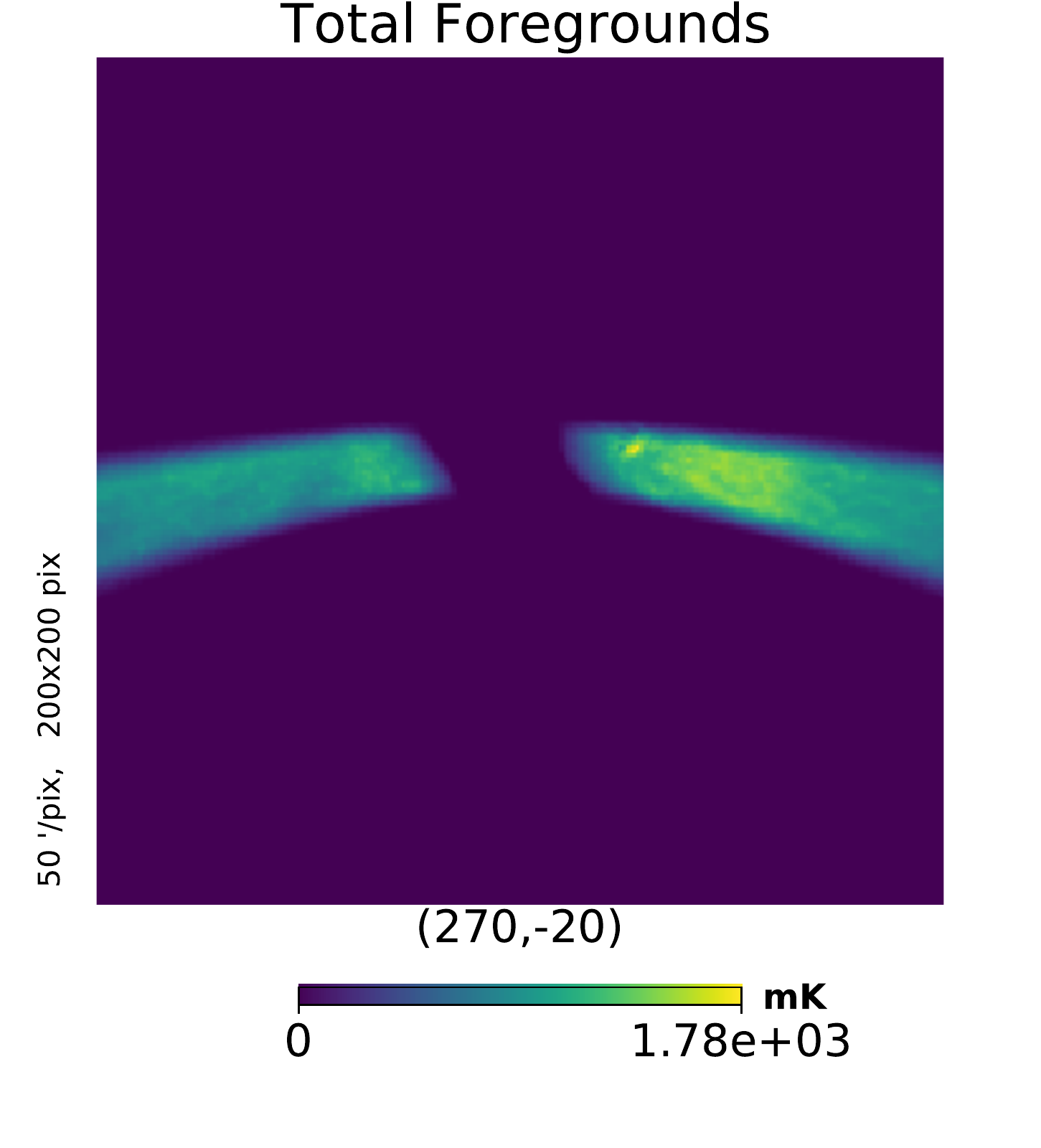}

\caption{ Effects of apodization in the maps. \textit{Top}: BINGO apodized mask format (5 degrees). The zeros in the scale correspond to non-observed spots and the transition between the observed and non-observed regions is done through the apodization (pixels with values $>0$ and $\le 1$). 
\textit{Bottom left}: Region centered $(\ell,b)= (270^{\circ}, -20^{\circ})$, where the mask was applied to the 21cm signal map (gnomonic projection).
\textit{Bottom right}: Same configuration as \textit{bottom left} for the total foreground map.
}
 \label{fig:maskmap3}

\end{figure}

After simulating full sky maps containing foregrounds and the 21cm signal, it is necessary to apply a mask to the maps in order to have the BINGO sky coverage assumed for this work. 
Our mask excludes pixels in the ranges of latitudes that are not observed by BINGO  (i.e., only the sky pixels for which $-22.5 \leq \delta \leq -7.5$ are kept). 



Inside the region observed by BINGO there are regions where the Galactic emission is too high to hope to detect the 21\,cm emission. We generate a Galactic mask using a smoothed version of the intensity map of the total Galactic emission in one of the BINGO channels. This mask is built using the following prescription: All observed pixels are sorted by decreasing Galactic emission amplitude, after smoothing with a ten-degree Gaussian beam. The brightest 10\%  are set to 0. The next 30\% are attenuated with a cosine apodization function $f(x)$ which smoothly increases from 0 when $x=10$\% to 1 when $x=40$\%.

Figure \ref{fig:maskmap3} shows the apodized mask map and gnomonic views of the foreground and 21cm cosmological signals after masking.

\subsection{BINGO noise simulation} 



As outlined by the instrument paper, paper II \citep{2020_instrument}, 
this work considers the BINGO Phase 1 configuration with 28 horns, where each horn observes the sky for a long period of time at a given elevation, which corresponds to a fixed declination. 
We assume that it is possible to change the elevation of the horns after a certain amount of observing time, as defined in papers III \citep{2020_optical_design} and IV \citep{2020_sky_simulation}
We also assume that each horn feeds two channels measuring $(I+V)/2$ and $(I-V)/2$, respectively, with a system temperature \mbox{$T_{\rm{sys}}=70$\,K} in each channel, where $I$ is the intensity and $V$ is the circular polarization. The value of 
$I$ is measured by summing the two outputs, and we analyze here only the $I$ maps.

The white noise level in the $I$ signal for a band of total width $\delta \nu$ is
 \begin{equation}\label{noise1}
 \begin{gathered}
 \sigma_{\rm{noise}} = K\,\frac{T_{\rm{sys}}}{\sqrt{\delta \nu}} \,s^{\frac{1}{2}}\,,
 \end{gathered}
 \end{equation}
 
\noindent where $K=\sqrt{2}$ for a correlation receiver and $\sigma$ can be related to the minimum detectable flux density for the telescope, see Eq. 3 in \citet{2020_instrument}, and the $s$ is in units of one over $\delta \nu$. 

Assuming that the observation time for $N_{\rm{horns}}$ horns is uniformly spread across the BINGO survey area, and a pixelized map with total number of pixels $N_{\rm{pix}}=12 N^{2}_{\rm{side}}$, with $N_{\rm{side}} =512$, the total white noise level per pixel for full observing time $\tau_{\rm{obs}}$ is
\begin{equation}\label{noise2}
\begin{gathered}
\sigma_{\rm{pix}} = \sigma_{\rm{noise}} \times \Bigg[\frac{f_{\rm{sky}} N_{\rm{pix}}}{N_{\rm{horns}} \tau_{\rm{obs}}}\Bigg]^{\frac{1}{2}}\,.
\end{gathered}
\end{equation}

We assume that for our survey $f_{\rm{sky}}=15\%$, $T_{\rm{sys}}=70$ K, $N_{\rm{horns}} = 28$, and $\tau_{\rm{obs}} = 1$ month gives a noise per pixel at \mbox{$N_{\rm{side}} =512$} (for white noise estimation), which  corresponds to a noise power in each of the 30 BINGO channels of
\begin{equation}\label{noise3}
\begin{gathered}
 N_{l} = 4\pi \frac{\sigma_{\rm{pix}}^{2}}{N_{\rm{pix}}} = 10.64\, \mu \text{K}^{2}\,
\end{gathered}
\end{equation}
across multipoles $\ell$.

We note  that in the limit where the number of data samples in the BINGO timeline is much larger than the number of pixels, each pixel value is obtained as the average of data samples uniformly spread over the pixel. The signal in each pixel is thus the integral of the signal coming from beams centered at points distributed over the pixel area. 
The signal is thus effectively convolved with the beam and with the pixel shape, and the power spectrum is multiplied by the square of the beam transfer function $B_{l}^{2}$ and the pixel window function $W_{l}$. The noise spectrum is directly proportional to $W_{l}$. 

Two complications must be taken into account for an observation with a non-ideal instrument. The first is the presence of low frequency (1/f)  noise in the timestreams. We do not consider  low frequency noise in this paper and leave it for future analysis. The second is the non-uniformity of the coverage because of the layout of the BINGO horns in the focal plane. This second issue has been solved with the vertical displacement of the horns in the focal plane as described in the optics and simulation papers \citep{2020_optical_design,2020_sky_simulation}. 



\begin{figure}[h]
\begin{center}
    \includegraphics[width=0.23\textwidth]{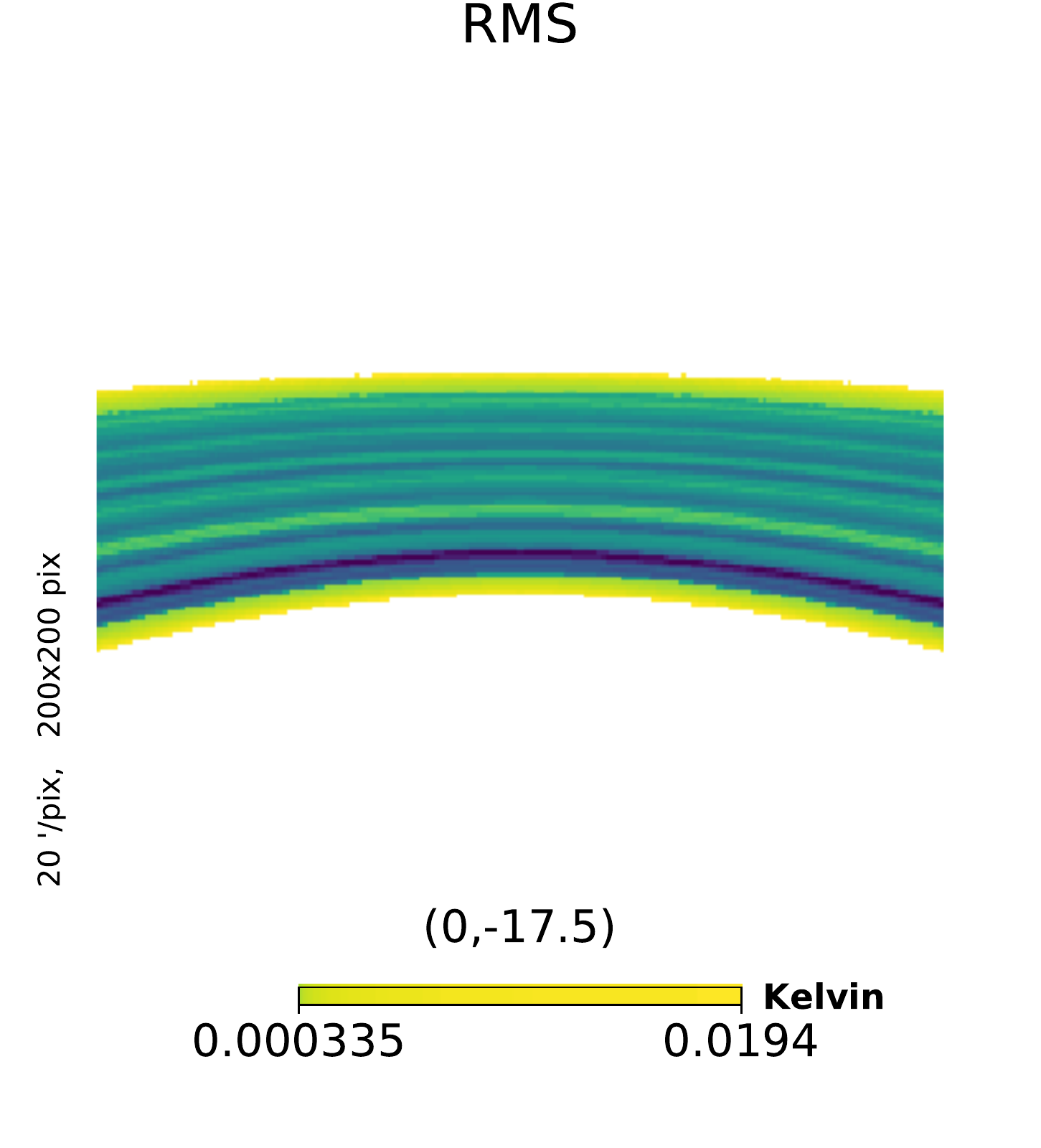}
    \includegraphics[width=0.23\textwidth]{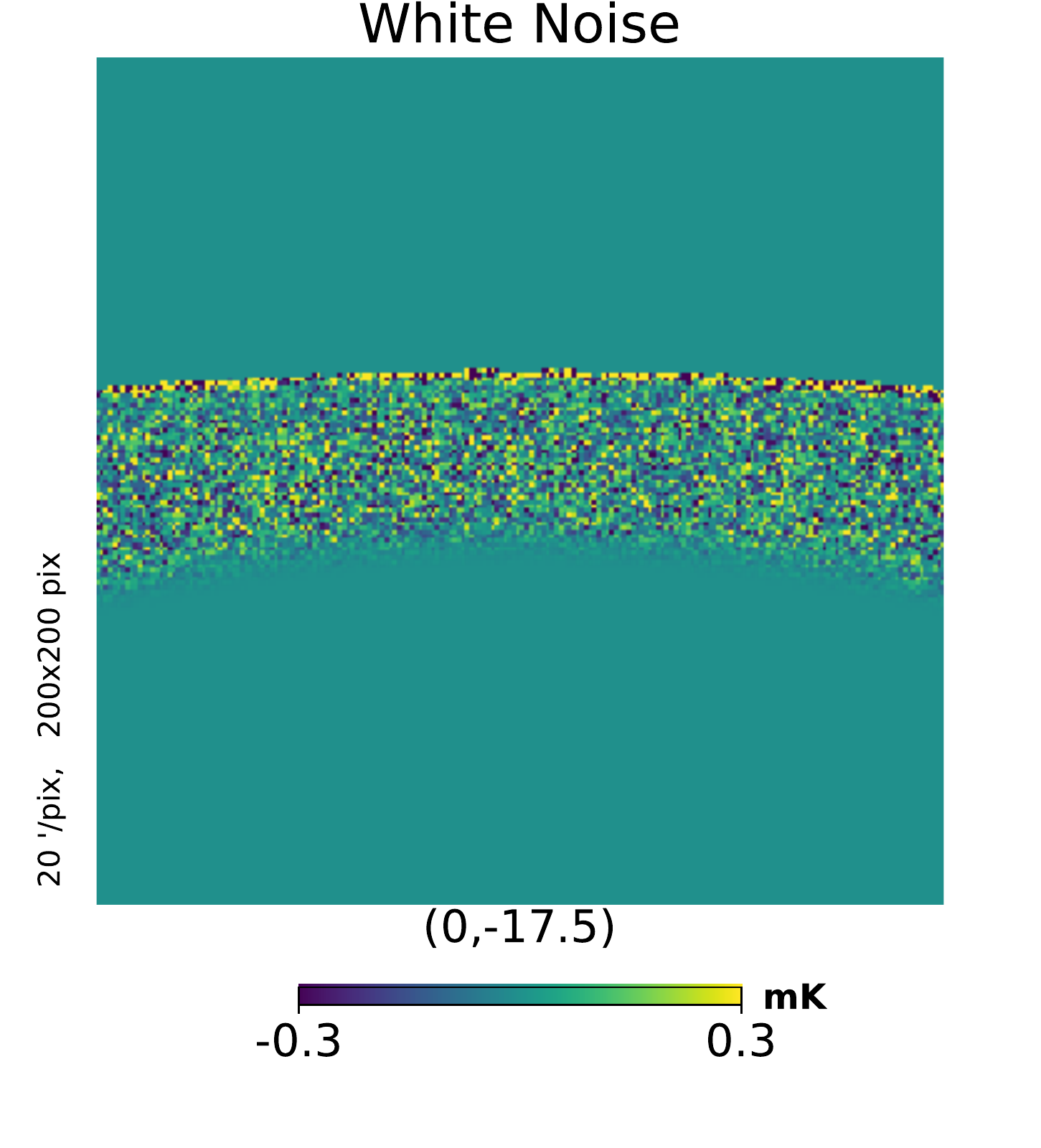}
\end{center}
\caption{
Noise realizations for the "double retangular" horn arrangements, in Healpix gnomonic coordinates centered in ($\alpha=0^{\circ}, \delta = -17.5^{\circ}$).
\textit{Left}: White noise realization after multiplying the r.m.s. by a Gaussian map, using $N_{\rm{side}}=256$. 
\textit{ Right}: R.m.s. realization for the double-rectangular configuration (with time spread between five different horn offsets) for two years of observation with a 70\,K system temperature at {\tt HEALpix} $N_{\rm{side}}=512$. (The r.m.s. maps are produced at a higher {\tt HEALpix} pixelization and then degraded to the working resolution discussed in the text). Map of the corresponding r.m.s. of the projected white noise part. The color scale is saturated at five times the r.m.s. of a map with homogeneous coverage and same sky fraction.   
} 
\label{fig:noise6}
\end{figure}

This work uses the  ``double rectangular'' horn arrangement described in \cite{2020_optical_design}, with 28 horns. Figure \ref{fig:noise6} shows an example noise realization for this configuration. Overall, we are able to produce reasonably homogeneous noise maps, apart from edge effects above the minimum and maximum declination covered with this horn arrangement, which should be masked in the final map analysis. 

\section{Component separation analysis}\label{sec:powerspectum}

Observations of the radio sky encompass a mixture of cosmological signals emitted in the early Universe (e.g., CMB and cosmological \hi\, signal), astrophysical sources emitting in the late Universe (e.g., Galactic foregrounds and extragalactic point sources), and instrumental signals (e.g., thermal noise). Component separation is a term that refers to any data processing that tries to disentangle these emissions by exploiting correlations in observations made at separate frequencies, external constraints, and/or physical models of the different sources of emission \citep{Delabrouille2007}.

Component separation techniques can be divided into two categories: parametric methods and blind (or nonparametric) methods.
Parametric methods assume a spectral model for the foregrounds, while blind methods use only the observed data to recover the cosmological signal, and therefore do not make any assumption about the foregrounds. We note that in this definition of blind methods only the use of prior information about the cosmological signal is allowed.

The component separation problem is particularly relevant to IM because the observed signal is dominated by astrophysical emission, both from our Galaxy and from extragalactic sources. The removal (or the mitigation) of the astrophysical foreground contamination is a fundamental step in IM data analysis. Given that the spectral signature of the cosmological signal is nontrivial, one of the difficulties for IM component separation is the preservation of the 21cm signal during foreground removal. Either an excess of foreground residuals in the 21cm maps or an oversubtraction of the foregrounds from the data will result in erroneous cosmological results.

Several component separation methods have been proposed in the literature to disentangle cosmological \hi\, 21cm emission and astrophysical foregrounds. Many of them rely on the assumption that foregrounds are spectrally smooth.
However, the smoothness assumption may be broken by instrumental effects of the telescope, such as standing waves and calibration uncertainties, although the design of the BINGO telescope has the goal of minimizing these effects. 

Slight departures from spectral smoothness of the observed foregrounds may result in biases on the detection of the faint \hi\ signal for some methods such as COMMANDER \citep{Eriksen:2004ss,Eriksen:2007mx} or wp-FIT \citep{2009MNRAS.397.1138H}, given the huge dynamic range in amplitude between foregrounds and \hi\, emission. 
In the case of the method GNILC \citep{Olivari2015} that we adopt in our analysis and describe below, no assumption is made on the spectral shape of the foregrounds. To disentangle the cosmological 21cm signal from the foreground emission, GNILC relies only on the property that the foreground emission is much more strongly correlated across frequencies than the 21cm signal is.



\subsection{GNILC methodology}

In this work we use {\tt GNILC}, a blind component separation method originally devised for CMB data analysis by \cite{Remazeilles2011} and extended to 21cm data analysis by \cite{Olivari2015}. The main idea of {\tt GNILC} is to use prior information on the power spectrum of the cosmological signal to either compensate for the lack of knowledge on the frequency dependence of the targeted signal, as  in the case for the 21cm signal \citep{Olivari2015}, or to overcome spectral degeneracies between components, as in the case for example between cosmic infrared background and thermal dust \citep{PlanckXLVIII}.


Hence, in our case, the only necessary ingredient to {\tt GNILC} is a theoretical \hi\ 21cm power spectrum across the redshift bins as a prior
to the algorithm:  $C_\ell^{\rm 21cm, prior}(\nu)$. We also include additional noise information in the {\tt GNILC} prior, so the prior can be described by a power spectrum given by  ${C_\ell^{\rm 21cm+noise,\, prior}(\nu)\equiv C_\ell^{\rm 21cm, \,prior}(\nu) + C_\ell^{\rm noise,\, prior}(\nu)}.$ No assumption is made about the foregrounds.


The observed data by BINGO $d_{\nu}(\hat{n})$ in any direction of the sky $\hat{n}$ (or pixel in the map) and at any frequency $\nu$ are a mixture of signal, foregrounds, and noise 
\begin{equation}\label{eq:debias_Eq1}
d_{\nu}(\hat{n})= s^{\rm 21cm}_{\nu}(\hat{n}) + s^{\rm fg}_{\nu}(\hat{n}) +n_{\nu}(\hat{n}) \,,
\end{equation}
where $s^{\rm 21cm}_{\nu}(\hat{n})$ is the 21cm signal at frequency $\nu$ that we aim to recover with {\tt GNILC}, $s^{\rm fg}_{\nu}(\hat{n})$ denotes the total foreground emission at that same frequency, and $n_{\nu}(\hat{n})$ is the instrumental noise in this channel. 

The formalism of {\tt GNILC} has been described in detail in the literature \citep{Remazeilles2011,PlanckXLVIII,Olivari2015}. We refer the reader to \cite{Olivari2015} for a full description of the method in the context of 21cm intensity mapping. We summarize it below.



We first define a set of functions in harmonic space, $h^{(j)}_\ell$ ($j=1,...,4$), called needlet windows, which work as bandpass filters to handle different ranges of angular scales in the maps independently for component separation. The four needlet windows are chosen to satisfy the constraint

\begin{equation}\label{eq:needlets}
\sum_{j=1}^4 \left(h^{(j)}_\ell\right)^2=1
\end{equation}
over the full multipole range in order to conserve the total power of the sky emission in the data processing. While choosing more than four needlet windows to partition the multipole range would allow more localized filtering in harmonic space, by the uncertainty principle the support of the filter would in contrast be less compact in pixel space and exceed the size of the small area of sky observed by BINGO. To allow sufficient localization of the filter in pixel space and capture the variations of the foreground contamination inside the BINGO stripe, we have to relax localization in harmonic space by limiting the partitioning to four needlet windows.

The spherical harmonic coefficients of each BINGO channel map $d_{\ell m}(\nu)$ are bandpass-filtered through each needlet window as $\tilde{d}^{\,(j)}_{\ell m}(\nu) = h^{(j)}_\ell\, d_{\ell m}(\nu)$. An inverse spherical harmonic transform of the bandpass-filtered coefficients thus provides four needlet maps 
$\tilde{d}^{(j)}_\nu(\hat{n})$ 
at each frequency. These needlet maps contain only temperature fluctuations of typical angular scales selected by the needlet window.

In the second step, for each needlet scale $(j)$ we compute the $n\times n$ data covariance matrix of the needlet maps in each pixel $\hat{n}$ for all pairs of frequencies $(\nu,\nu')$ as
\begin{equation}\label{eq:datacov}
\widehat{\textbf{R}}^{(j)}_{\nu\nu'}(\hat{n}) = \sum_{\hat{n}' \in \mathcal{D}(\hat{n})}  d^{(j)}_\nu(\hat{n}') d^{(j)}_{\nu'}(\hat{n}')\,, 
\end{equation} 
where $\mathcal{D}(\hat{n})$ is a domain of pixels centred around pixel $\hat{n}$, chosen in such a way to avoid artificial correlations between the signal and the foregrounds at large angular scales where the statistics is poor (the so-called ILC bias, see \citealt{delabrouille2009full}).

Third, using our prior estimate of the 21cm (+ noise) power spectrum, $C_\ell^{\rm 21cm+noise,\, prior}(\nu)$, we simulate 21cm (+ noise) maps $y_\nu(\hat{n})$ for each channel $\nu$. These simulated 21cm (+ noise) maps $y_\nu(\hat{n})$ receive the same needlet bandpass filtering as the data, thus leading to four needlet maps $y^{(j)}_\nu(\hat{n})$ for each frequency. As in Eq.~\ref{eq:datacov}, for each needlet scale $(j)$ we compute the prior signal (+ noise) covariance matrix as
\begin{equation}\label{eq:prior}
\widehat{\textbf{R}}^{(j)}_{\mathrm{S} \, \nu\nu'} (\hat{n}) = \sum_{\hat{n}' \in \mathcal{D}(\hat{n})}  y^{(j)}_\nu(\hat{n}') y^{(j)}_{\nu'}(\hat{n}')\,.
\end{equation} 
We note that this prior is independent of the particular realization of the 21cm signal in the observed sky.

In step four, as evident from Eq.~\ref{eq:debias_Eq1}, the data covariance matrix (Eq.~\ref{eq:datacov}) must receive contributions from signal, foregrounds, and noise covariance matrices as
\begin{eqnarray}
\widehat{\textbf{R}} & = & \textbf{R}_{\rm 21cm} + \textbf{R}_{\rm noise} + \textbf{R}_{\rm fg} \nonumber \\
& = & \textbf{R}_{\mathrm{S}} + \textbf{R}_{\rm fg}\,,
\end{eqnarray} 
where we omitted the implicit pixel, frequency, and needlet-scale indices to reduce the amount of notations, and defined ${\textbf{R}_{\mathrm{S}}\equiv \textbf{R}_{\rm 21cm} + \textbf{R}_{\rm noise}}$ as the covariance matrix of 21cm signal plus noise.

Hence, the whitened data covariance matrix, defined as ${\widehat{\textbf{R}}_{\mathrm{S}}^{-1/2}\, \widehat{\textbf{R}}\, \widehat{\textbf{R}}_{\mathrm{S}}^{-1/2}}$, where $\widehat{\textbf{R}}_{\mathrm{S}}$ is the prior covariance matrix (Eq.~\ref{eq:prior}), reduces to
\begin{equation}\label{eq:whitecov}
\widehat{\textbf{R}}_{\mathrm{S}}^{-1/2}\, \widehat{\textbf{R}}\, \widehat{\textbf{R}}_{\mathrm{S}}^{-1/2} \simeq \widehat{\textbf{R}}_{\mathrm{S}}^{-1/2}\, \textbf{R}_{\rm fg}\, \widehat{\textbf{R}}_{\mathrm{S}}^{-1/2}\,+\,\textbf{I}\,,
\end{equation} 
where $\textbf{I}$ is the identity matrix. The eigenvalue decomposition of matrix Eq.~\ref{eq:whitecov} thus yields
\begin{equation}\label{eq:eigen}
\widehat{\textbf{R}}_{\mathrm{S}}^{-1/2}\, \widehat{\textbf{R}}\, \widehat{\textbf{R}}_{\mathrm{S}}^{-1/2} \simeq \textbf{U}_N\, \textbf{D}_N\, \textbf{U}_N^{\rm T}\, +\, \textbf{U}_S\, \textbf{U}_S^{\rm T}\,, 
\end{equation} 
where $\textbf{D}_N$ collects the $m$ largest eigenvalues departing from unity, $\textbf{U}_N$ the corresponding eigenvectors, and $\textbf{U}_S$ the $(n - m)$ eigenvectors whose eigenvalue is close to unity. 

The decomposition in Eq.~\ref{eq:eigen} enables {\tt GNILC} to identify the signal and foreground subspaces in the data, since the $m$ eigenvalues collected in the $m\times m$ diagonal matrix $\textbf{D}_N$ indicate significant power from the foregrounds in the data, while the $n-m$ eigenvalues of matrix that are close to unity correspond to power from the 21cm signal (+noise) in the data. Hence, the $m$ eigenvectors collected in the $n\times m$ matrix $\textbf{U}_N$ form the principal components of the foreground subspace, while the $n-m$ eigenvectors collected in the $n\times (n-m)$ matrix $\textbf{U}_S$ form the independent components of the targeted 21cm signal subspace.

In step five, unlike PCA, the effective dimension $m$ of the foreground subspace is not pre-defined in an ad hoc 
manner by {\tt GNILC}, but estimated directly from the data using Eq.~\ref{eq:eigen} and minimizing the Akaike information criterion (AIC), which in the {\tt GNILC} formalism reduces to solving \citep[e.g.,][]{Olivari2015}
\begin{equation}\label{eq:aic}
\mathrm{min} \; \left(2m + \sum_{i=m+1}^{n} [\mu_i - \log \mu_i - 1] \right) \;\;\; \mathrm{with} \;\;\; m \in [1, n]\,, 
\end{equation} where $\mu_i$ are the eigenvalues of matrix $\widehat{\textbf{R}}_{\mathrm{S}}^{-1/2}\, \widehat{\textbf{R}}\, \widehat{\textbf{R}}_{\mathrm{S}}^{-1/2}$ (Eq.~\ref{eq:whitecov}). The foreground dimension $m$, which minimizes the AIC criterion, is denoted $m_\mathrm{AIC}$ in the next sections.

The foreground dimension $m_\mathrm{AIC}$ is estimated by the AIC locally across the sky and across the angular scales thanks to needlet decomposition. Hence, unlike PCA, {\tt GNILC} also allows the effective dimension of the foreground subspace $m_\mathrm{AIC}\equiv m_\mathrm{AIC}^{(j)}(\hat{n})$ to vary with the needlet scale $(j)$ and the position in the sky $\hat{n}$ depending on the local signal-to-foregrounds ratio in the data (Eq.~\ref{eq:whitecov}). For the same reason, the $n\times (n-m_\mathrm{AIC})$ matrix $\textbf{U}_S\equiv \textbf{U}_S^{(j)}(\hat{n})$, which spans the targeted 21cm signal subspace, varies across the sky and the scales.

In the sixth step, for each needlet scale $(j)$, each pixel $\hat{n}$, and each frequency $\nu$, an estimate of the 21cm signal $\widehat{s}_\nu^{\,{\rm 21cm} \, (j)}(\hat{n})$ is obtained by applying a multi-dimensional ILC filter to the data
\begin{align}
\widehat{s}_\nu^{\,{\rm 21cm} \, (j)}(\hat{n}) &= \sum_{\nu'} \textbf{W}^{(j)}_{\nu \nu'}(\hat{n})\, d^{(j)}_{\nu'}(\hat{n})\,,
\end{align}
where the expression for the {\tt GNILC} weights $\textbf{W}\equiv\textbf{W}^{(j)}_{\nu \nu'}(\hat{n})$ is given by
\begin{align}
\label{eq:filter_ilc}
\textbf{W}=\widehat{\textbf{S}} \,(\widehat{\textbf{S}}^{\rm T}\, \widehat{\textbf{R}}^{-1}\, \widehat{\textbf{S}} )^{-1}\, \widehat{\textbf{S}}^{\rm T} \widehat{\textbf{R}}^{-1}
\end{align} 
and the signal mixing matrix is given by
\begin{align}
\label{mixmat}
\widehat{\textbf{S}} = \widehat{\textbf{R}}_{\mathrm{S}}^{1/2} \textbf{U}_S\,. 
\end{align} 
The mixing matrix (Eq.~\ref{mixmat}) is the only information needed to implement the {\tt GNILC} filter (Eq.~\ref{eq:filter_ilc}). 

The exact amplitude of the 21cm prior $\widehat{\textbf{R}}_{S}$ is not critical for {\tt GNILC} because it could be multiplied by a constant factor while leaving Eq. \ref{eq:filter_ilc} unchanged. This is only true as long as this constant factor is not large enough to modify the dimension of the matrix $\textbf{U}_S$. 

Finally, we synthesize the needlet-map estimates $\widehat{s}_\nu^{\,{\rm 21cm} \, (j)}(\hat{n})$ of the 21cm signal to form the complete 21cm map that includes all scales. First, we compute the spherical harmonic coefficients $\widehat{s}_{\ell m}^{\,{\rm 21cm} \, (j)}(\nu)$ of the 21cm map estimates. These coefficients are again bandpass filtered by the needlet windows (this step guarantees the normalization of the maps), and the filtered coefficients are transformed back to real space by inverse spherical harmonic transform. This operation gives one reconstructed 21cm map per needlet scale and per frequency channel. The reconstructed 21cm map per needlet scales are finally co-added to give the complete {\tt GNILC} 21cm map, $\widehat{s}_\nu^{\,{\rm GNILC}}(\hat{n})$ , for each frequency channel.






To perform our analysis, we generated a cube with 30 BINGO channels and/or redshift bins for the simulation of the 21cm signal, with the 21cm signal having a different seed for each frequency channel. Simulated foregrounds and noise at BINGO frequency channels were then added to the corresponding redshift slices of the 21cm cube.
This cube is labeled as $L_0$ simulation. We also generated 100 cubes with different realizations of 21cm signal and noise for our debiasing procedure, which is  described in Sect.~\ref{subsec:debiasing}.

We apply {\tt GNILC} to the BINGO channel maps of the $L_0$ simulation to extract maps of the 21cm emission at each frequency $\widehat{s}_\nu^{\,{\rm GNILC}}$ with reduced foreground contamination, and reconstruct the 21cm power spectrum for each redshift bin $C_\ell^{\rm GNILC}(\nu)$. 
Error bars on the reconstructed power spectra are computed analytically for each channel $\nu$
\citep[e.g.,][]{Tristram:2004if}
\begin{equation}
\label{eq:errorbar}
\sigma_\ell(\nu) = \sqrt{ \frac{2}{  (2\ell +1)\Delta\ell f_{\rm sky} }} C_\ell^{\rm GNILC}(\nu)\,,
\end{equation}
where $\ell$ is the central multipole of the bin, 
$\Delta\ell$ is the bin size, $f_{\rm sky}$ is sky fraction of the BINGO survey, and $C_\ell^{\rm GNILC}(\nu)$ is the binned power spectrum of the {\tt GNILC} 21cm map in that channel. The uncertainty calculated by Eq.~\ref{eq:errorbar} thus includes cosmic variance from the 21cm signal plus contributions from residual foregrounds and noise.







\subsection{Additive and multiplicative errors}

The set of reconstructed 21cm maps by {\tt GNILC} across BINGO channels $\mathbf{s}^{\rm GNILC}=\{s_\nu^{\rm GNILC}(\hat{n})\}_\nu$  contains residual contamination by foregrounds and noise
\begin{align}\label{eq:debias_Eq3}
\mathbf{s}^{\rm GNILC}&=\mathbf{W}\,\mathbf{d}\,\cr
&=\textbf{W}\,\textbf{s}^{\rm 21cm} + \textbf{W}\,\textbf{s}^{\rm fg}+ \textbf{W}\,\textbf{n}\,,
\end{align}
where $\textbf{W}\,\textbf{s}^{\rm FG}$ is the residual foreground contribution and $\textbf{W}\,\textbf{n}$ is the residual noise contribution to the reconstructed 21cm maps after component separation. These additive errors are in principle minimized by the {\tt GNILC} filter Eq.~\ref{eq:filter_ilc}. In addition, the {\tt GNILC} filter Eq.~\ref{eq:filter_ilc} is built to ensure that $\textbf{W}\textbf{s}^{\rm 21cm} \simeq \textbf{s}^{\rm 21cm}$, so that $\tt GNILC$ fully recovers the 21cm signal with minimum foregrounds and noise:



\begin{equation}\label{eq:debias_Eq5}
\textbf{s}^{\rm GNILC} \simeq \textbf{s}^{\rm 21cm} + \textbf{W}\,\textbf{s}^{\rm fg}+ \textbf{W}\,\textbf{n}\,.
\end{equation}
However, a small part of the 21cm signal can be removed along with the foregrounds by the filtering because the signal and foreground subspaces are not fully orthogonal in the eigenvector decomposition outlined in Eq.~\ref{eq:eigen}. 


Hence, in practice the reconstructed 21cm signal may suffer from a small multiplicative error or bias $b$ (i.e., $\textbf{W}\,\textbf{s}^{\rm 21cm} \simeq b\,\textbf{s}^{\rm 21cm}$) and
\begin{equation}\label{eq:debias_Eq6}
\textbf{s}^{\rm GNILC} \simeq b\, \textbf{s}^{\rm 21cm} + \textbf{W}\,\textbf{s}^{\rm fg}+ \textbf{W}\,\textbf{n}\,,
\end{equation}
with $b < 1$. The risk of partial loss of the 21cm signal is common to all 21cm foreground removal techniques.

Therefore, the reconstructed power spectrum of the {\tt GNILC} 21cm map $C_{\ell}^{\rm GNILC}(\nu)$, at a given frequency $\nu$, may have a small multiplicative error on the 21cm signal through a suppression factor $S_\ell < 1$
\begin{equation}\label{eq:debias_Eq7}
C_{\ell}^{\rm GNILC} = S_\ell\, C_{\ell}^{\rm 21cm} + C_{\ell}^{\rm fg,\, proj}+ C_{\ell}^{\rm noise,\, proj}\,,
\end{equation}
along with additive errors due to projected noise ${C_\ell^{{\rm noise, proj}} \equiv C_\ell (\textbf{W}\,\textbf{n})}$ and residual foreground contamination  ${C_\ell^{{\rm fg, proj}} \equiv C_\ell(\textbf{W}\,\textbf{s}^{\rm fg})}$.

\subsection{Debiasing the power spectrum from noise bias and 21cm signal loss}\label{subsec:debiasing}

We first correct the {\tt GNILC} map power spectrum $C^{{\rm GNILC}}_\ell(\nu)$ for the residual noise bias by estimating the projected noise power spectrum in Eq.~\ref{eq:debias_Eq7} as follows.


Using the BINGO specifications, we generated $100$ realizations of white noise maps $n_{\nu}^{(i)}(\hat{n})$ ($1 \leq i \leq 100$) for each channel $\nu$. We found that using 100 simulations was enough for a suitable bias subtraction to be obtained for the purposes of this paper; however, we cannot guarantee that a covariance arising from such simulations would not have errors that are large enough for the purposes of parameter fitting. It is beyond the scope of this paper to assess the covariance errors from any lack of simulations.


We  compute the projected noise realizations $n_{\nu}^{(i),\,\text{proj}}(\hat{n})$ by applying the {\tt GNILC} filter Eq.~\ref{eq:filter_ilc} of the fixed sky realization $L_0$ to the white noise map realizations:
\begin{equation}
\label{eq:debias_Eq16}
n_{\nu}^{(i), \,\text{proj}}(\hat{n}) =\sum_{\nu^{\prime}}W_{\nu,\nu^{\prime}}\,n_{\nu^{\prime}}^{(i)}(\hat{n})\,.
\end{equation}
We compute the power spectra of the projected noise realizations (Eq.~\ref{eq:debias_Eq16}) (i.e., $\widehat{C}_\ell^{\rm noise\, (i), proj}(\nu) \equiv C_\ell(n_\nu^{\rm (i), proj})$) for each realization $(i)$ and each frequency channel $\nu$. We then average over all the $N_\mathrm{SIM}$ realizations to get an estimate of the projected noise power spectrum in Eq.~\ref{eq:debias_Eq7}:
\begin{equation}
\label{eq:debias_Eq17}
\widehat{C}_{\ell}^{\,\text{noise},\,\text{proj}}(\nu) = \langle \widehat{C}_{\ell}^{\,\text{noise\,(i)},\,\text{proj}}(\nu) \rangle = {1\over N_\mathrm{SIM}}\sum_{i=1}^{N_\mathrm{SIM}} \widehat{C}_{\ell}^{\,\text{noise\,(i)},\,\text{proj}}(\nu)\,.
\end{equation}
Finally, we subtract the estimated projected noise power spectrum (Eq.~\ref{eq:debias_Eq17}) to the {\tt GNILC} power spectrum (Eq.~\ref{eq:debias_Eq7}) as
\begin{equation}
\label{eq:debias_Eq18}
\widehat{C}_{\ell}^{\,\rm GNILC}(\nu) = C_{\ell}^{\rm GNILC}(\nu) - \widehat{C}_{\ell}^{\,\text{noise},\,\text{proj}}(\nu)\,.
\end{equation}
The resulting power spectrum $\widehat{C}_{\ell}^{\,\rm GNILC}(\nu)$ is thus corrected for the noise bias, so that
\begin{align}
\label{eq:denoised}
\widehat{C}_{\ell}^{\,\rm GNILC}(\nu) &\simeq S_\ell\, C_{\ell}^{\rm 21cm}(\nu) + C_{\ell}^{\rm fg,\, proj}(\nu),\cr
& \simeq S_\ell\, C_{\ell}^{\rm 21cm}(\nu)\,,
\end{align}
where in the second line of Eq.~\ref{eq:denoised} we neglect the residual foreground power that is already strongly mitigated by {\tt GNILC}.

We then need to correct the reconstructed 21cm power spectrum $\widehat{C}_{\ell}^{\,\rm GNILC}(\nu)$ for the multiplicative bias as
\begin{equation}
\label{eq:debias_Eq19}
    \widetilde{C}^{{\rm GNILC}}_\ell(\nu) = {\widehat{C}^{{\rm GNILC}}_\ell(\nu) \over \widehat{S}_\ell} \equiv {C^{{\rm GNILC}}_\ell(\nu) - \, \widehat{C}_\ell^{\,{\rm noise, proj}}(\nu) \over \widehat{S}_\ell}\,,
\end{equation}
where $\widehat{S}_\ell$ is an estimate of the 21cm suppression factor $S_\ell$, which we compute as follows.

Using the prior on the 21cm power spectrum, we generate $100$ pure 21cm map realizations $s_\nu^{\rm 21cm\, (i)}(p)$ ($1\leq i \leq 100$) for each channel $\nu$, which we then pass through the {\tt GNILC} filter of the fixed sky realization $L_0$ to obtain the projected 21cm map realizations
\begin{equation}\label{eq:debias_Eq20}
s_\nu^{\rm 21cm\, (i), proj} = \sum_{\nu'} W_{\nu \nu'}\, s_{\nu'}^{\rm 21cm\,(i)}\,.
\end{equation}
For each realization $(i)$ we compute the ratio of the power spectra of the projected to the input 21cm realizations
\begin{equation}
S^{(i)}_\ell(\nu) =  {C_\ell^{\rm 21cm\, (i), proj}(\nu) \over C_\ell^{\rm 21cm\, (i)}(\nu)}\,,
\end{equation}
in other words, the suppression factors for all realizations $(i)$ and frequency channels $\nu$. By averaging over all realizations we then get an overall estimate of the 21cm suppression factor for a given channel
\begin{equation}
 \widehat{S}_\ell(\nu) = \langle S^{(i)}_\ell(\nu) \rangle\,,
\end{equation}
which we use to renormalize the noise-debiased {\tt GNILC} power spectrum $\widehat{C}_{\ell}^{\rm GNILC}(\nu)$ as ${\widetilde{C}^{{\rm GNILC}}_\ell(\nu) = \widehat{C}_{\ell}^{\rm GNILC}(\nu) / \widehat{S}_\ell(\nu)}$ following the prescription in Eq.~\ref{eq:debias_Eq19}.

\begin{figure*}[t]
    \centering
    \includegraphics[width=17.5cm]{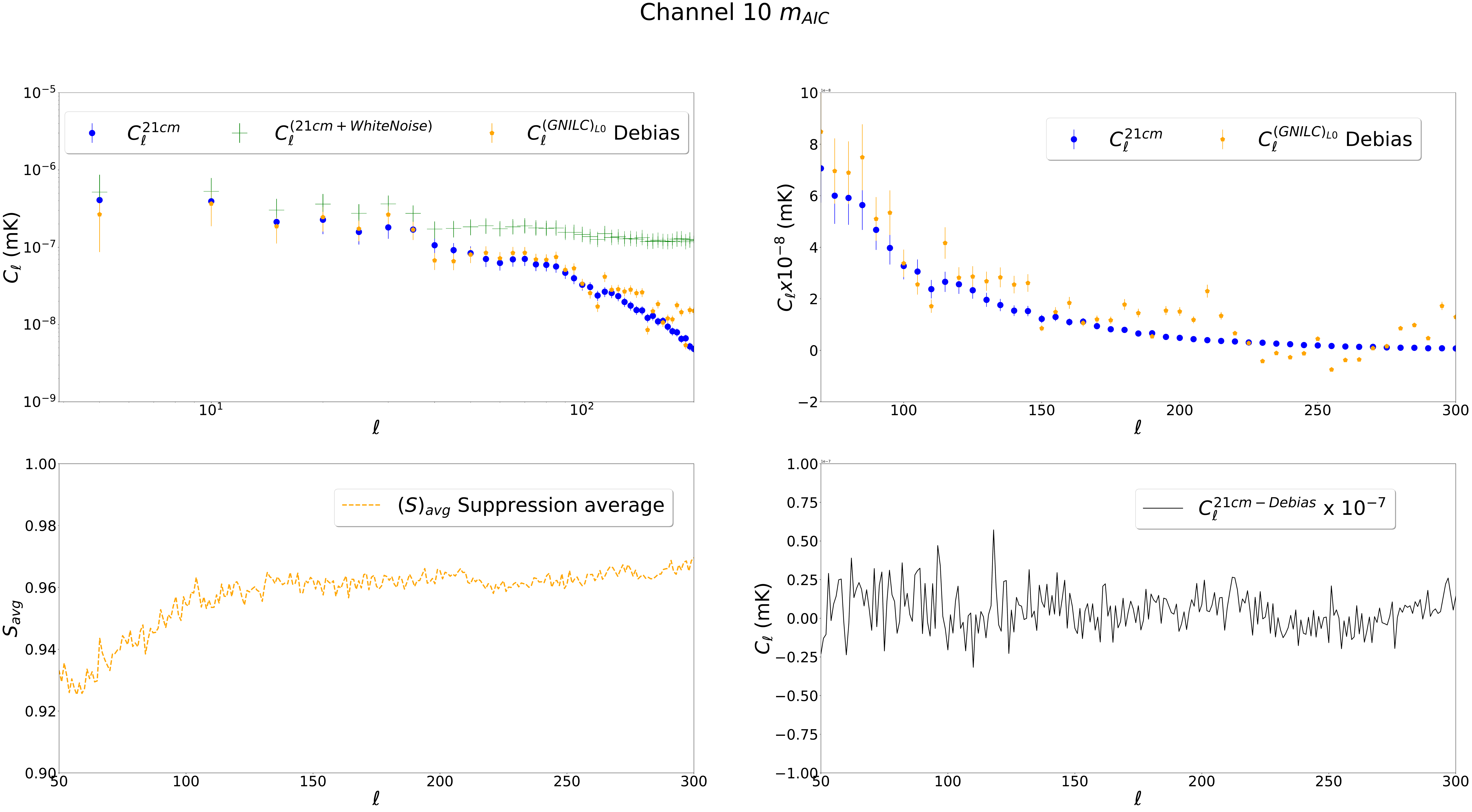}
\caption{
Different aspects concerning the debias procedure for BINGO channel 10 ($1070-1081$ MHz).}
\emph{Top}: Reconstructed 21cm power spectrum (yellow dots) for the BINGO channel 10 ($1070-1081$ MHz) in logarithmic scale (\emph{left}) and linear scale (\emph{right}), after foreground cleaning with {\tt GNILC} and debiasing. \emph{Bottom right}: Difference between the reconstructed and input 21cm signal power spectra. \emph{Bottom left}: Estimate of the suppression factor on the 21cm signal across multipoles.

\label{fig:debias_figall_channel10}
\end{figure*}

\begin{figure*}[!h]
    \centering
    \includegraphics[width=17.5cm]{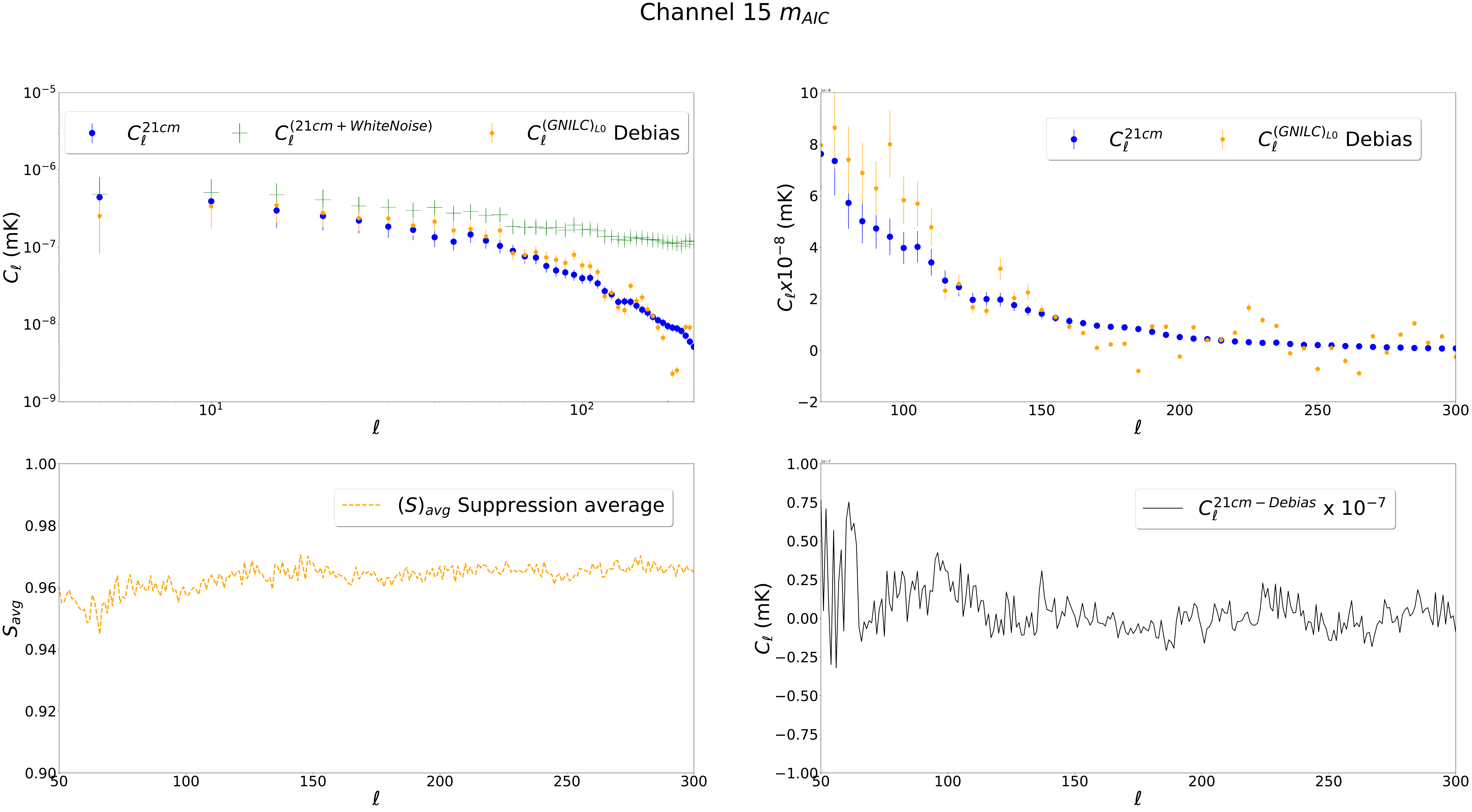}
\caption{
Different aspects concerning the debias procedure for the BINGO channel 15 ($1125-1136$ MHz).
\emph{Top}: Reconstructed 21cm power spectrum (yellow dots) for the BINGO channel 15 ($1125-1136$ MHz) in logarithmic scale (\emph{left}) and linear scale (\emph{right}), after foreground cleaning with {\tt GNILC} and debiasing.  \emph{Bottom left}: Estimate of the suppression factor on the 21cm signal across multipoles. \emph{Bottom right}: Difference between the reconstructed and input 21cm signal power spectra.
}
\label{fig:debias_figall_channel15}
\end{figure*}

\begin{figure*}[!htb]
    \centering
    \includegraphics[width=18cm]{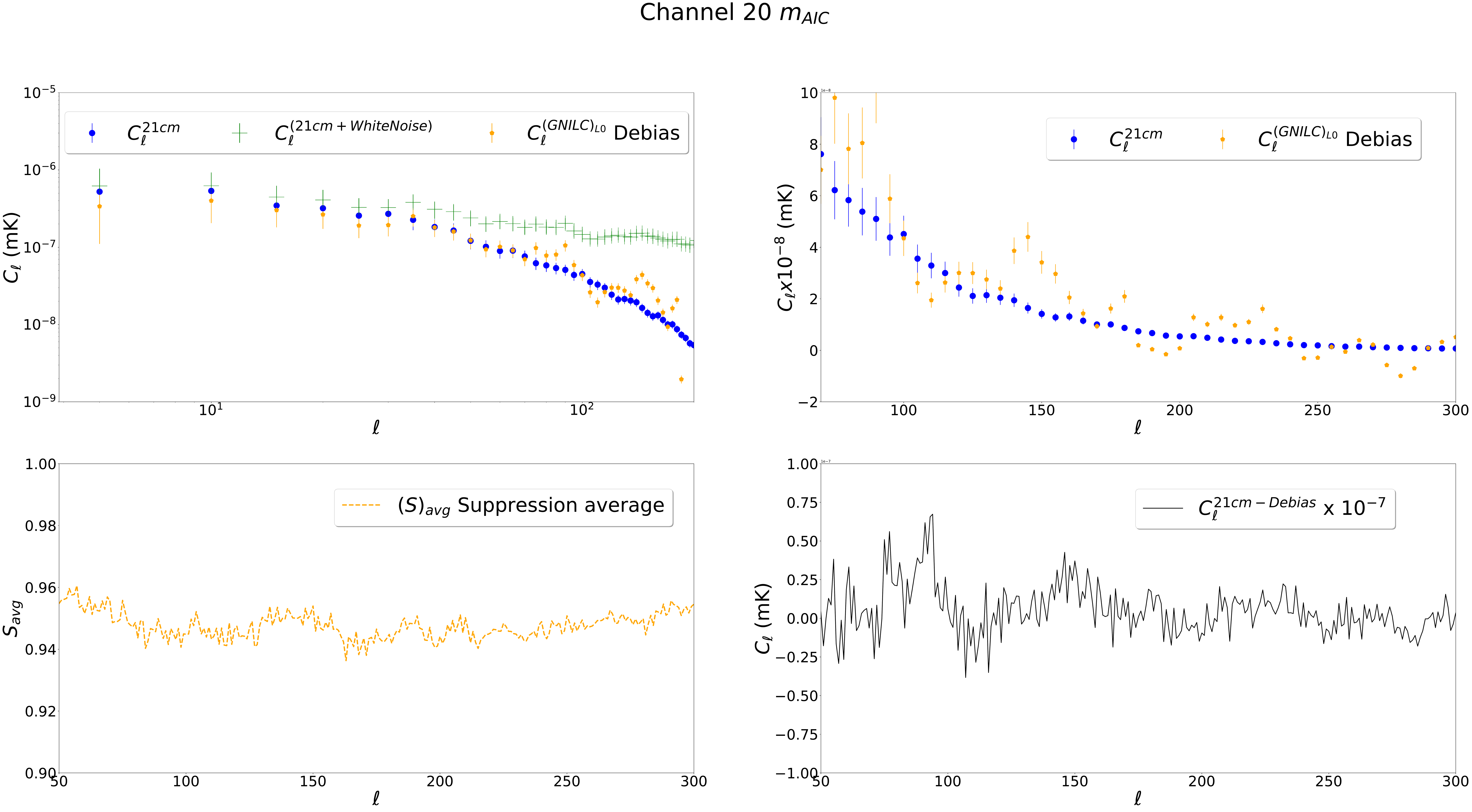}
\caption{
Different aspects concerning the debias procedure for the BINGO channel 20 ($1180-1190$ MHz).
\emph{Top}: Reconstructed 21cm power spectrum (yellow dots) for the BINGO channel 20 ($1180-1190$ MHz) in logarithmic scale (\emph{left}) and linear scale (\emph{right}), after foreground cleaning with {\tt GNILC} and debiasing.  \emph{Bottom left}: Estimate of the suppression factor on the 21cm signal across multipoles. \emph{Bottom right}: Difference between the reconstructed and input 21cm signal power spectra.
}
\label{fig:debias_figall_channel20}
\end{figure*}

The top panels of Figures \ref{fig:debias_figall_channel10}, \ref{fig:debias_figall_channel15}, and \ref{fig:debias_figall_channel20} show, for three different BINGO channels, the reconstructed 21cm power spectrum after foreground cleaning by {\tt GNILC} and debiasing. The green bins correspond to the prior signal (+ noise) power spectrum used in the analysis. We can see that for each channel the recovered 21cm power spectrum (yellow bins) matches relatively closely  the power spectrum of the input 21cm signal (blue bins) of the $L_0$ simulation across multipoles. The difference between the recovered 21cm power spectrum and the input 21cm power spectrum is shown in the bottom right panels of each figure, highlighting an unbiased recovery of the 21cm signal at multipoles $20\lesssim \ell \lesssim 800$. In the bottom left panels of each figure we show our estimate of the suppression factor $\widehat{S}_\ell$ after foreground cleaning, showing a $2\%$ to $6\%$ loss (depending on multipole and channel) of the 21cm signal before correction.



Our results show that our debiasing procedure applied to {\tt GNILC} is quite successful in removing the additive noise bias and correcting for the small multiplicative bias on the 21cm signal, in addition to performing efficient foreground cleaning.
We have obtained similar results for the other BINGO channels. It is beyond the scope of this paper to investigate the cosmological implications of these projections and biases; however, this will be incorporated into the final BINGO pipeline so that we take into account the effects that such residual biases have on the estimation of cosmological parameters.



In the next sections we investigate the residual foreground contamination in the reconstructed 21cm maps by looking at the power spectrum of the projected foreground components (Sect.~\ref{sec:gnilcresiduals}) and the bispectrum of the reconstructed 21cm maps (Sect.~\ref{sec:bispectrum}).


\section{Bispectrum analysis as a test of residual foreground contamination}\label{sec:bispectrum}

The core science aim of BINGO is to use the angular power spectrum of the fluctuation of the redshifted \hi\, 21cm radiation in order to measure the BAO and redshift-space distortions \citep{2020_project}. However, using higher order statistics to characterize the data can bring important information that cannot be probed with the power spectrum alone.

If the signal is Gaussian, the power spectrum contains all the information. However, NGs
(i.e., deviations in our maps from Gaussian statistics) will be present in our data at all redshifts, but most notably in the low redshift 21cm signal. They might come from the early-time evolution of the Universe (what we call primordial NG),  will be imprinted onto the 21cm maps by gravity itself~\citep{Bernardeau:2001qr,dePutter:2018jqk}, and/or will get imprinted on the 21cm maps when these maps are cleaned via residuals from the Galactic foreground distribution.

From the CMB anisotropy measurements~\citep{Ade:2015xua} we know that our Universe evolved from adiabatic initial fluctuations that are very close to Gaussian. Inflation is the most accepted paradigm for the early evolution of the Universe that generated such initial fluctuations. 
 There are a plethora of models of inflation, and also alternatives to inflation models, that explain the evolution of the Universe at early times and that are in accordance with current cosmological observations. 
 However, these models might predict different observables like primordial NGs (PNGs) (for reviews, see~\citealp{Bartolo:2004if,Liguori,Chen:2010xka}). 
 In this way NGs represent a distinctive signature of these models, and measuring them will allow us to exclude and differentiate models and might teach us about the physics that happened at earlier times of our Universe. The presence of PNGs in the initial fluctuations that seed the density perturbations will also be imprinted in the 21cm anisotropies measurement at late times. 
 
However, PNGs are not the only source of NGs in the late time 21cm \hi\, signal. As we already discussed, measuring the cosmological 21cm signal is a daunting task since the 21cm \hi\, data are dominated by foregrounds coming from cosmological and astrophysical sources. These foregrounds together with radio frequency interference might be limiting factors of 21cm experiments, if not cleaned and mitigated correctly. Foreground cleaning techniques and RFI mitigation are some of the main techniques applied to 21cm maps in order to recover the cosmological 21cm \hi\, signal. However, these techniques might introduce NG features in the residual maps, even if NGs are not present initially. In addition, any residual foreground contamination left over in the 
reconstructed maps of the 21cm radiation would also create a large non-Gaussian imprint on the signal. These systematic effects are inherent in the 21cm residual maps that will be analyzed since foreground removing is always necessary.

Given that the 21cm \hi\, data at late times has a NG component, we need to use higher order statistics to characterize the data and estimate these signatures.
 The use of the bispectrum in the 21cm data analysis is not so unusual  \citep[see,
e.g.,][]{Cunnington:2021czb, Jolicoeur:2020eup,Durrer:2020orn}. In \cite{Cunnington:2021czb} there is a good description of this technique over the years. In this work we follow mainly the description found in \cite{Liguori, Komatsu} and \cite{Zaldarriaga}
in order to develop the tool for identifying the NG. For the contour plot and bispectrum analysis we follow \cite{Regan:2009hv}.

As we mention  above, the bispectrum output is used as a double-check for the cleaning procedure; it does not provide  cosmological information  on $f_{nl}, g_{nl}$ (the third- and fourth-order amplitudes, respectively, of non-Gaussianity. See \citep{Komatsu} for more details), or any other parameter.


In this sense, no cosmological analysis of information related to these parameters took part here. In addition, we are not arguing that the bispectrum of the 21cm radiation is actually zero in the configurations measured, but we are arguing that in the lognormal simulations that we are producing it is close enough to zero for the assumptions made in this paper to be valid.

\subsection{Angular bispectrum}
\label{sec:ang_bispectrum}

We wish to compute higher angular correlation functions for the first-order brightness temperature fluctuations. We focus here on the three-point correlation function or on its Fourier transform, called the bispectrum, given that higher order correlators are usually subdominant. Since our aim is to determine the angular bispectrum, similarly to what was done for the angular power, we decompose the brightness temperature fluctuation $\frac{\Delta T(\mathbf{\hat{n}})}{T}$ in spherical harmonics
\begin{equation}\label{bisp12}
    a_{\ell m} = \int d^{2}\mathbf{\hat{n}}\frac{\Delta T (\mathbf{\hat{n}})}{T}Y_{lm}^{\ast}(\mathbf{\hat{n}}) \,,
\end{equation}
where the hats denote unit vectors. Given this, the angular three-point correlation function with the flat sky approximation is given by \citep{Liguori, Komatsu, Zaldarriaga}
\begin{equation}
     \langle a_{\ell_{1}m_{1}} a_{\ell_{2}m_{2}} a_{\ell_{3}m_{3}} \rangle \equiv  B_{\ell_{1}\ell_{2} \ell_{3}}^{m_{1} m_{2} m_{3}} = \langle B_{\ell_{1}\ell_{2} \ell_{3}} \rangle \, \Bigg (\begin{matrix} \ell_{1} & \ell_{2} & \ell_{3} \\ m_{1} & m_{2} & m_{3} \end{matrix}\Bigg )\,,
 \end{equation}
 where $B_{\ell_{1}\ell_{2} \ell_{3}}^{m_{1} m_{2} m_{3}}$ is the angular bispectrum and $B_{\ell_{1}\ell_{2} \ell_{3}}$ is the averaged angular bispectrum given by
 \begin{eqnarray}\label{bisp14}
 B_{\ell_{1}\ell_{2} \ell_{3}} =  \sum_{m}\Bigg (\begin{matrix} \ell_{1} & \ell_{2} & \ell_{3} \\ m_{1} & m_{2} & m_{3} \end{matrix} \Bigg) \cdot B_{\ell_{1}\ell_{2} \ell_{3}}^{m_{1} m_{2} m_{3}}\,.
\end{eqnarray}
 The matrix denotes the Wigner-$3j$ symbol, representing the azimuthal  angle dependence of the bispectrum, which is invariant under permutations. It describes three angular momenta that form a triangle $\mathbf{L}_1+\mathbf{L}_2+\mathbf{L}_3=0$, where $m_1+m_2+m_3 =0$, which implies that the matrix is only nonzero if the triangle conditions are satisfied: $|\ell_{i} -\ell_{j} | \leq \ell_{k} \leq \ell_{i} + \ell_{j}$.  The angular correlation function is invariant under parity, which implies that $\ell_{1} + \ell_{2} + \ell_{3}$ is even. In this way, $B_{\ell_{1}\ell_{2} \ell_{3}}$ is only nonvanishing if the above triangle and parity conditions are met.
 
We note here that we neglect the contributions from the shot noise in the theoretical and simulated estimation of the bispectrum and the power spectrum. We assume that the contribution from the shot noise is small assuming that it arises from the number density of galaxies which emit in HI \citep{Olivari:2017}. It is beyond the scope of this analysis to check if this assumption is accurate and this term can be neglected or if there are conditions under which this term can be large; however,  there are simulations that show that this can in fact be larger than the value suggested by \cite{Olivari:2017} and in Section 4 of \cite{Zhang_2020}.

 The angular three-point correlation is invariant under rotations, thus the bispectrum can be written as
 \begin{equation}
      B_{\ell_{1}\ell_{2} \ell_{3}}^{m_{1} m_{2} m_{3}} = G_{\ell_{1}\ell_{2} \ell_{3}}^{m_{1} m_{2} m_{3}} \, b_{\ell_{1}\ell_{2} \ell_{3}}\,,
     \label{eq.:bispectrum}
 \end{equation}
 where $b_{\ell_{1}\ell_{2} \ell_{3}}$ is the reduced bispectrum, which is a real and symmetric function of $\ell_1$, $\ell_2$, and $\ell_3$, and $G_{\ell_{1}\ell_{2} \ell_{3}}^{m_{1} m_{2} m_{3}}$ is the Gaunt integral defined as
 \begin{align}\label{bisp15}
& G_{\ell_{1}\ell_{2} \ell_{3}}^{m_{1} m_{2} m_{3}}  =
 \int d^{2}\mathbf{\hat{n}}Y_{\ell_{1}m_{1}}(\mathbf{\hat{n}})Y_{\ell_{2} m_{2}}(\mathbf{\hat{n}})Y_{\ell_{3}m_{3}}(\mathbf{\hat{n}})\nonumber \\
& =\sqrt{\frac{(2\ell_{1}+1)(2\ell_{2}+1)(2\ell_{3}+1)}{4\pi}} \,
 \Bigg (\begin{matrix} \ell_{1} & \ell_{2} & \ell_{3} \\ 0 & 0 & 0 \end{matrix} \Bigg) \, \Bigg (\begin{matrix} \ell_{1} & \ell_{2} & \ell_{3} \\ m_{1} & m_{2} & m_{3} \end{matrix} \Bigg)\,.
\end{align}
The Gaunt integral obeys the conditions mentioned above. As all the dependencies 
on the Wigner-$3j$ symbol appears only in the Gaunt integral, it is easier to study the physical properties of the bispectrum using $b_{l_{1}l_{2} l_{3}}$. We can then rewrite the angular averaged bispectrum $B_{\ell_{1}\ell_{2} \ell_{3}}$ in terms of the reduced bispectrum $b_{\ell_{1}\ell_{2} \ell_{3}}$:

\begin{eqnarray}\label{bisp17}
 B_{\ell_1 \ell_2 \ell_3}  =
 \sqrt{\frac{(2\ell_{1}+1)(2\ell_{2}+1)(2\ell_{3}+1)}{4\pi}} \,
 \Bigg (\begin{matrix} \ell_{1} & \ell_{2} & \ell_{3} \\ 0 & 0 & 0 \end{matrix} \Bigg) \, b_{\ell_{1}\ell_{2} \ell_{3}} \, .
\end{eqnarray}

 The bispectrum can form different triangles depending on the relation between $\ell_1,\, \ell_2,\, \ell_3$. The type of triangle describes the shape of the bispectrum, with the following known shapes~\citep{Jeong_Komatsu}: equilateral  $\ell_1 = \ell_2 = \ell_3$,  squeezed $\ell_1 \eqsim \ell_2 \gg \ell_3$, isosceles $\ell_2 = \ell_3$, elongated $\ell_1=\ell_2+\ell_3$, and folded $\ell_1=2\ell_2=2\ell_3$.
 We  use certain combinations of the above-defined bispectrum in later sections to show how its measurements can impose strict tests to the nature and level of the foreground residuals in the simulations and data extraction we presented at the beginning of this paper.
 
 For the purposes of this work, different than the forms cited above, it is useful to define a specific subset of bispectrum values where $\ell_1+\ell_2+\ell_3 = N$, where the sum of the $\ell$s equals a given factor. We call this shape ``equisize.'' This configuration corresponds to triangles of roughly similar size in $\ell$ space. We define the quantity

\begin{eqnarray}\label{bisp18}
 B_{\ell_1+\ell_2+\ell_3=N}  =
 \sum_{\ell_1=1}^{N}\sum_{\ell_2=1}^{N}\sum_{\ell_3=1}^{N}{B_{\ell_{1} \ell_{2} \ell_{3}}\delta_{\ell_1+\ell_2+\ell_3,N}}\,,
\end{eqnarray}
\noindent which is  the sum of the entries of the nonzero bispectrum measurements, where the configurations obey $\ell_1+\ell_2+\ell_3=N$ from the Kronecker delta. We note that the specific case where $N$ is divisible by three, one of the above-mentioned configurations will correspond to the equilateral configuration where $\ell_1 = \ell_2 = \ell_3$.

\subsection{GNILC foreground residuals}\label{sec:gnilcresiduals}

The AIC criterion Eq.~\ref{eq:aic} is used by {\tt GNILC} to determine the effective dimension $m_\mathrm{AIC}\equiv m_\mathrm{AIC}^{(j)}(\hat{n})$ of the foreground subspace (i.e., the number of principal components of matrix Eq.~\ref{eq:whitecov}) favored by the data among the class of models $1\leq m \leq n$, depending on the local signal-to-foreground ratio (Eq.~\ref{eq:whitecov}) across the sky and the scales. Without the AIC penalty the maximum likelihood solution would tend to overestimate the dimension of the foreground subspace (i.e., $m\lesssim n$, where $n$ is the number of available channels), and thus would tend to remove the 21cm signal from the data along with the foreground contamination. The nominal setting for {\tt GNILC} is to use the AIC selected value $m_\mathrm{AIC}$ as the default value.

In this subsection and in the next  we purposefully run {\tt GNILC} with sub-optimal settings for $m_\mathrm{AIC}$ in order to have sub-optimal foreground subtraction to quantify our ability to assess the quality of the foreground subtraction method via the bispectrum, as defined in the previous subsection. We aim to establish that we have tools available to measure and assess the systematic effects in such residual maps as obtained with the {\tt GNILC} method, as described in this paper, although this approach can also be used for other component separation methods. 

Because of the finite number of available frequency channels, but with a high dimensionality of both foreground and 21cm components, component separation methods face a trade-off between two directions: removing as much foreground contamination as possible  from the data without losing much 21cm signal, or conserving as much  21cm signal as possible at the cost of accepting more residual foreground contamination. The AIC enables {\tt GNILC} to find a sweet spot in this trade-off. However, the $n$-dimensional space of the data is not a direct sum of the foreground and 21cm subspaces, hence the frontier between the two subspaces is not strict.




In order to test the robustness of our results, in this and the following subsections we test {\tt GNILC} with three different options for the dimension parameter of the foreground subspace: $m_\mathrm{AIC} -1$, $m_\mathrm{AIC}$ as selected by the AIC, and $m_\mathrm{AIC}+1$ (i.e., the AIC selected dimension increased by one).
It is reasonable to assume that by increasing the number of dimensions by one or two, we would have increasingly more aggressive foreground mitigation strategies, which in turn require a larger correction factor for the \mbox{21cm} signal loss, but also a smaller projection of foregrounds in the respective 21cm reciprocal space. 

When we artificially impose one less dimension for the foreground subspace in {\tt GNILC} (i.e., $m_\mathrm{AIC}-1$) the power spectrum of the recovered 21cm map in channel 15 shows higher residual foreground contamination at all multipoles. Similarly, when we impose an extra dimension, the residuals   decrease. 

\begin{figure}[!h]
    \centering
    
    \includegraphics[width=8cm]{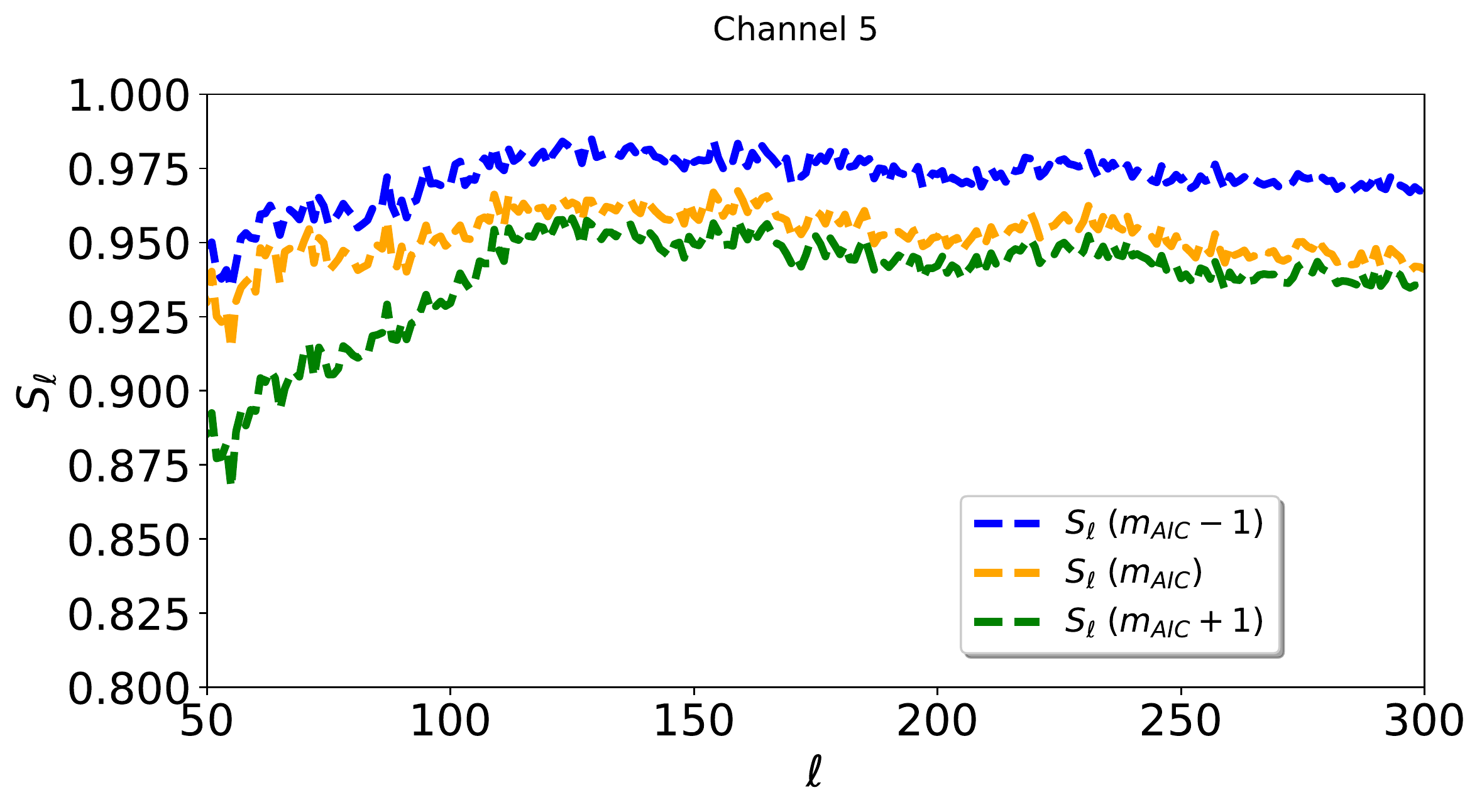}
    \includegraphics[width=8cm]{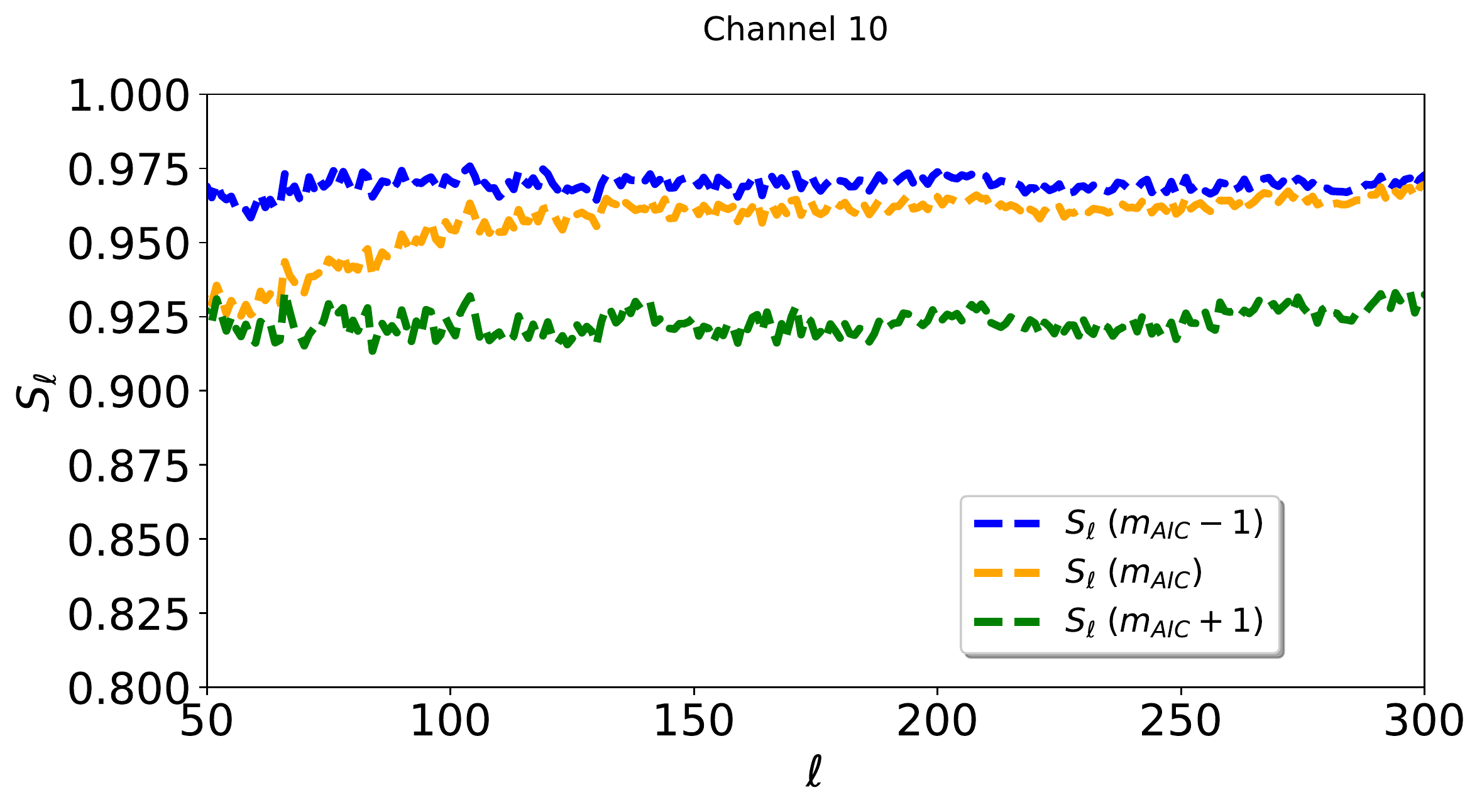}
    \includegraphics[width=8cm]{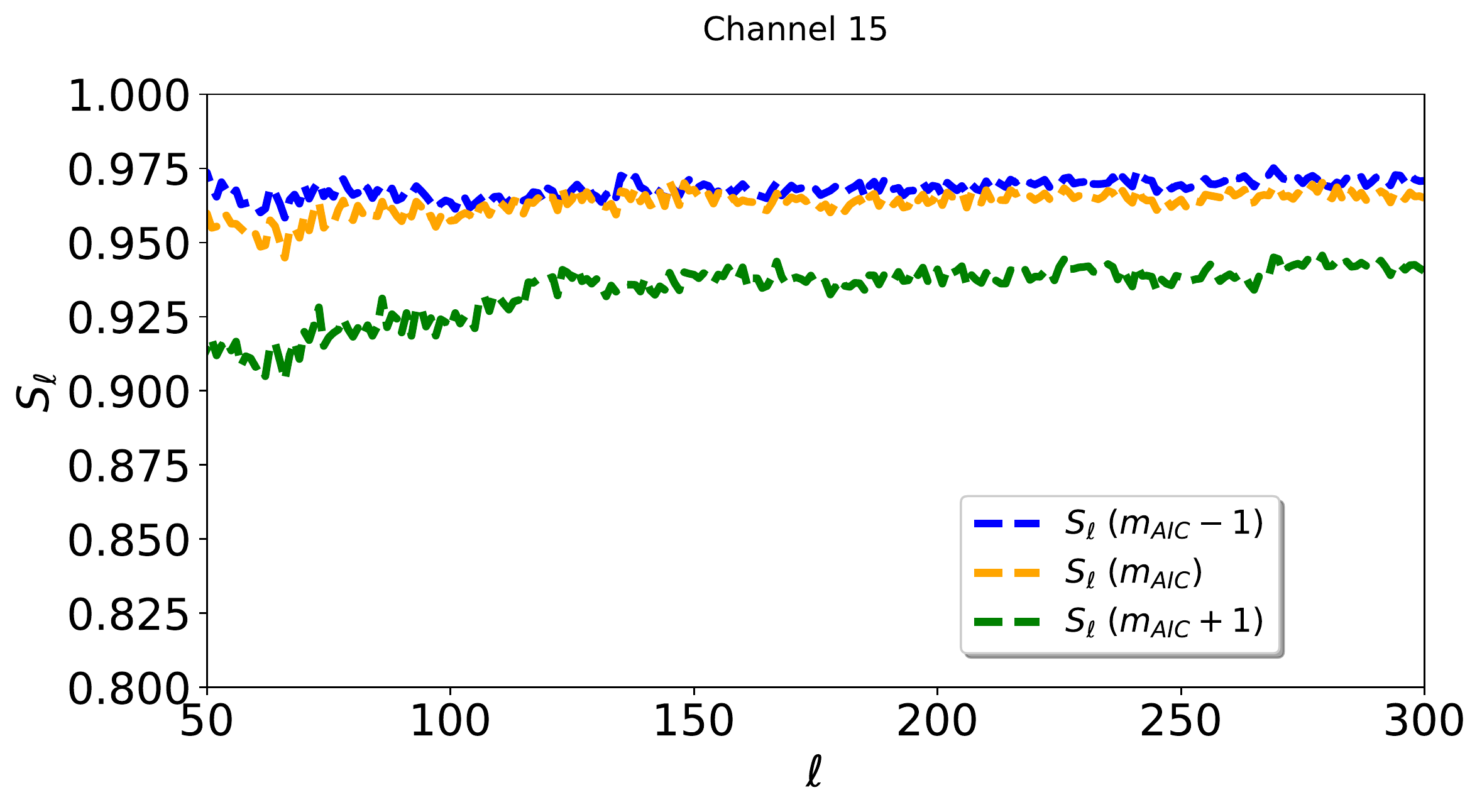}
    \includegraphics[width=8cm]{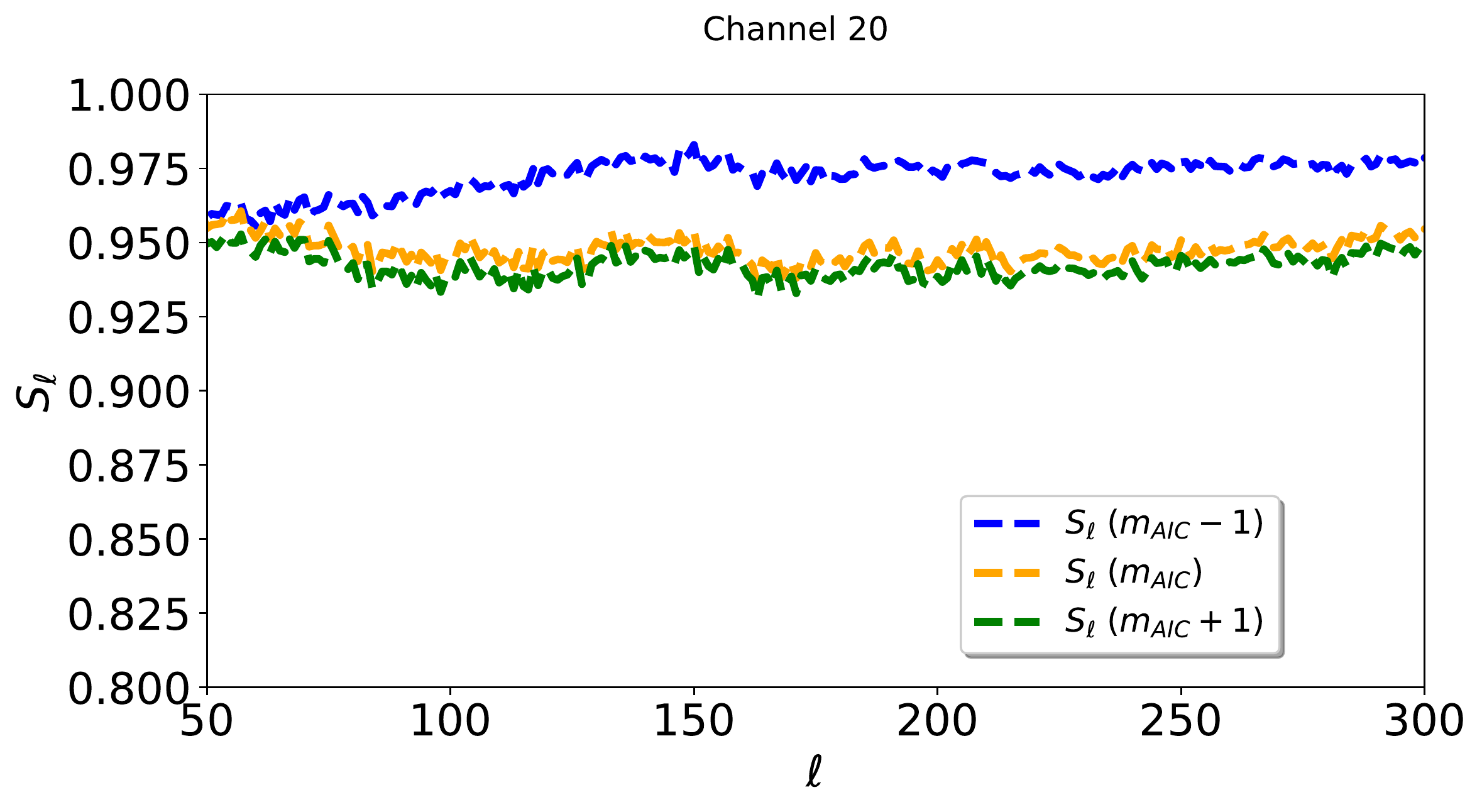}
    \includegraphics[width=8cm]{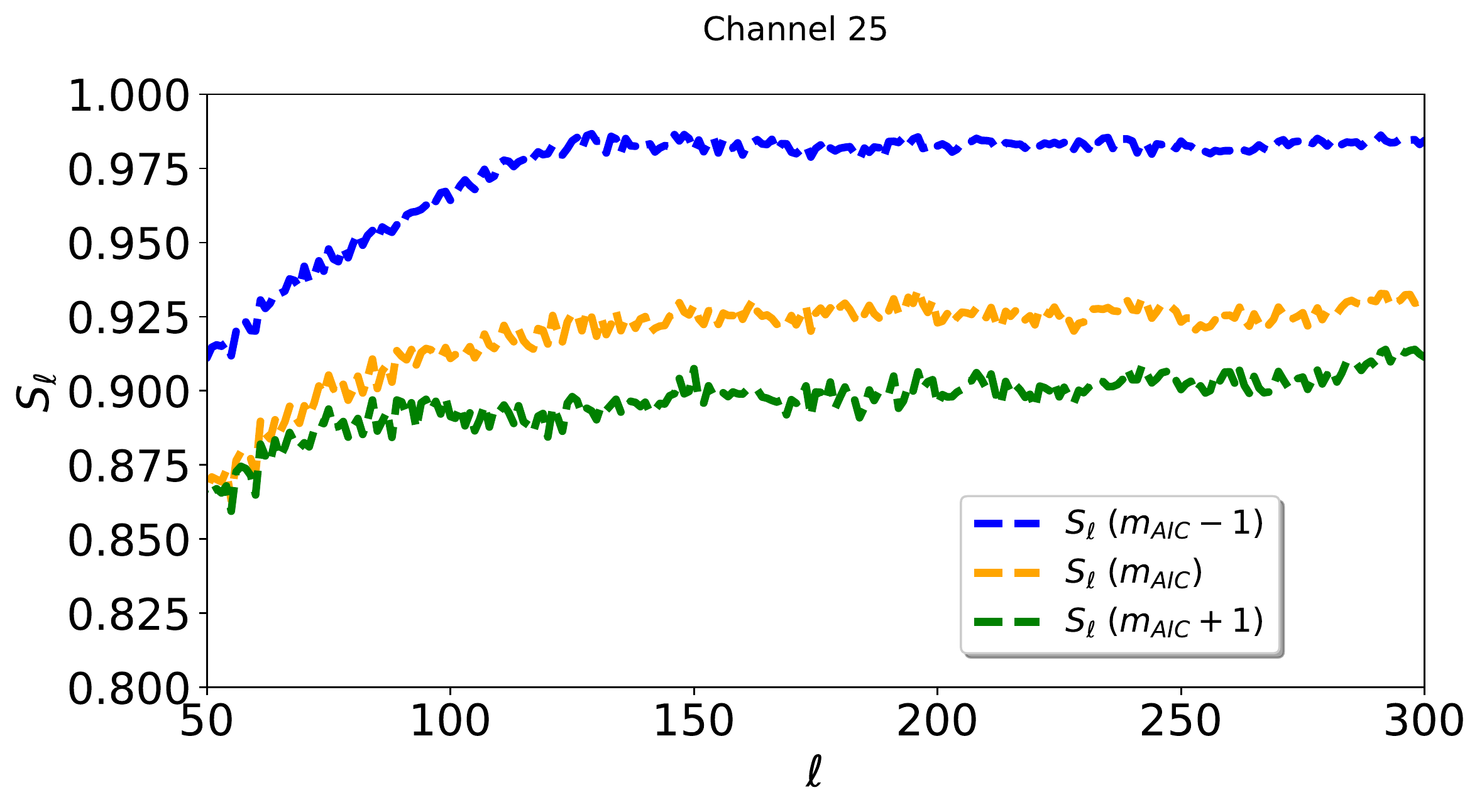}

\caption{Suppression factor of the 21cm signal for different values of the foreground dimension: $(m_\mathrm{AIC}-1)$ (blue), $m_\mathrm{AIC}$ (yellow), and $(m_\mathrm{AIC}+1)$ (green).}
\label{fig:signal_bias} 
\end{figure}

In contrast, as shown in Fig.~\ref{fig:signal_bias}, the more aggressive foreground strategy increases the loss of 21cm signal across multipoles, while the default AIC selection $m_\mathrm{AIC}$ guarantees minimal loss of 21cm signal. We find, however, that the signal loss on the 21cm field decreases slowly, ranging from 2\% to 10\% (depending on multipoles) in the ranges of strategies that we have analyzed. 

\begin{figure}
    \centering
    \includegraphics[width=8cm]{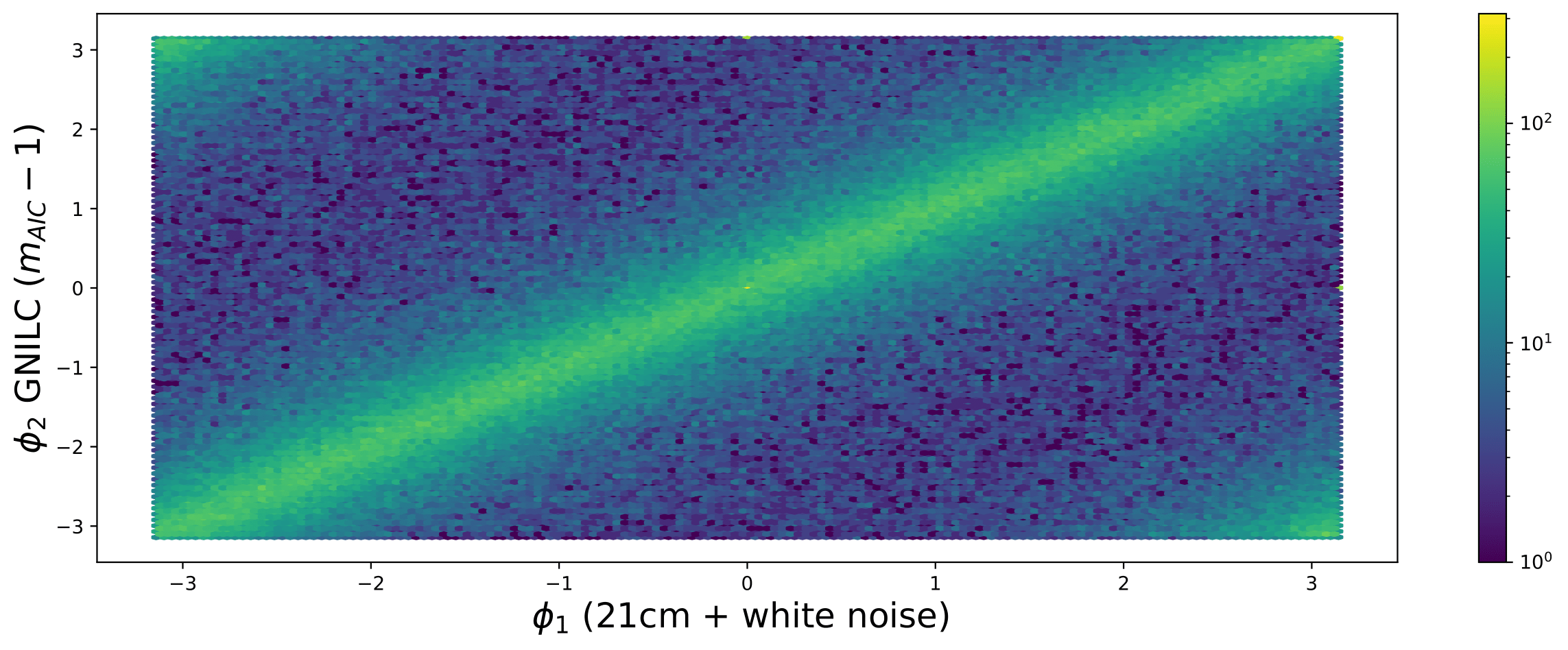}
    \includegraphics[width=8cm]{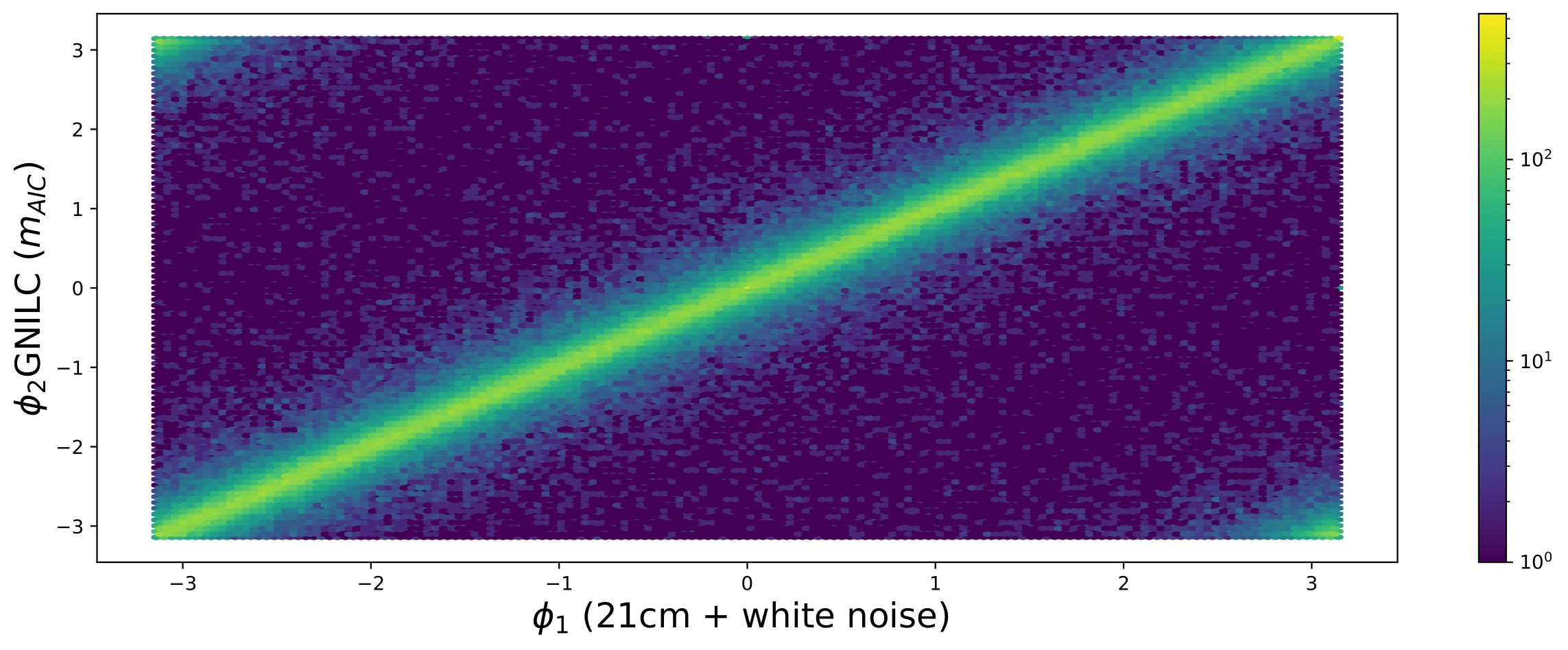}
    \includegraphics[width=8cm]{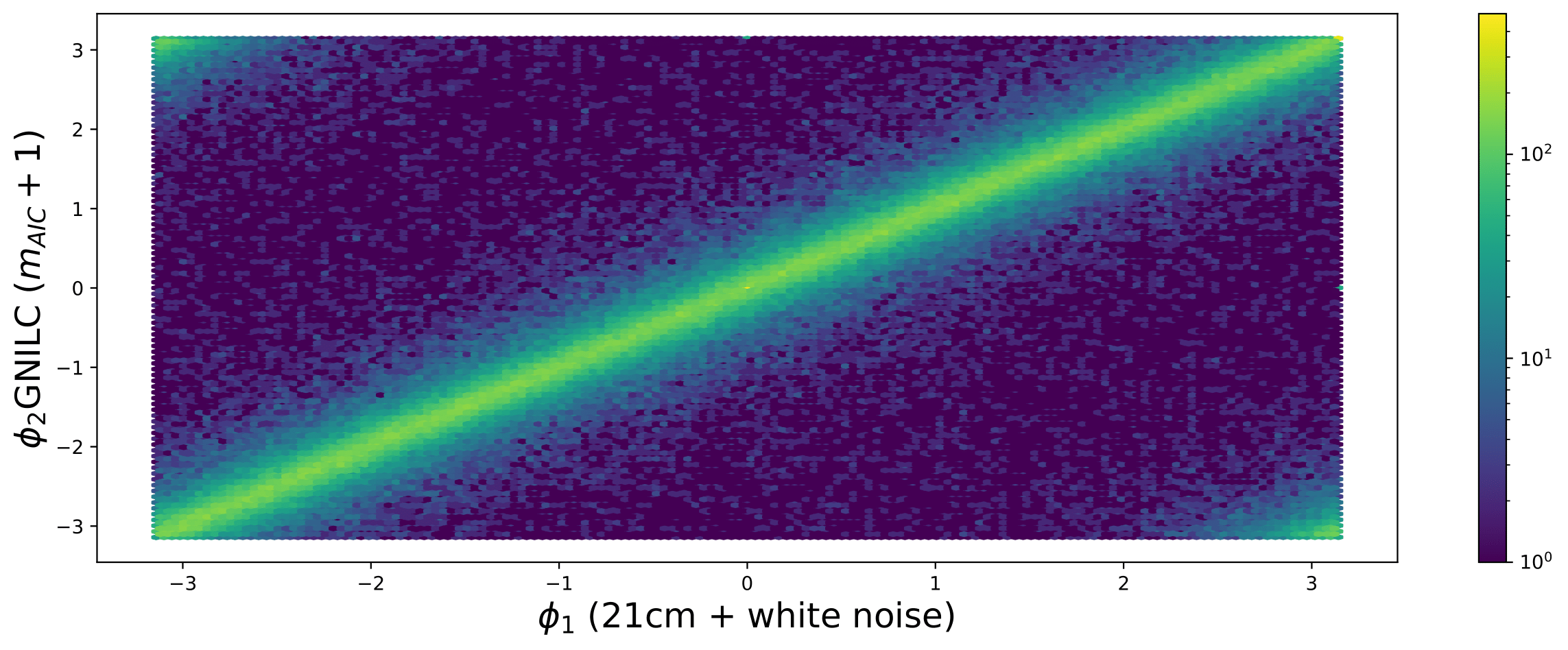}
\caption{Phase comparison between the reconstructed 21cm signal and the prior (21cm signal added to white noise) for different values of the foreground dimension: $m_\mathrm{AIC}-1$ (\textit{top}), $m_\mathrm{AIC}$ (\textit{middle}), and $m_\mathrm{AIC}+1$ (\textit{bottom}). Shown here are results for Channel 15.}
\label{fig:phases}  
\end{figure}

In addition, 
in Fig.~\ref{fig:phases} we present the correlation between the phases of the recovered and input 21cm maps for the three cases studied here. 
Increasing the default dimension of the foreground subspace clearly reduces the scatter due to foregrounds in the correlation. However, the correlation coefficient for pure 21cm (i.e., no noise) also degrades because of the increasing loss of signal according to Fig.~\ref{fig:signal_bias}.

\begin{figure*}
   \centering
    \includegraphics[width=8cm]{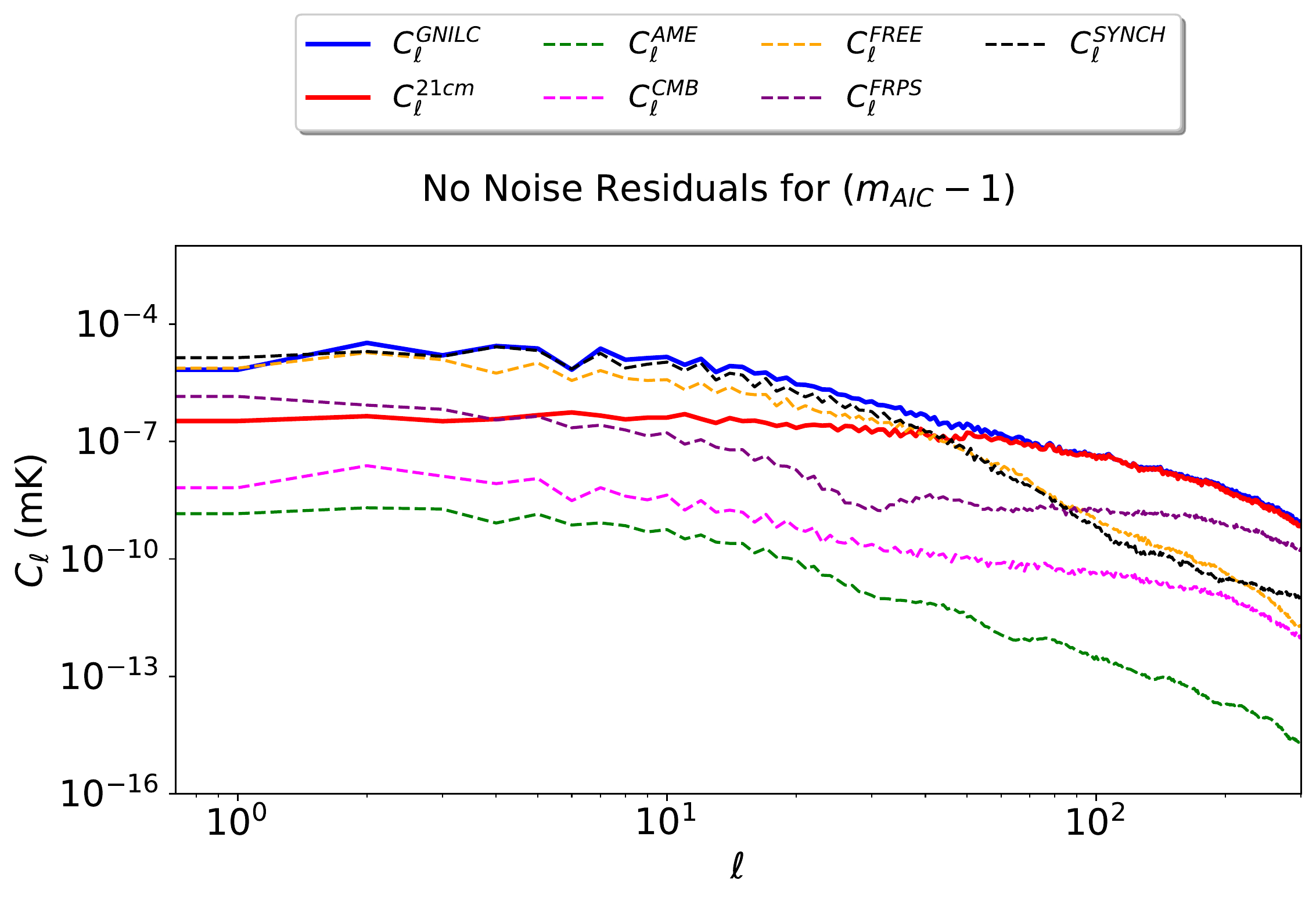}
    \includegraphics[width=8cm]{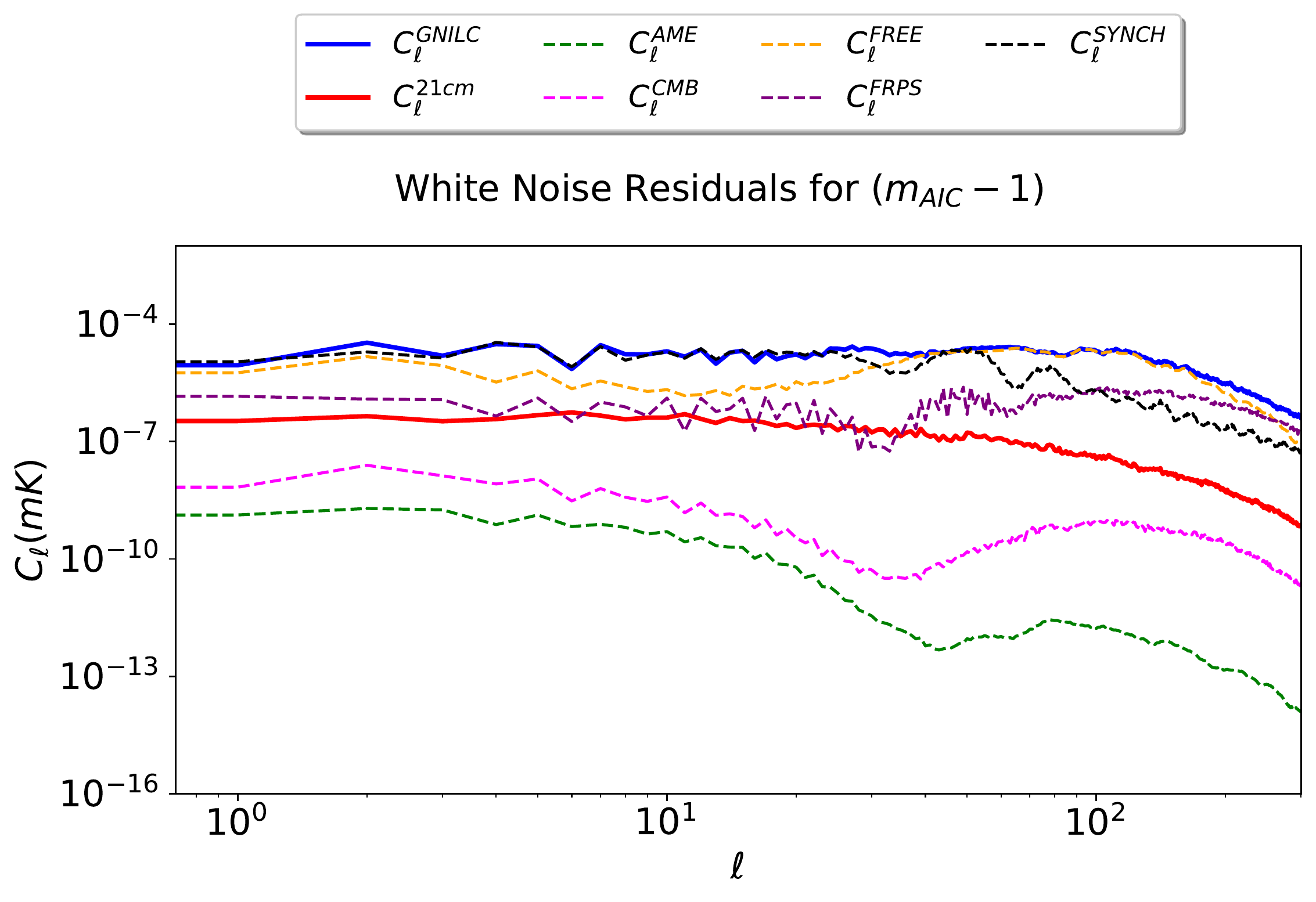}\hfill
    \includegraphics[width=8cm]{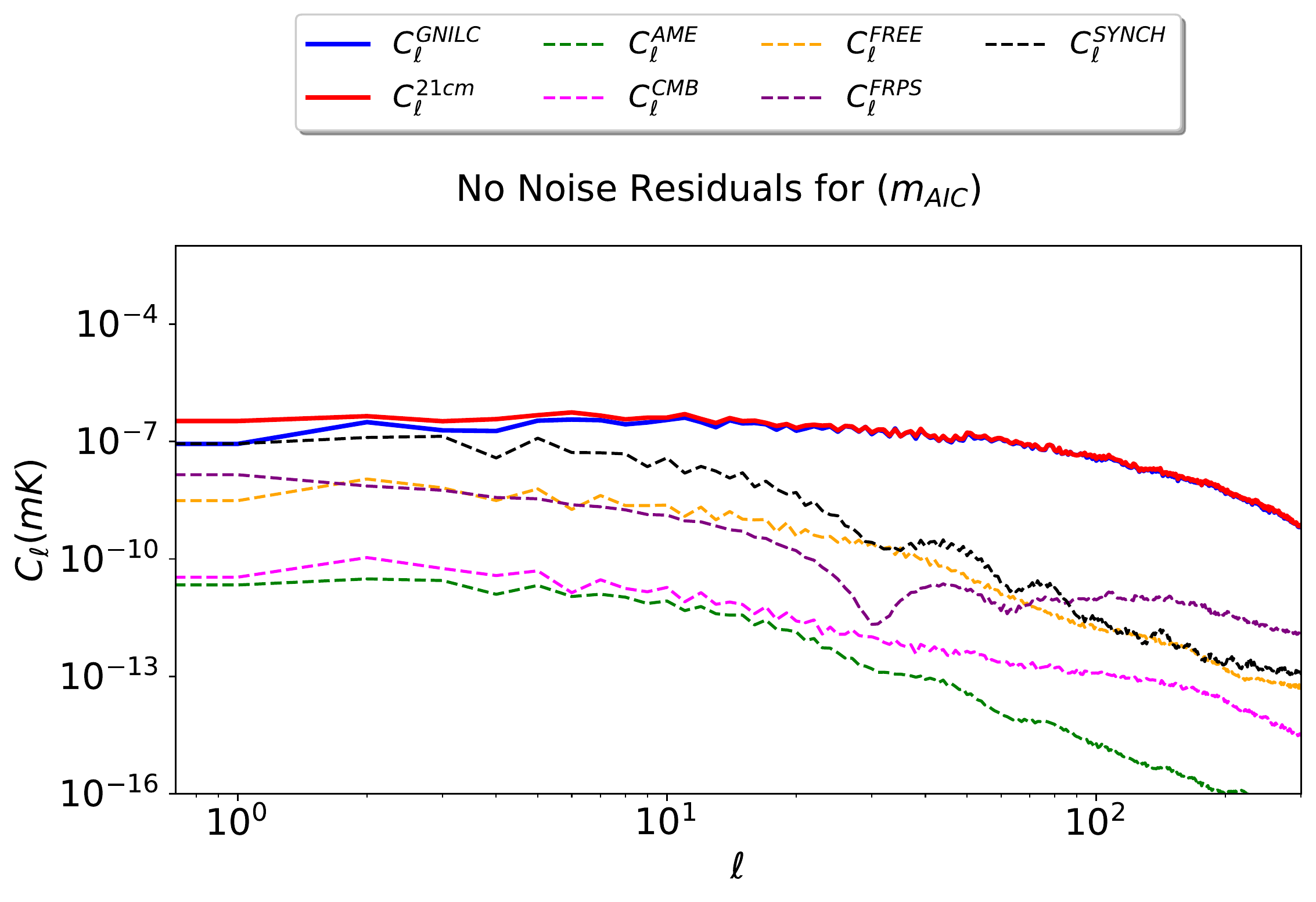}
    \includegraphics[width=8cm]{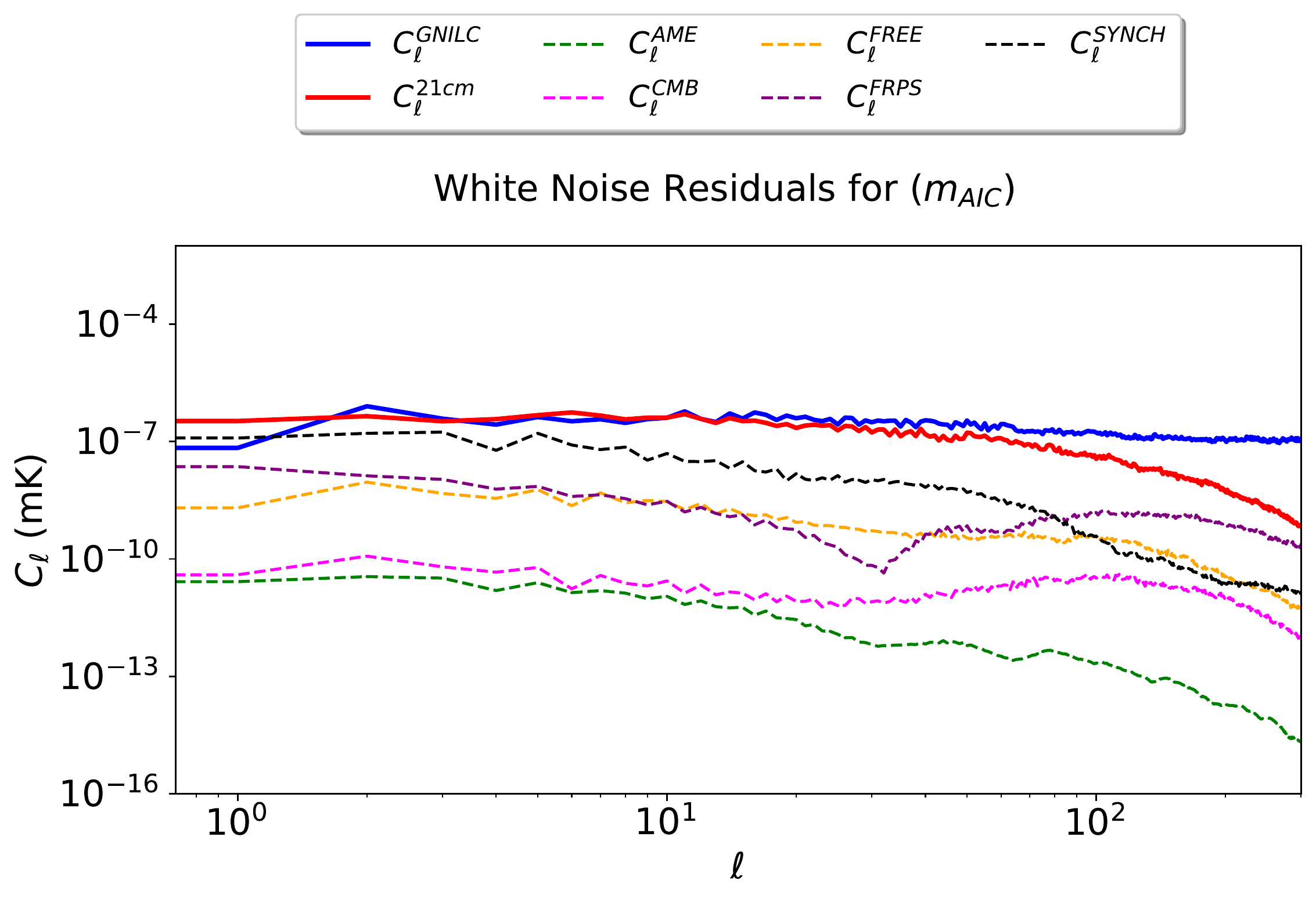}\hfill
    \includegraphics[width=8cm]{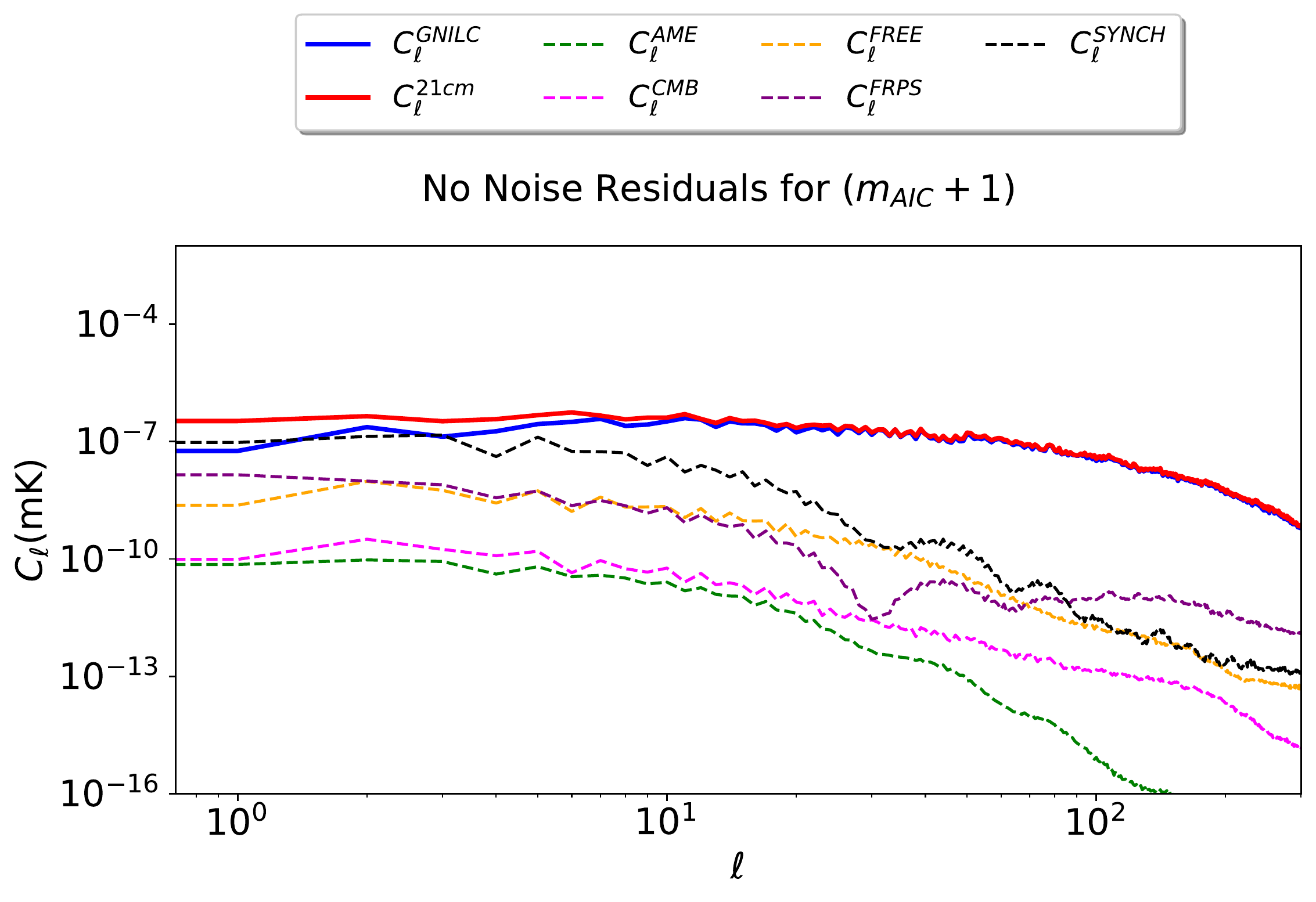}
    \includegraphics[width=8cm]{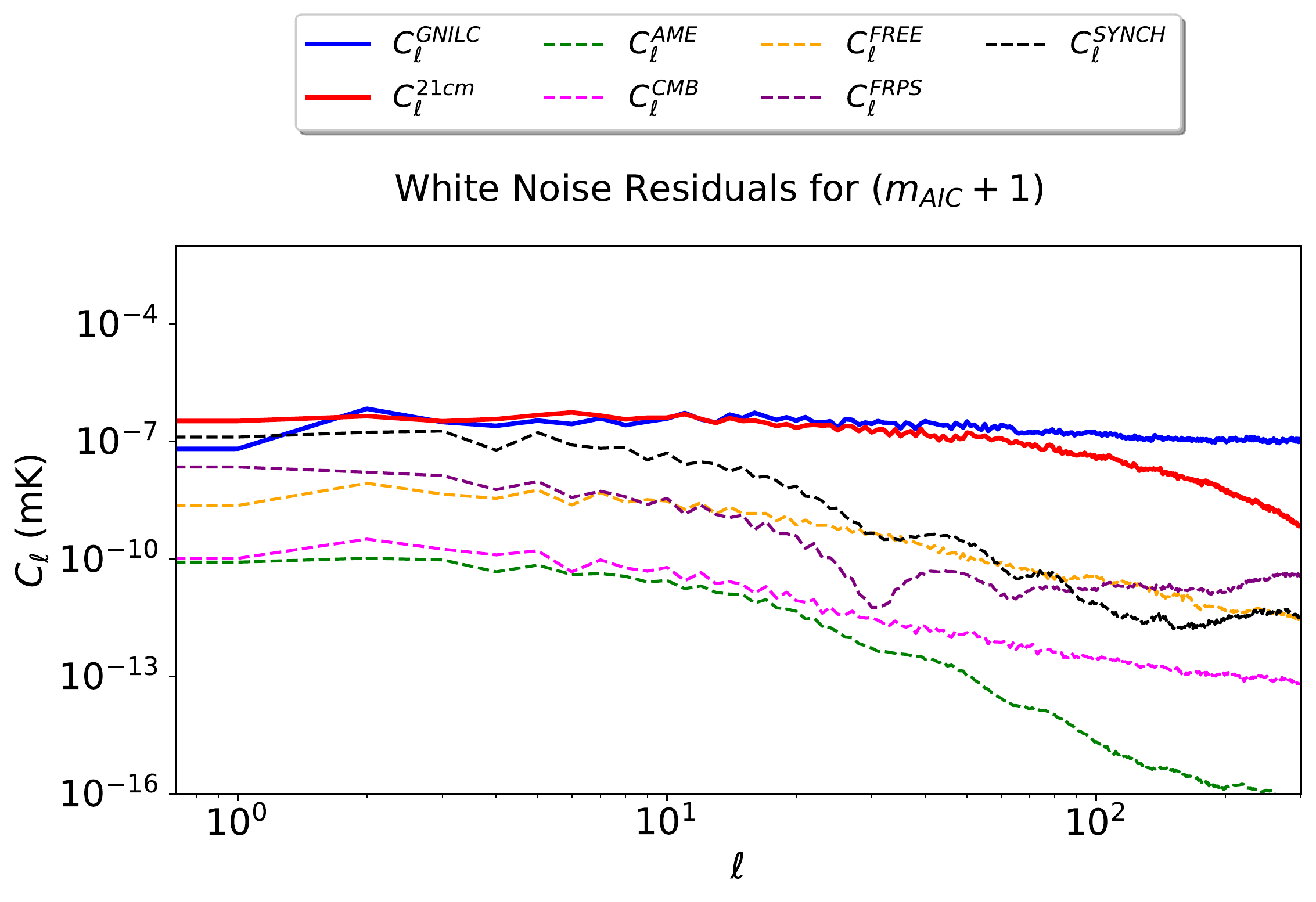}\hfill

\caption{Angular power spectra of the projected foreground components for the default dimension $(m_\mathrm{AIC}-1)$ as selected by AIC (\emph{top}), $(m_\mathrm{AIC})$ (\emph{middle}), and $(m_\mathrm{AIC}+1)$ (\emph{bottom}) in the absence of noise (\emph{left column}) or in the presence of white noise (\emph{right column}) for channel 15. Here the residuals are calculated with mask and convolved with a 40 arcmin beam. } 

\label{fig:residuals}
\end{figure*}

Figure~\ref{fig:residuals} shows the angular power spectrum of the residual foreground components projected in the recovered 21cm signal for the three different options of the foreground dimension parameter in {\tt GNILC}. The left column shows the results for noise-free simulations, while the right column shows the results in the presence of white noise and the use of 21cm plus the noise prior by {\tt GNILC}.

\begin{figure*}
\centering
\includegraphics[width=3.2cm]{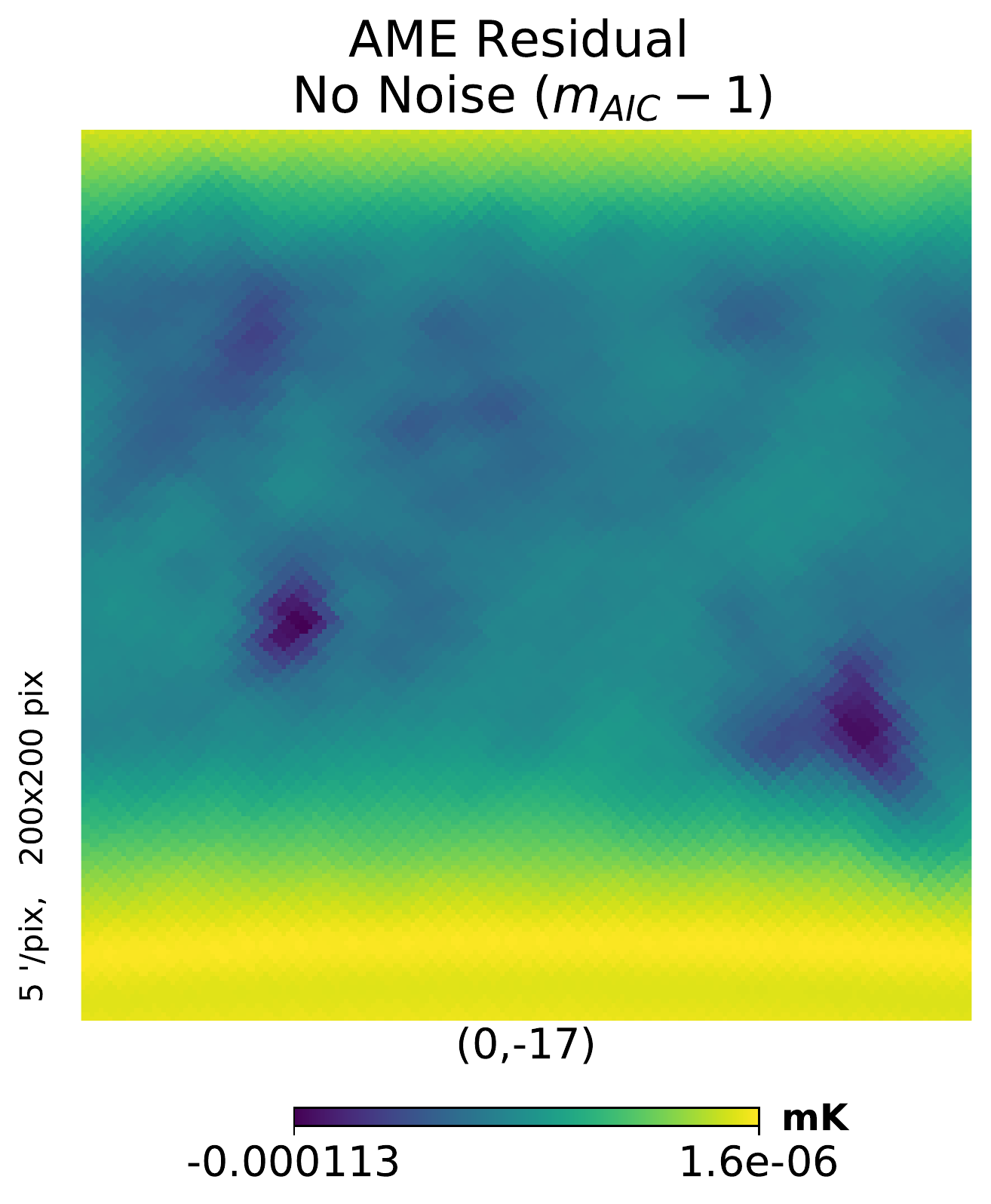}
\includegraphics[width=3.2cm]{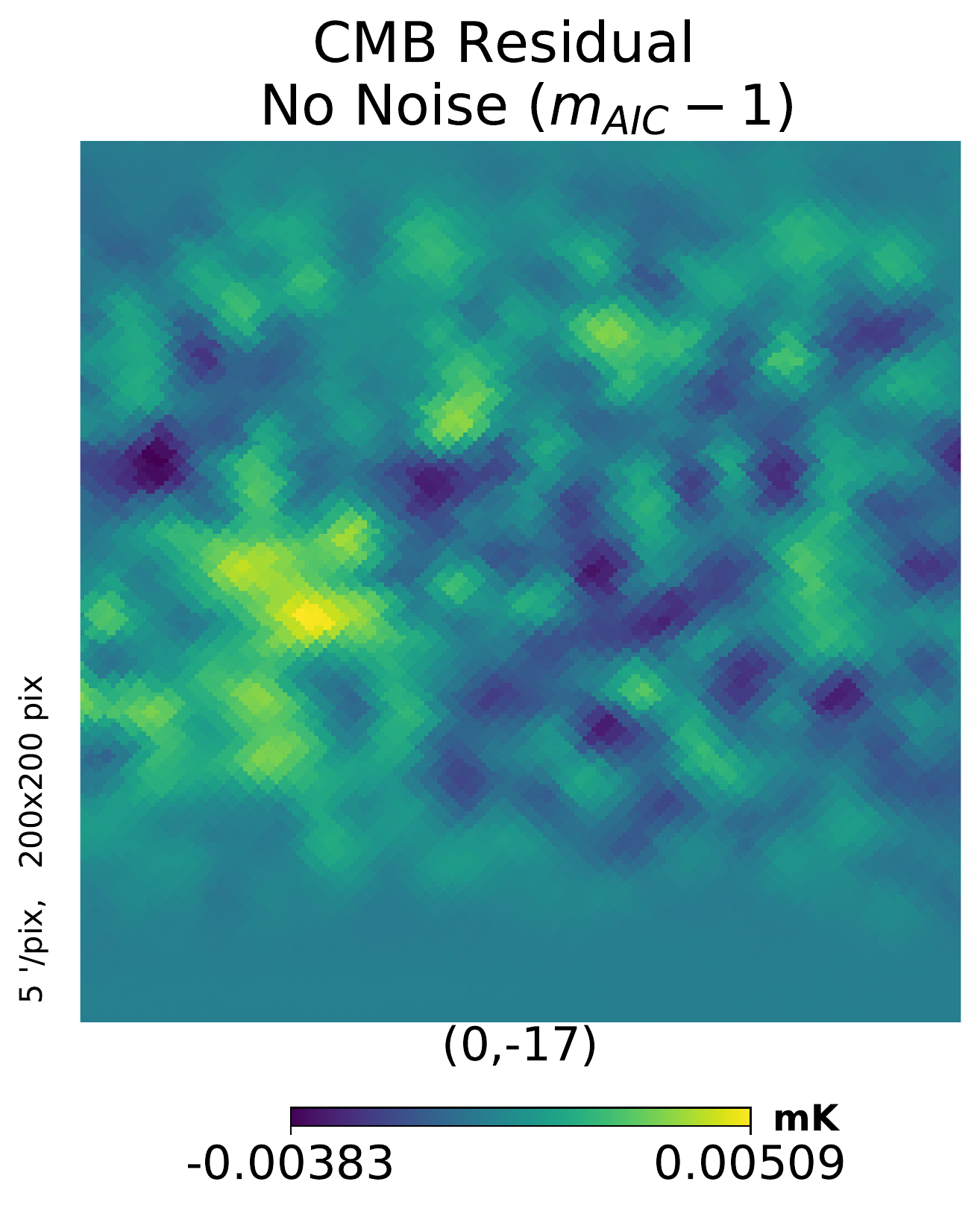}
\includegraphics[width=3.2cm]{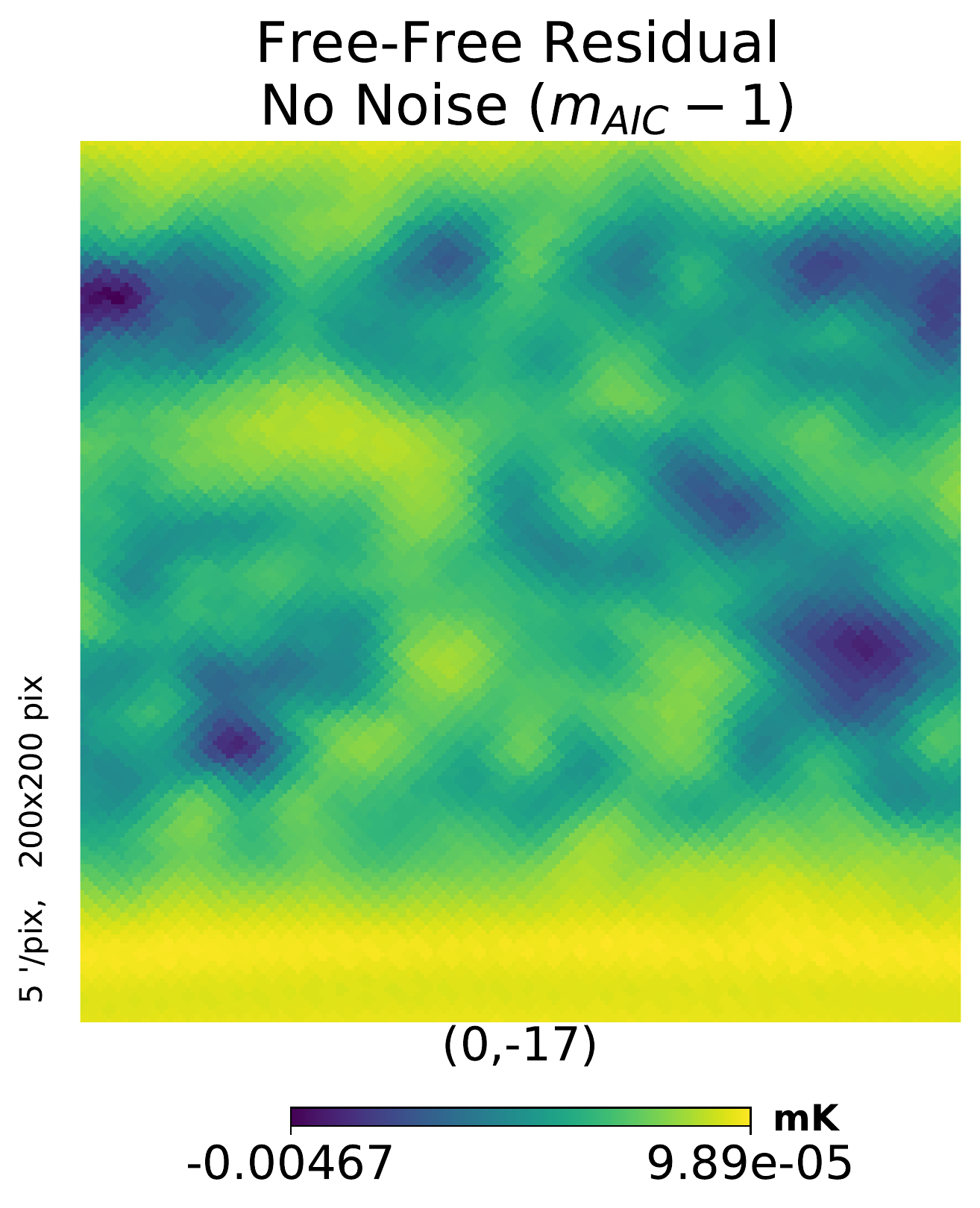}
\includegraphics[width=3.2cm]{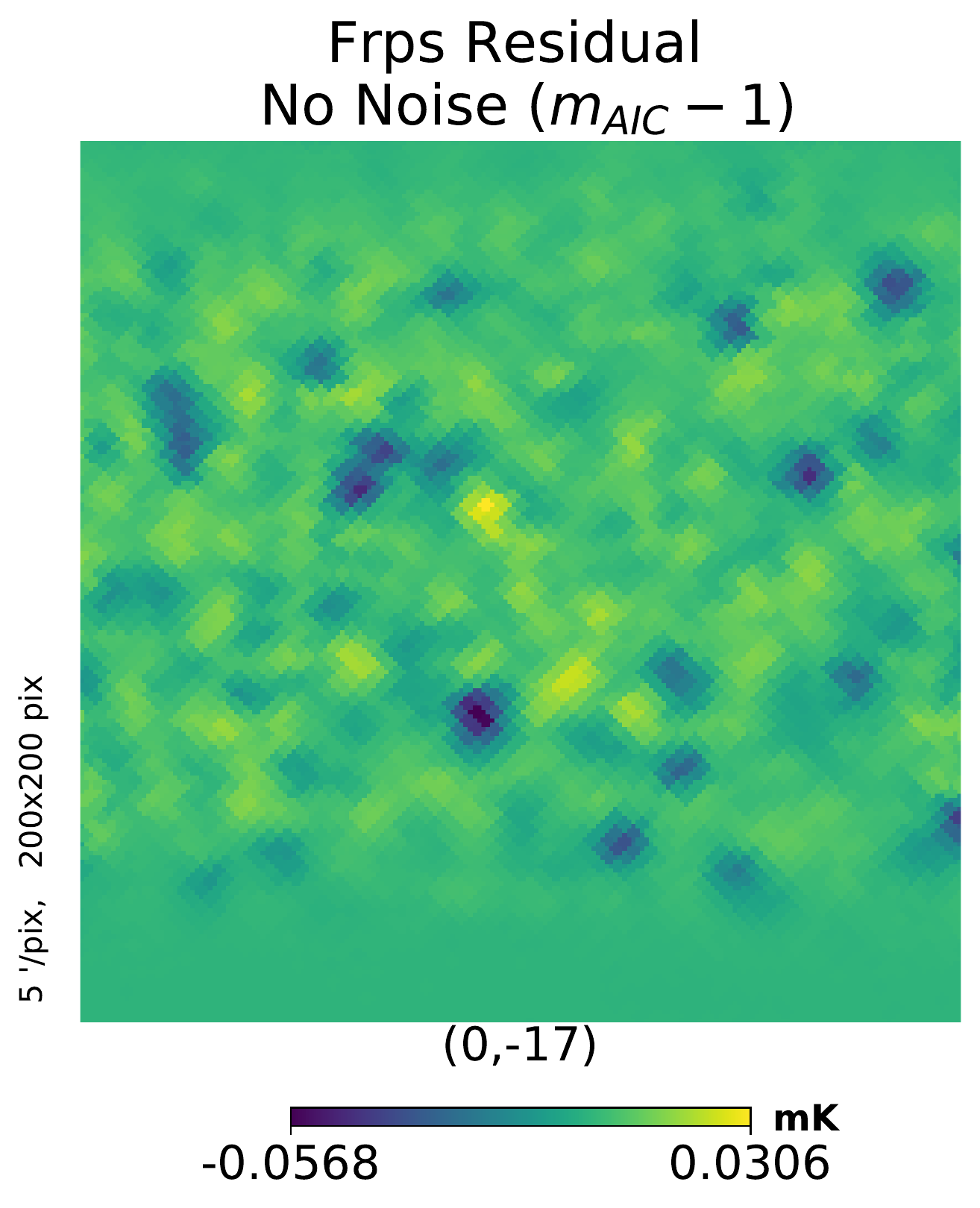}
\includegraphics[width=3.2cm]{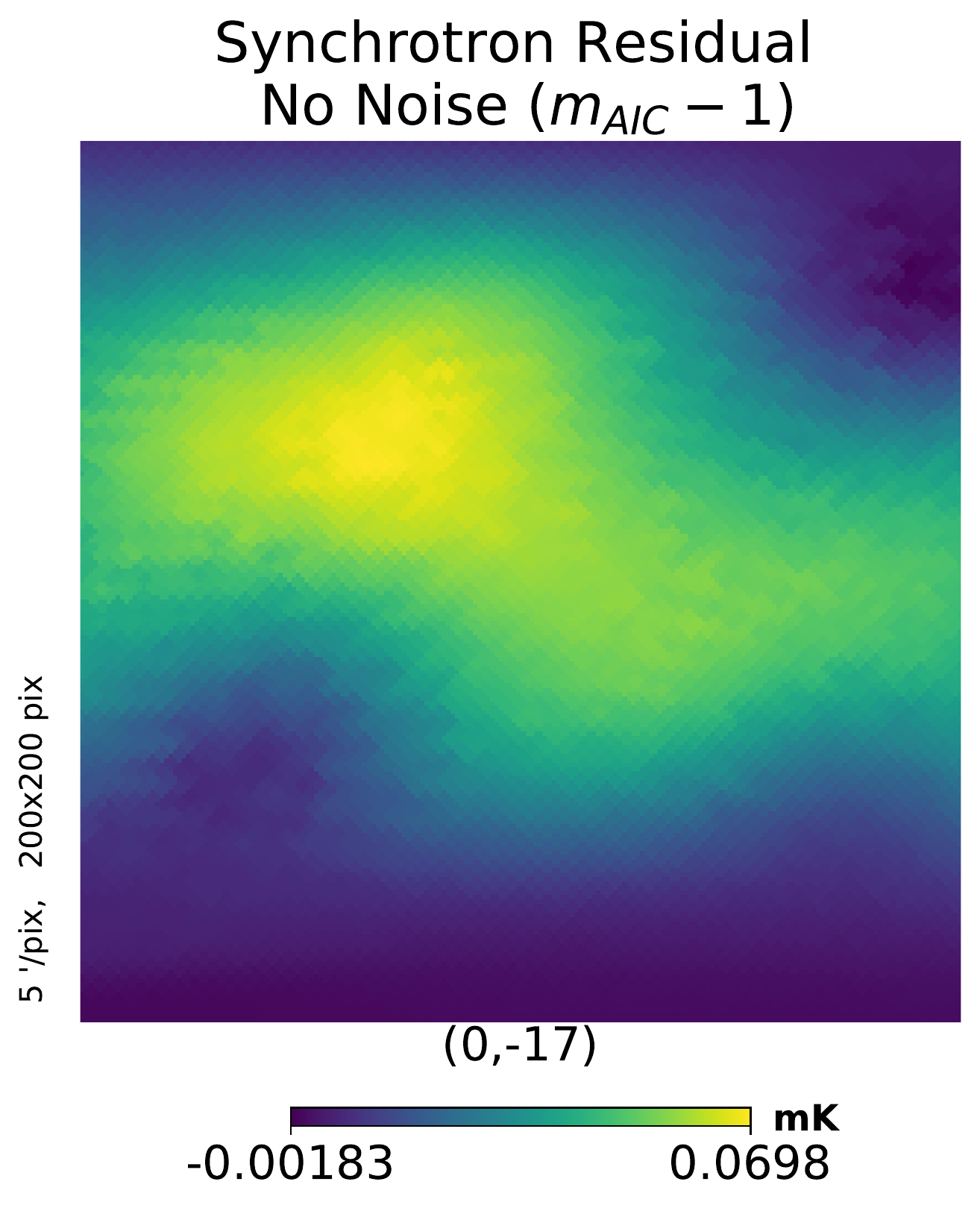} 

\includegraphics[width=3.2cm]{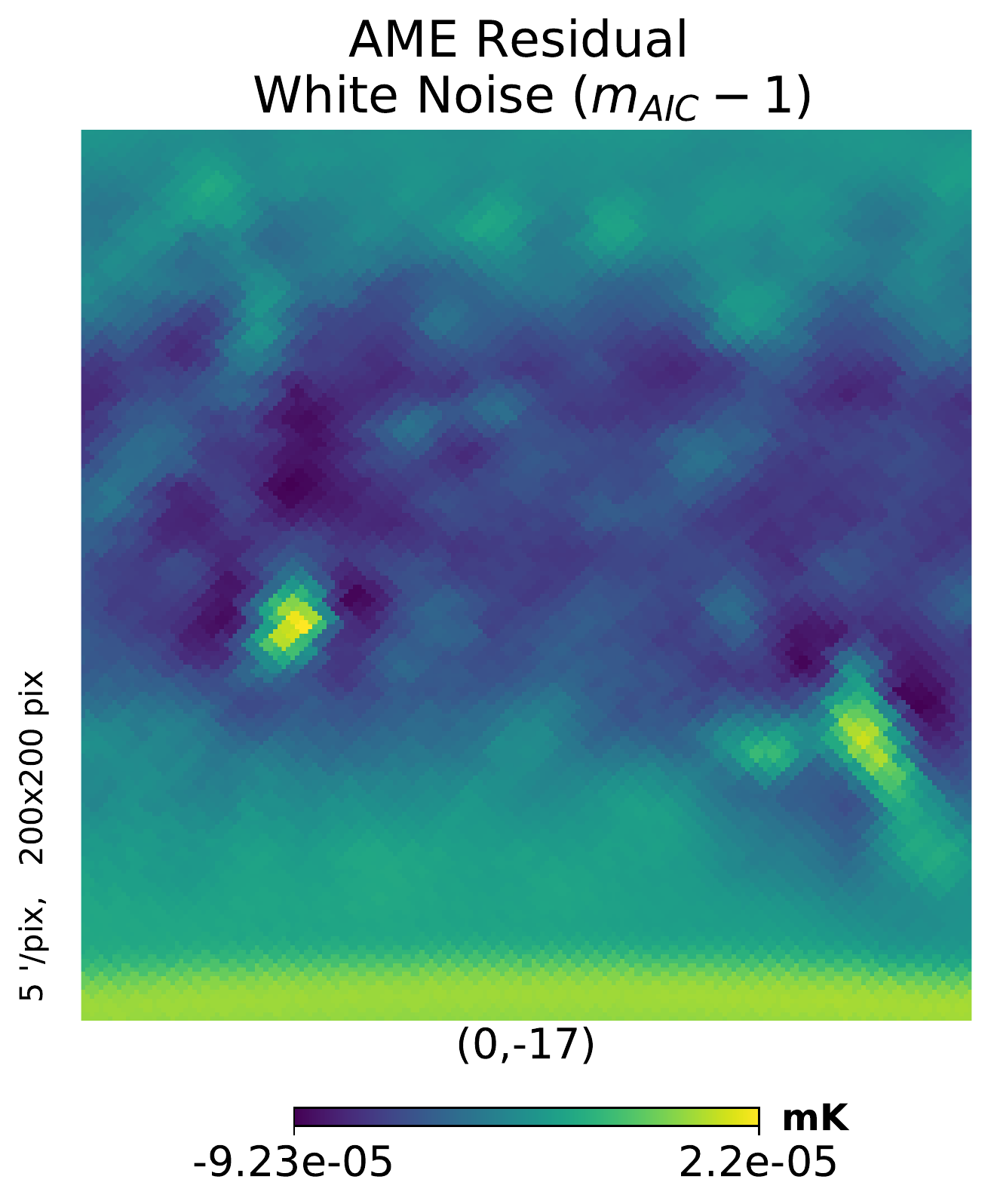}
\includegraphics[width=3.2cm]{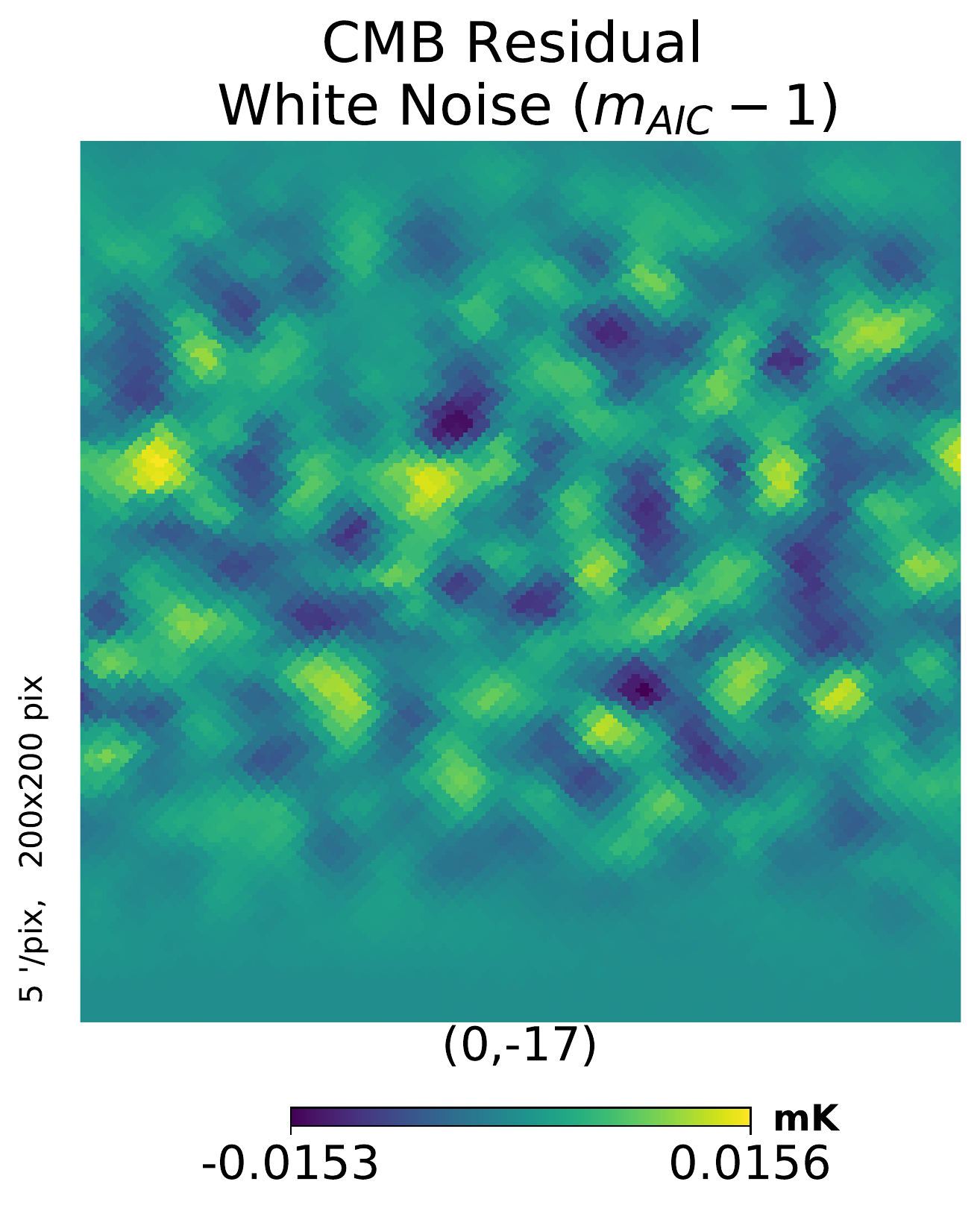}
\includegraphics[width=3.2cm]{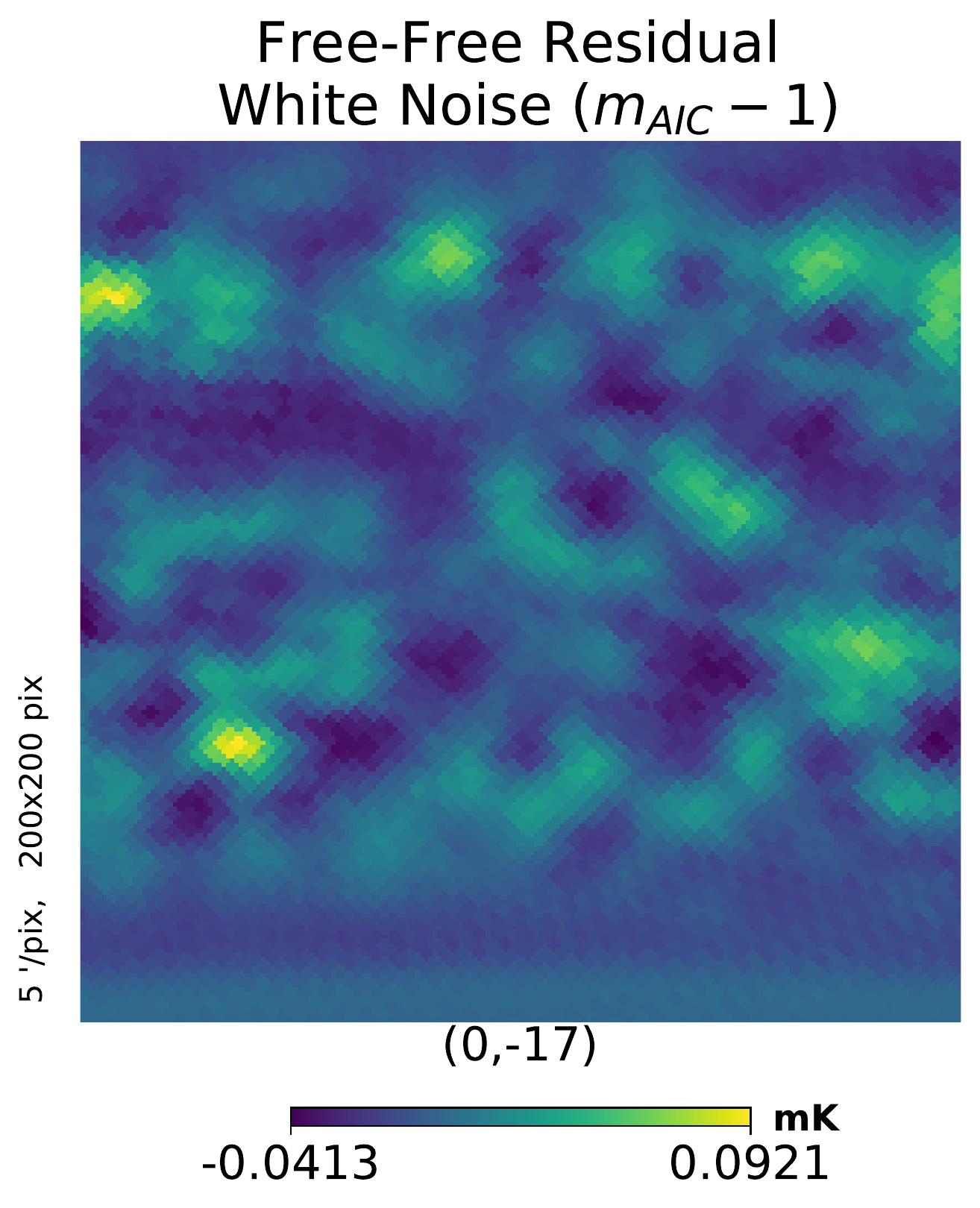}
\includegraphics[width=3.2cm]{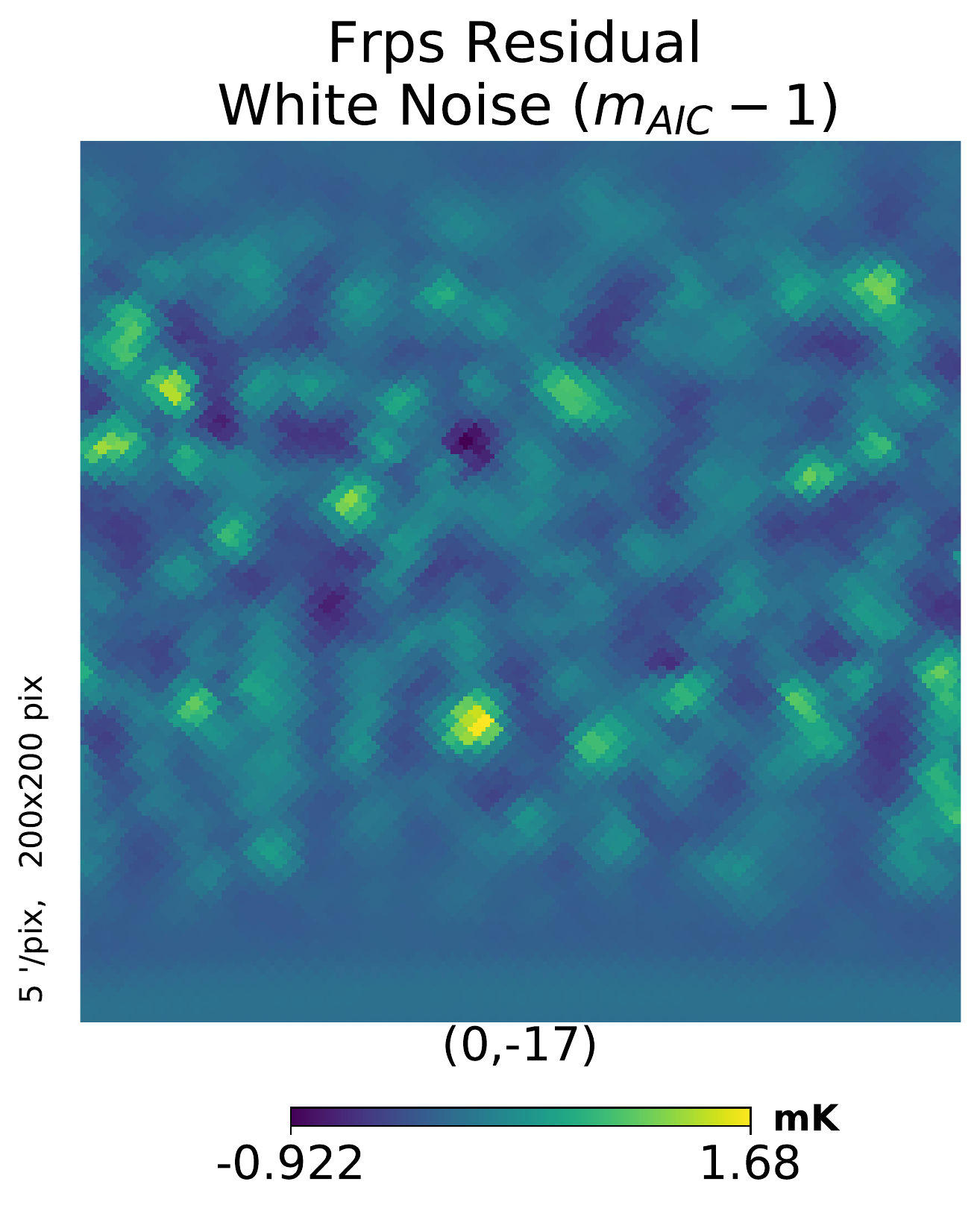}
\includegraphics[width=3.2cm]{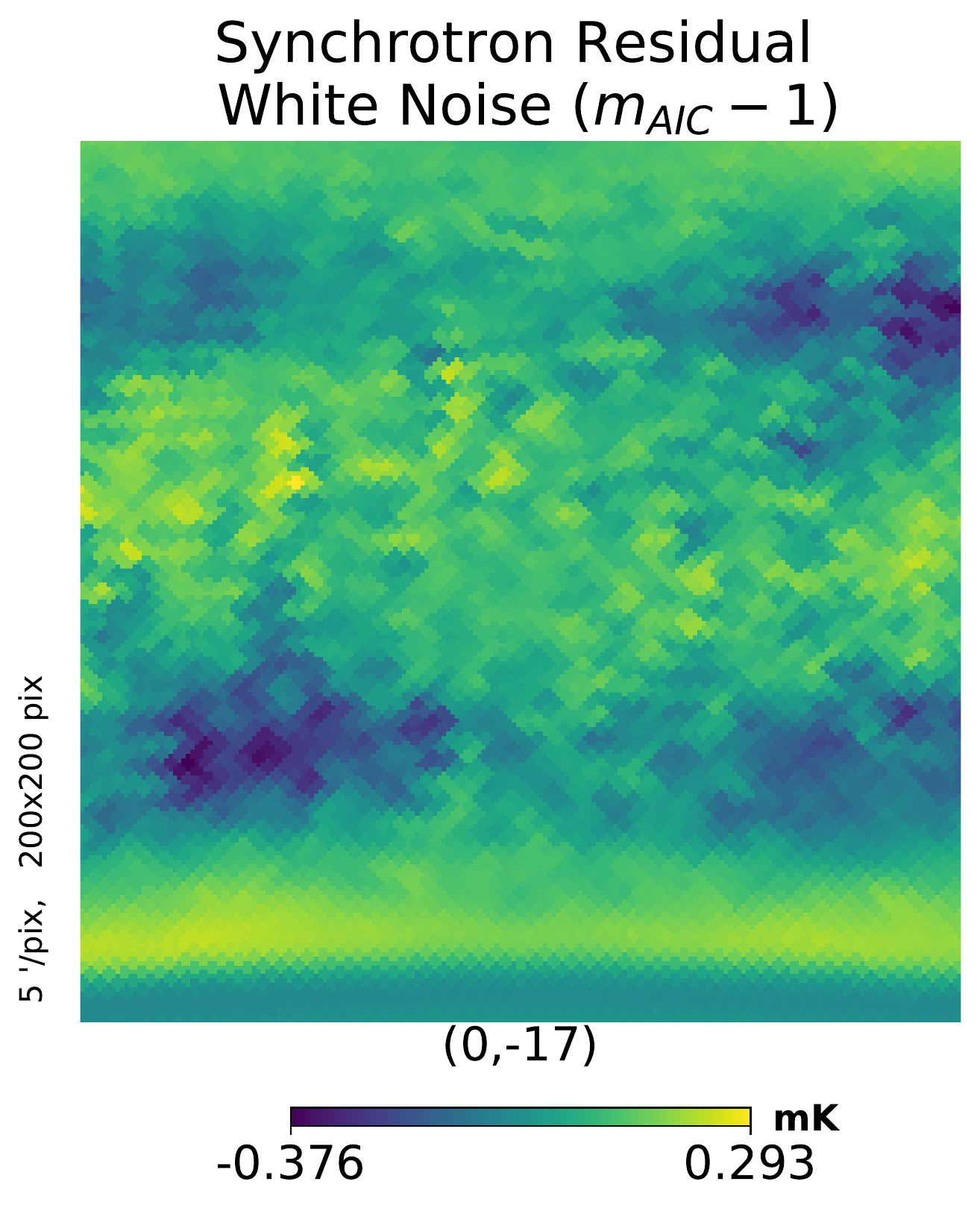} 

\includegraphics[width=3.2cm]{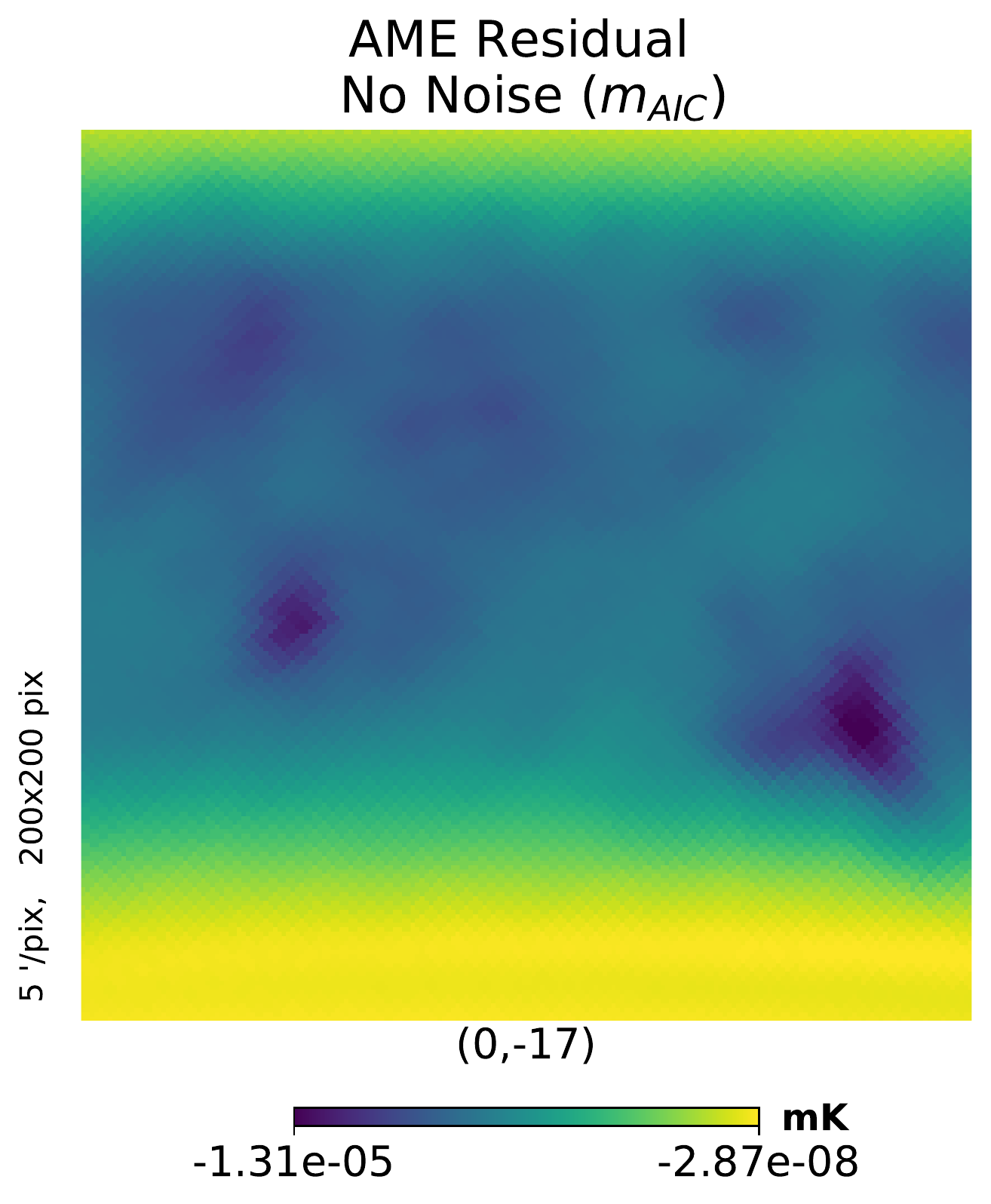}
\includegraphics[width=3.2cm]{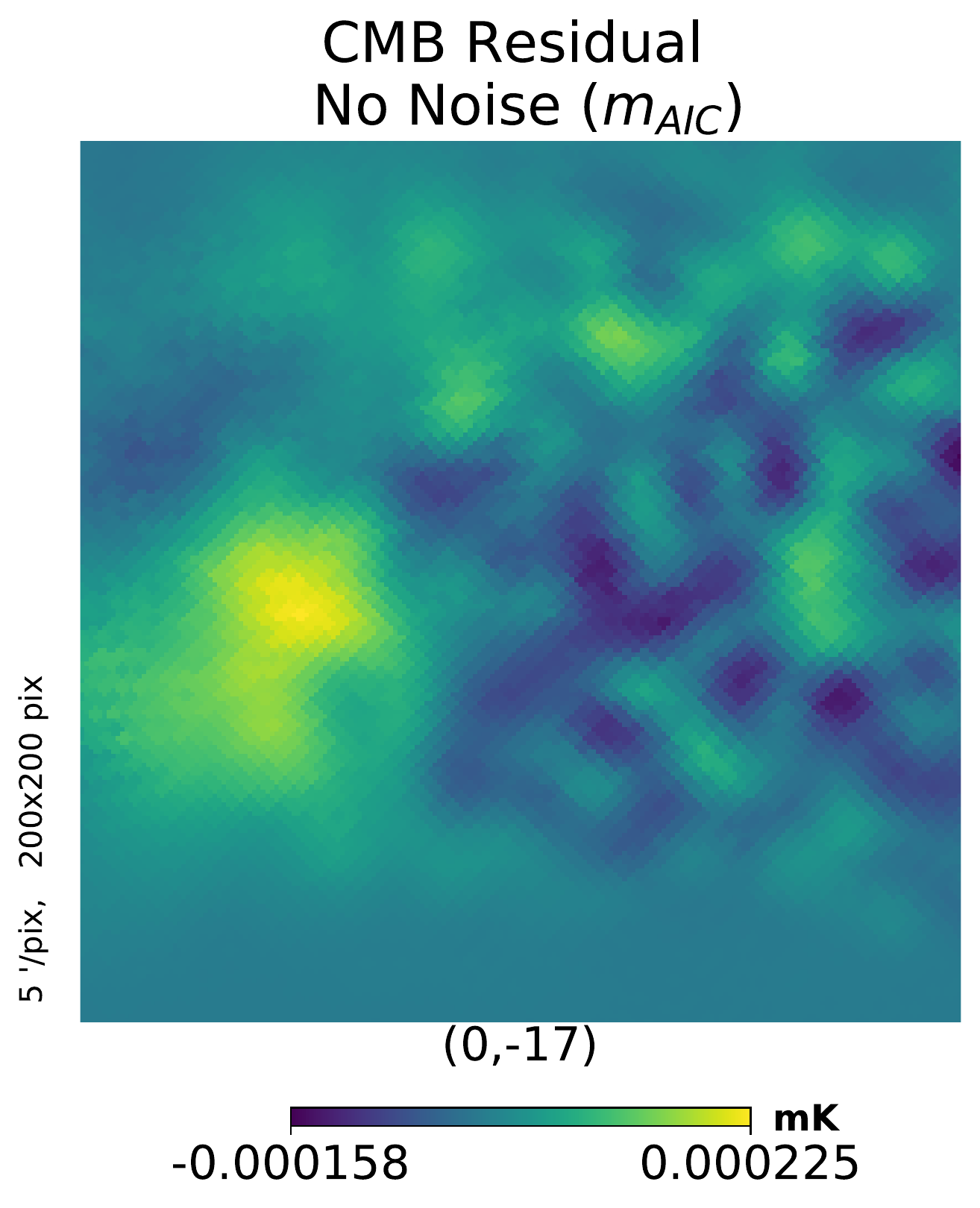}
\includegraphics[width=3.2cm]{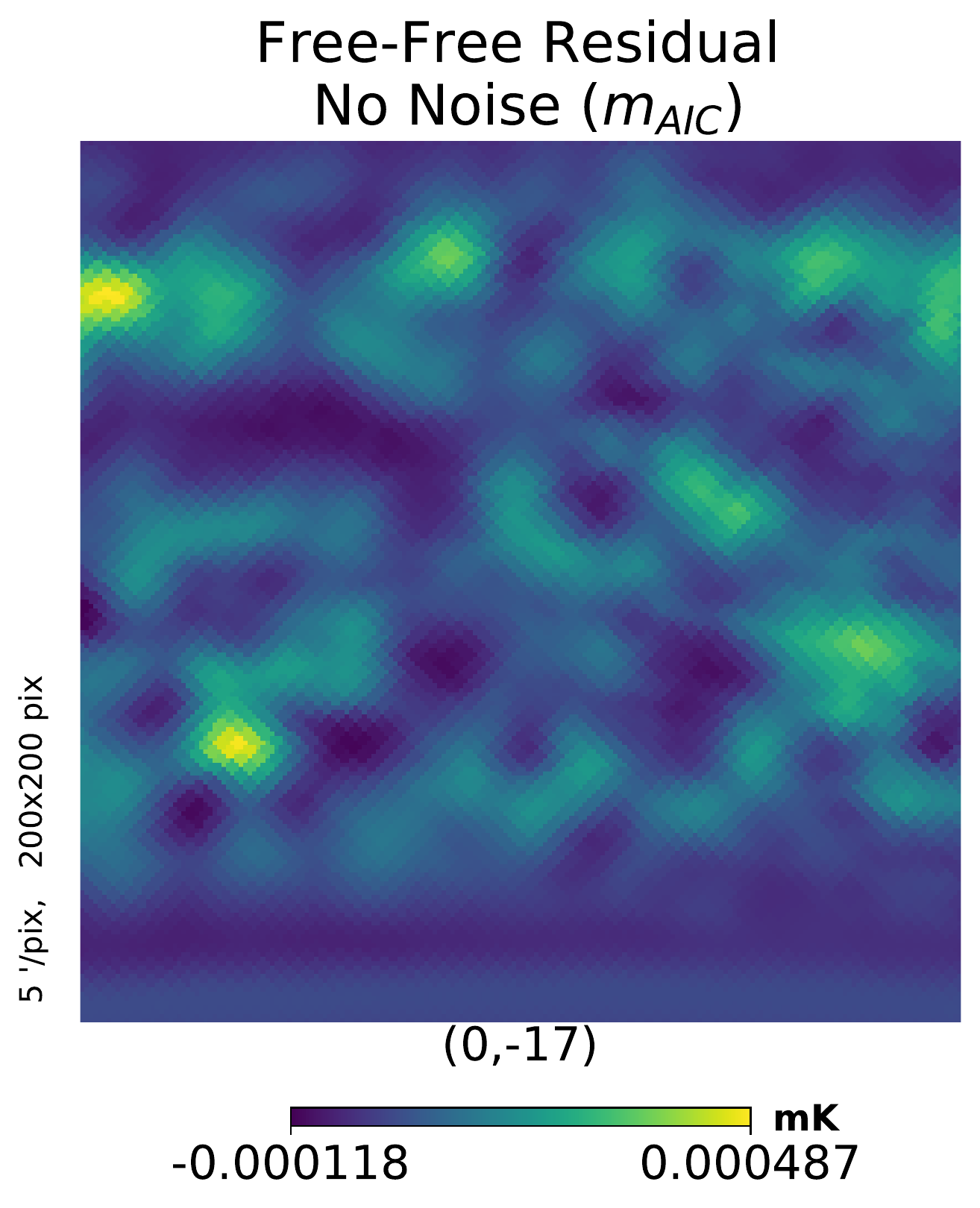}
\includegraphics[width=3.2cm]{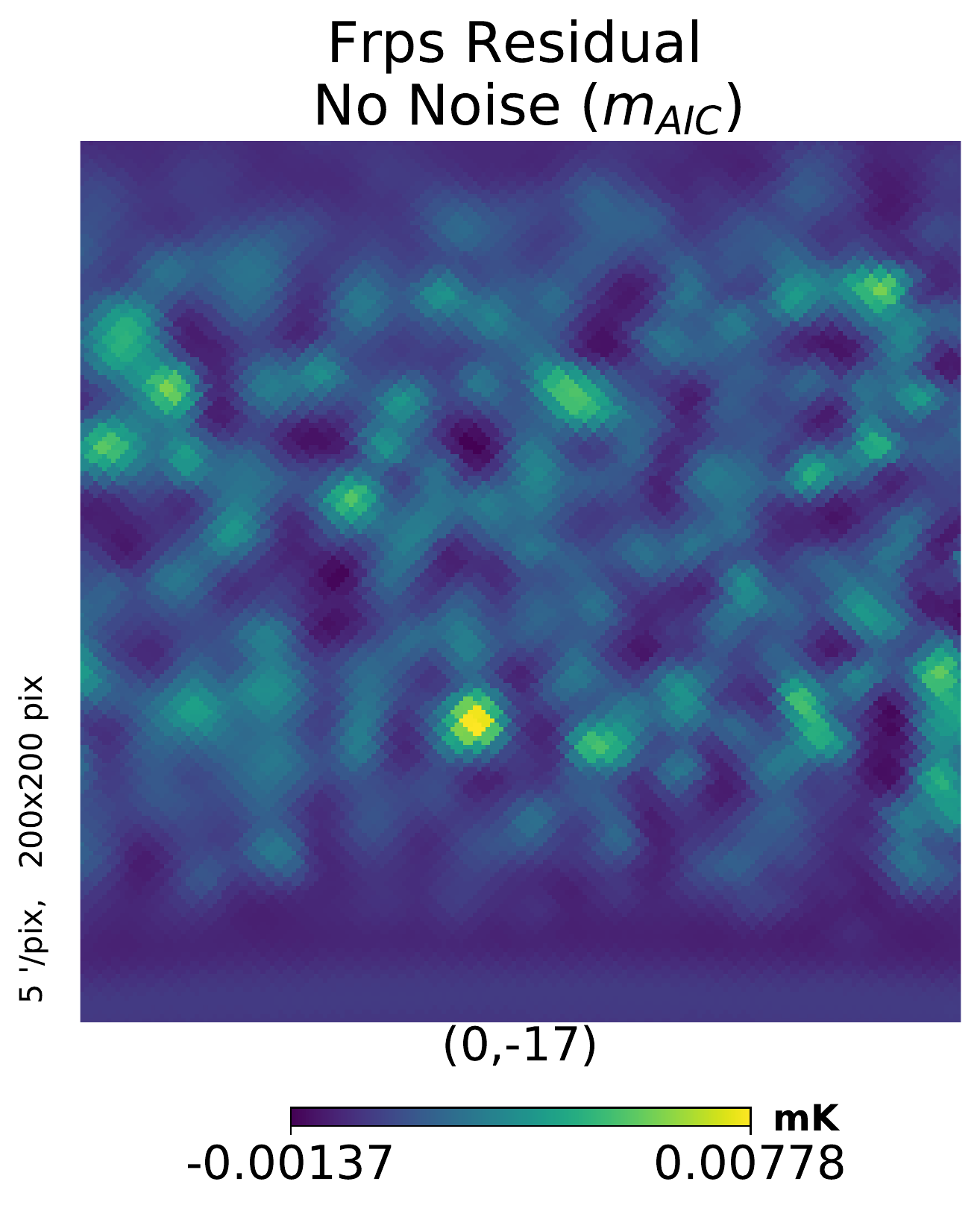}
\includegraphics[width=3.2cm]{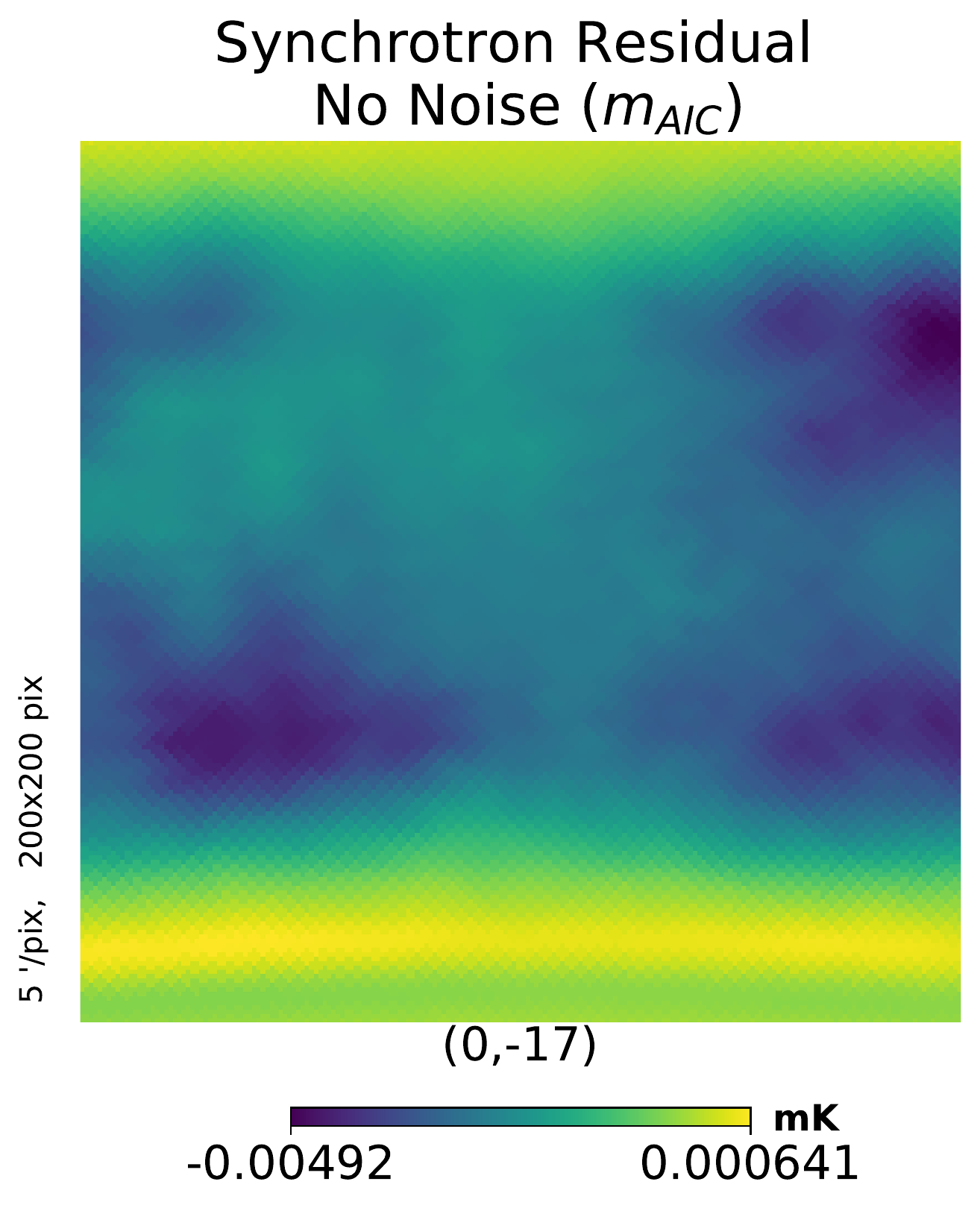} 

\includegraphics[width=3.2cm]{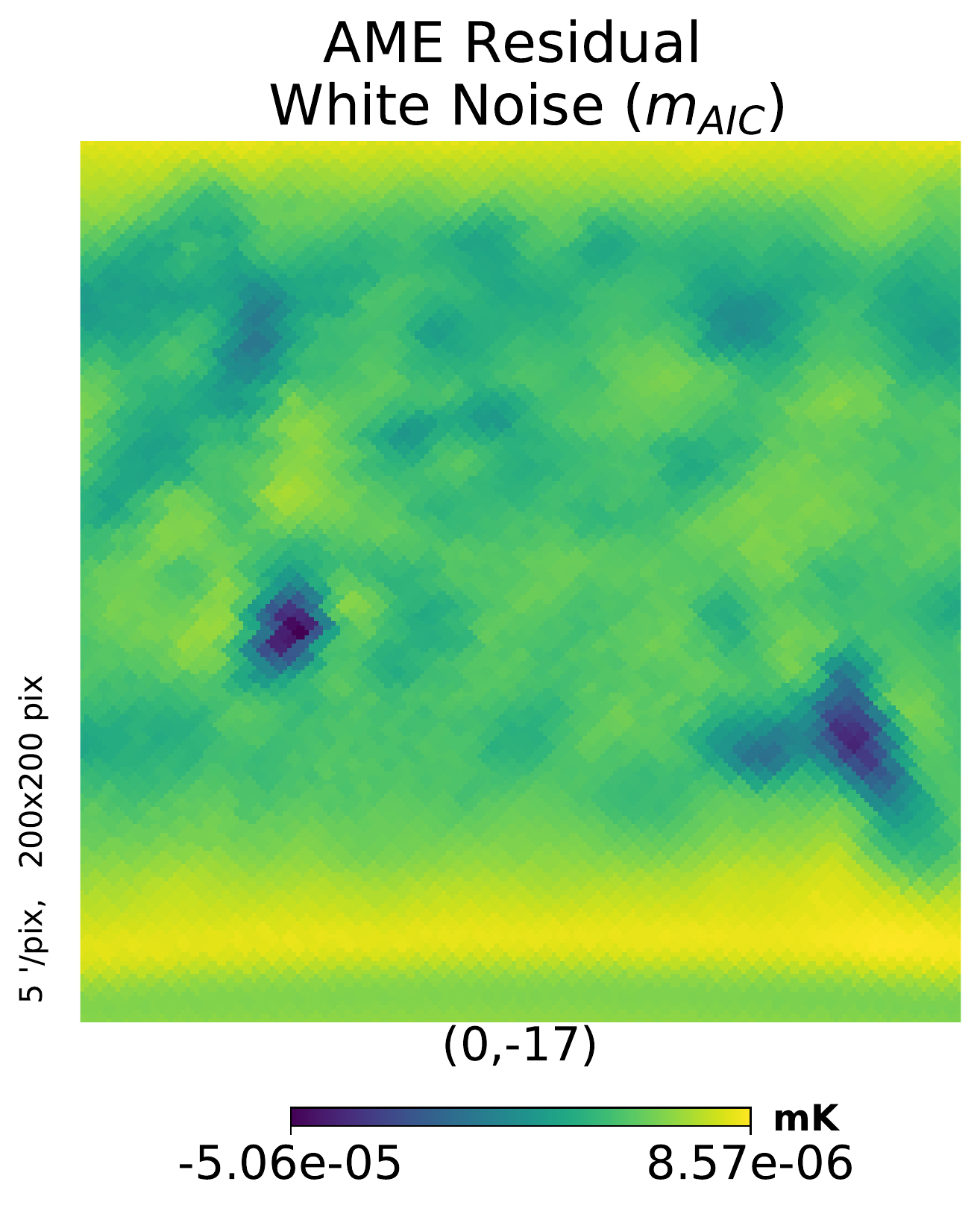}
\includegraphics[width=3.2cm]{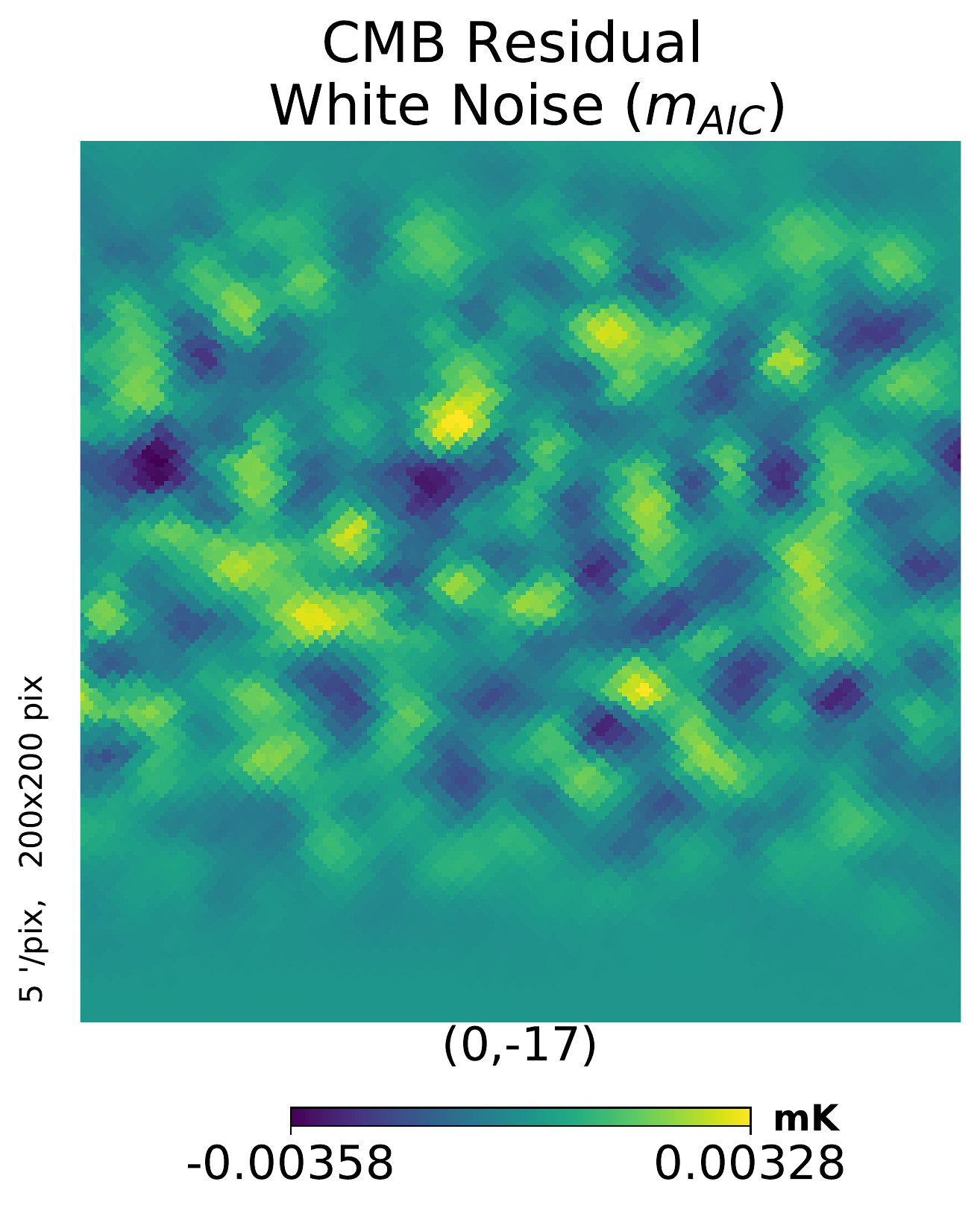}
\includegraphics[width=3.2cm]{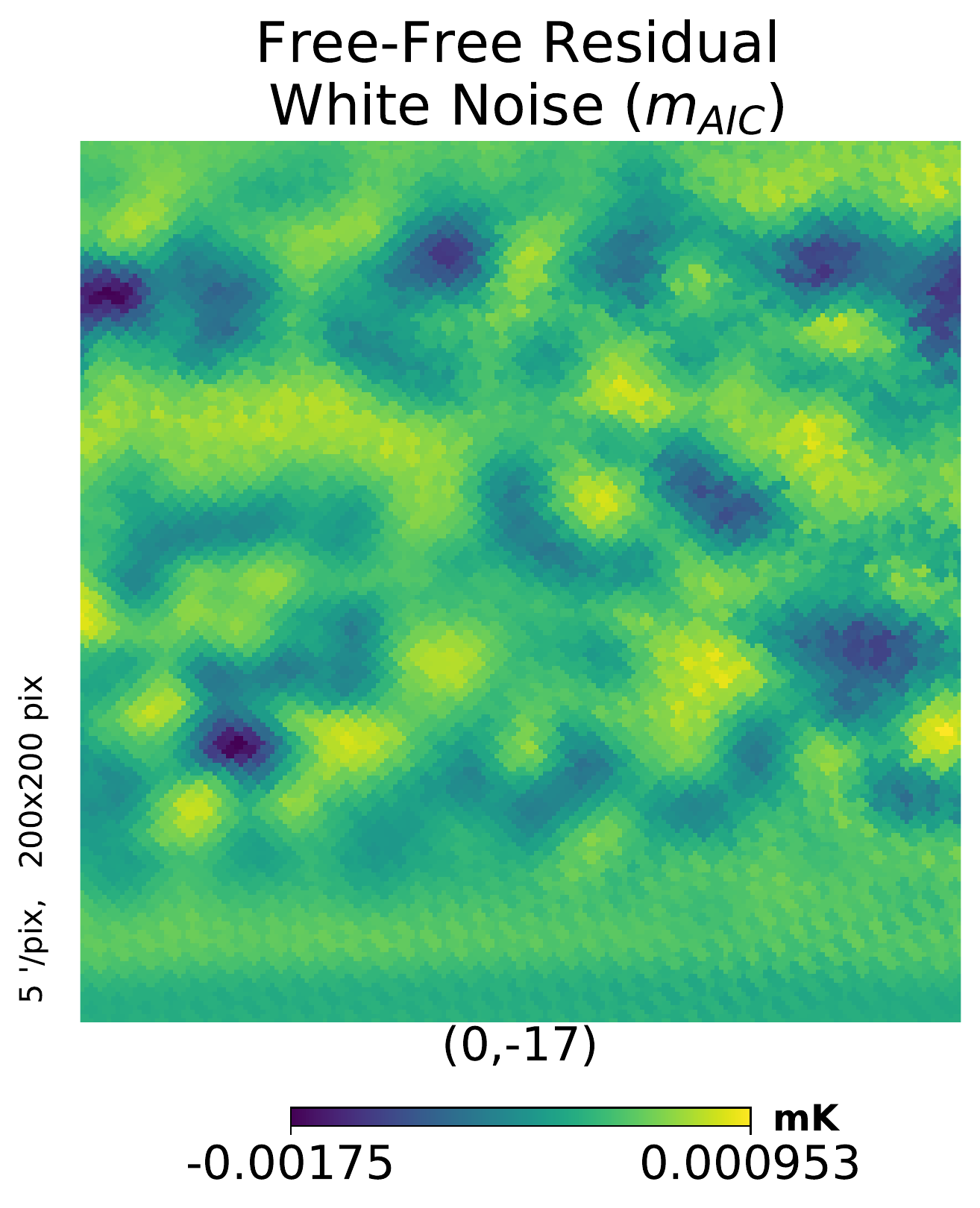}
\includegraphics[width=3.2cm]{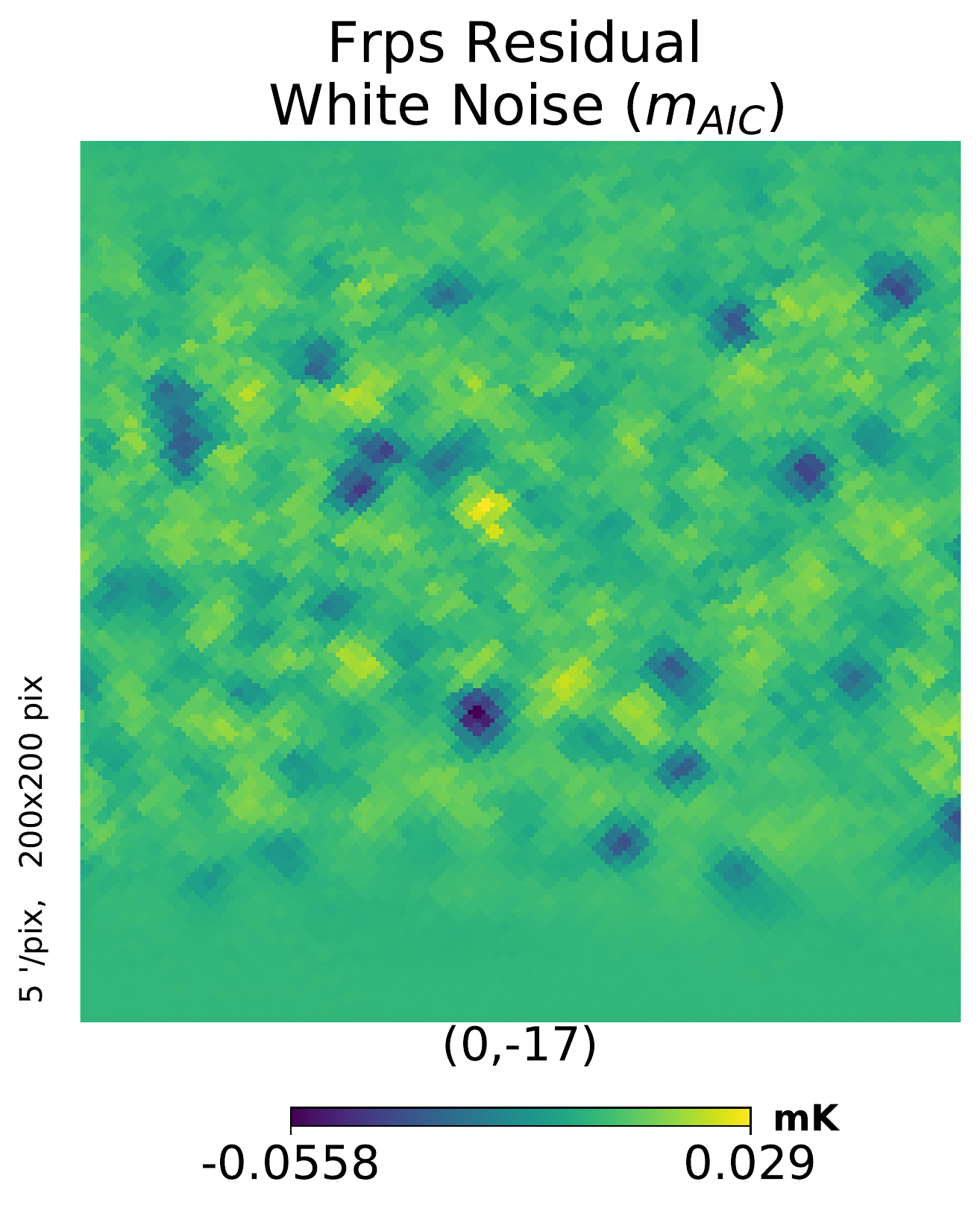}
\includegraphics[width=3.2cm]{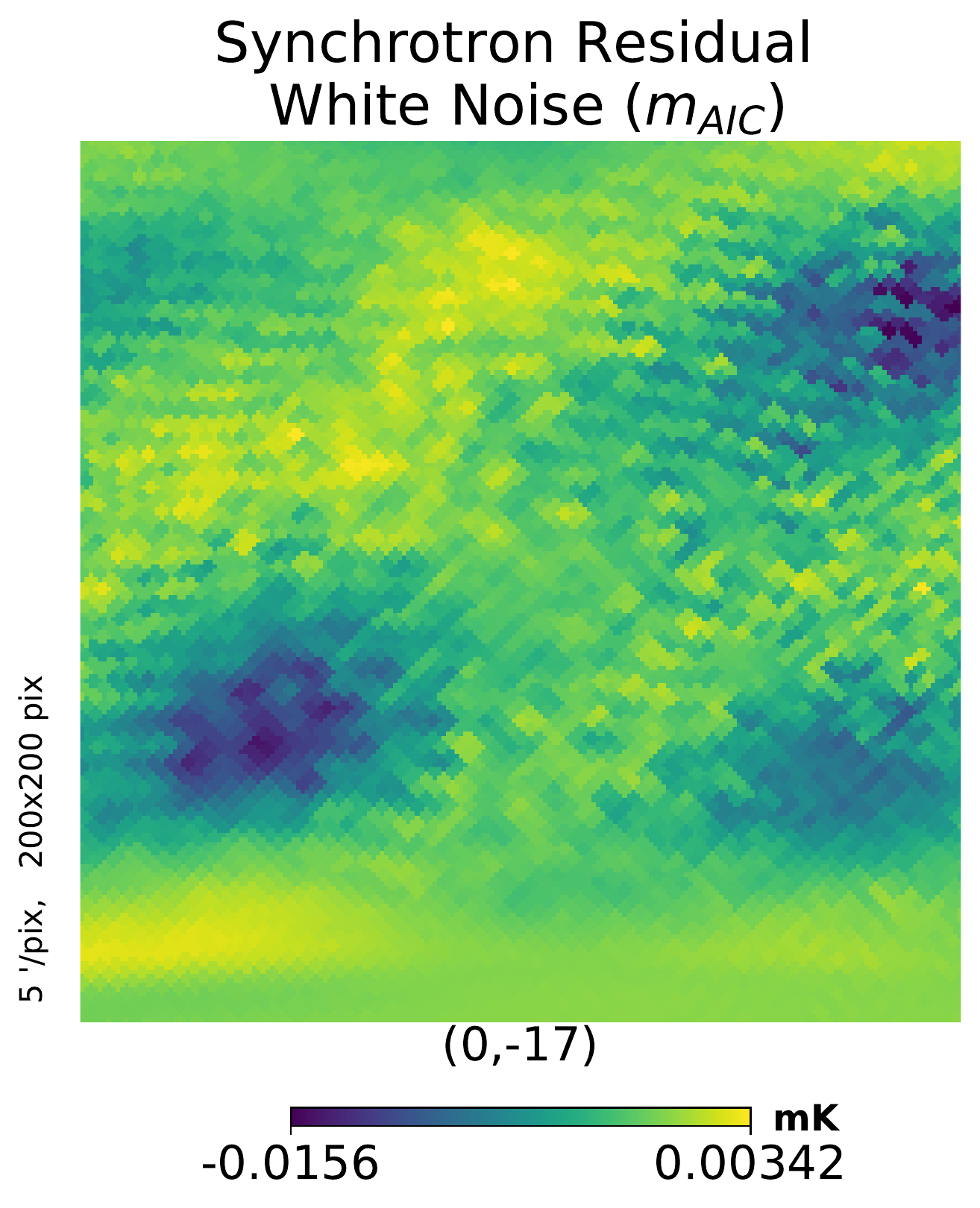} 

\includegraphics[width=3.2cm]{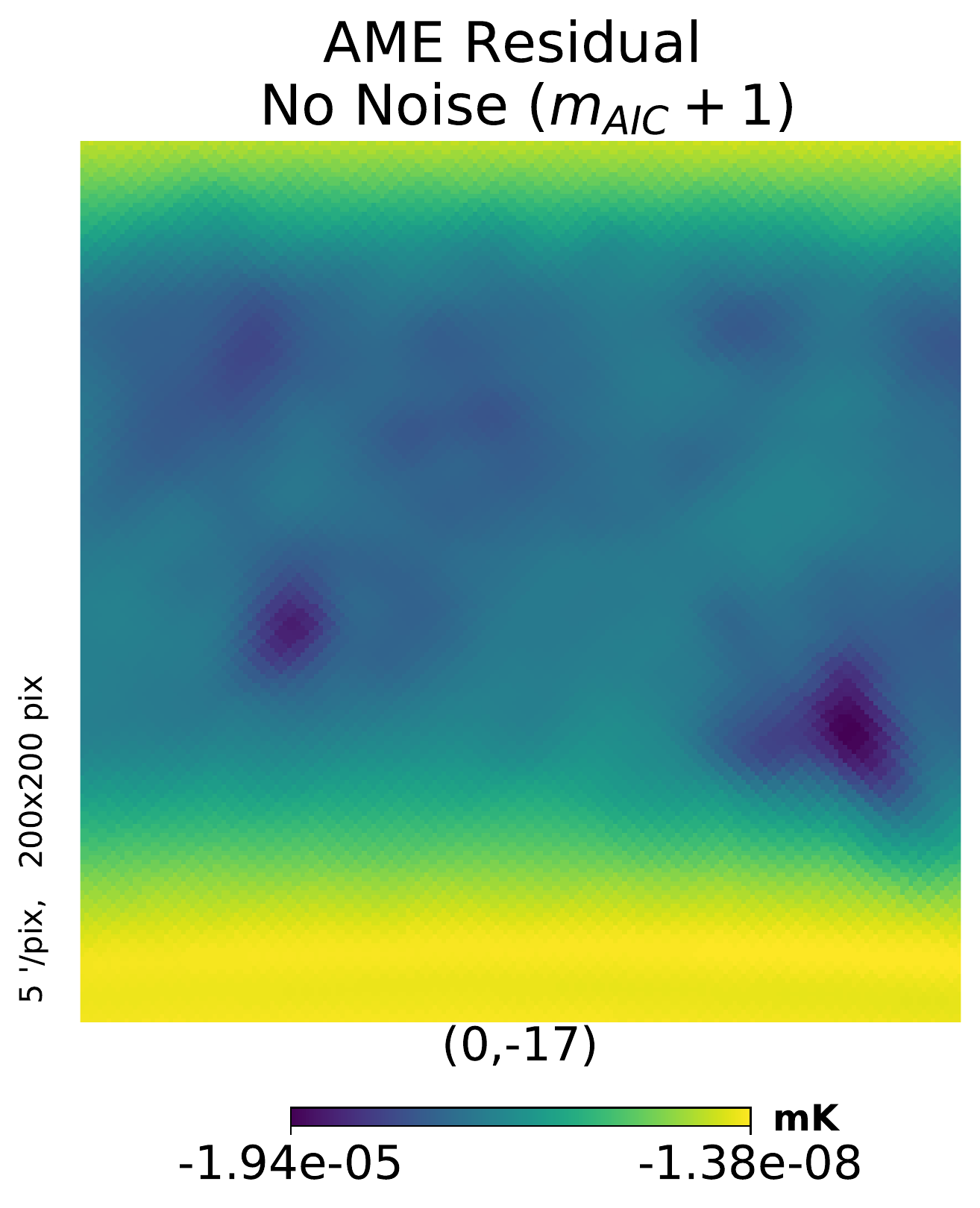}
\includegraphics[width=3.2cm]{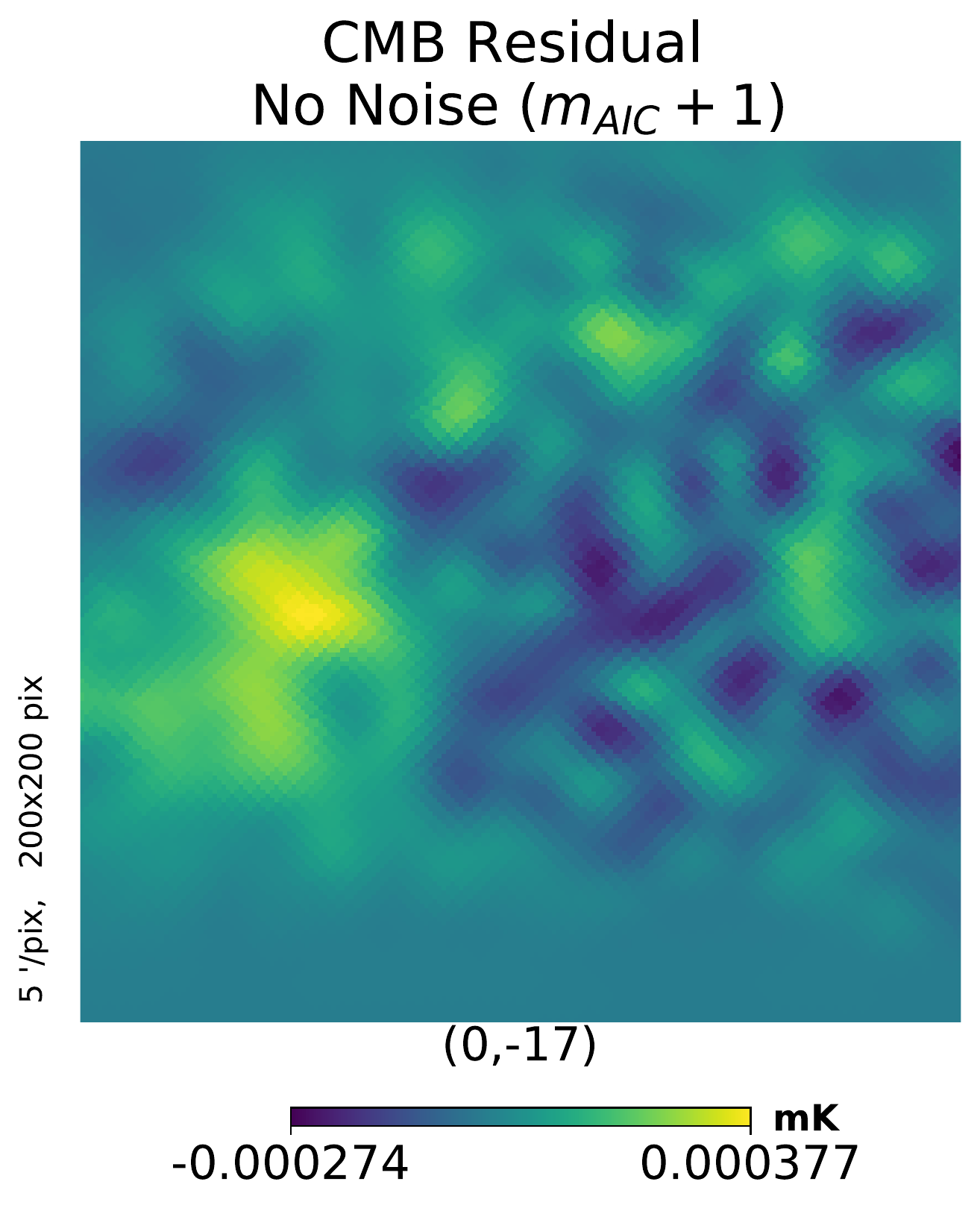}
\includegraphics[width=3.2cm]{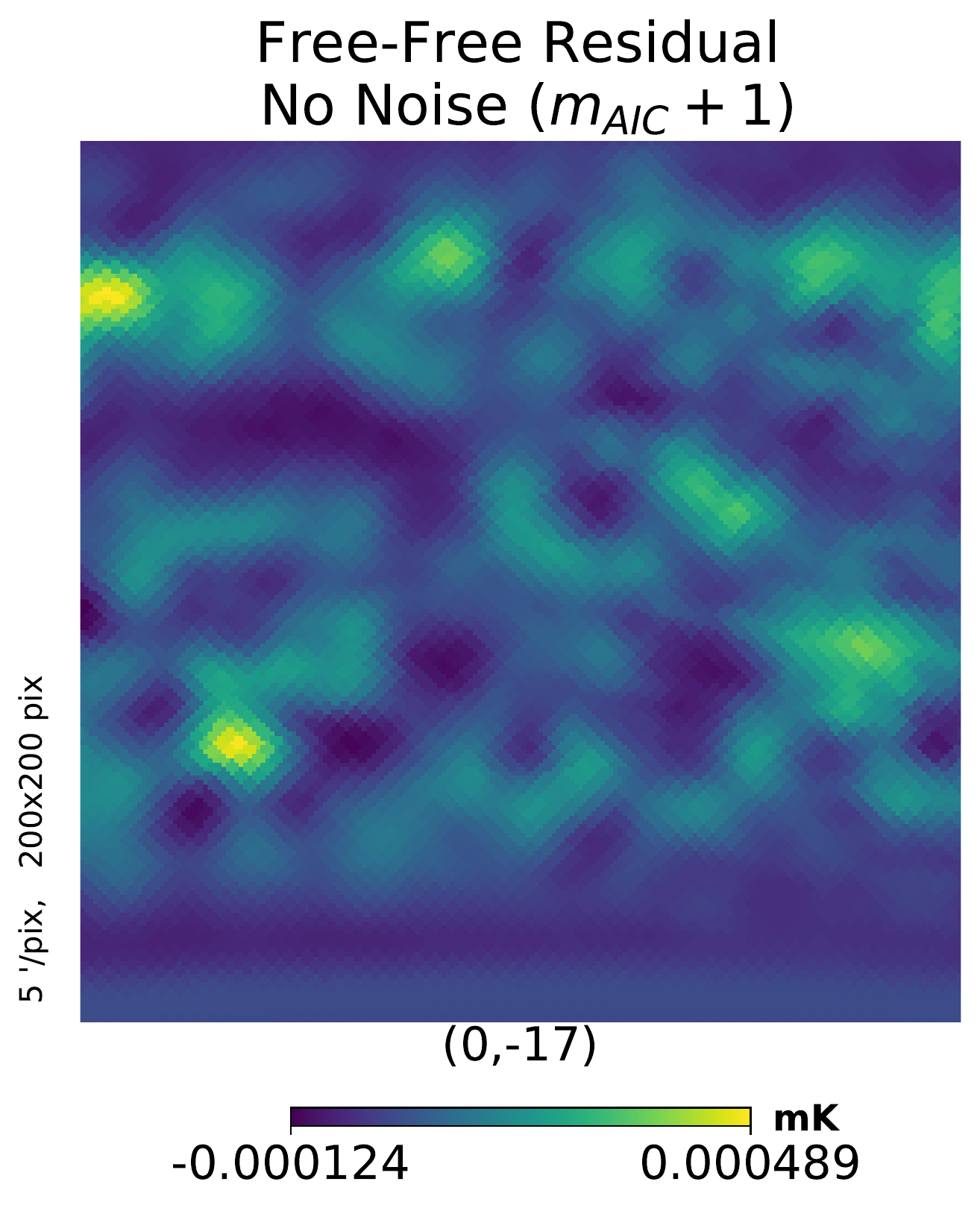}
\includegraphics[width=3.2cm]{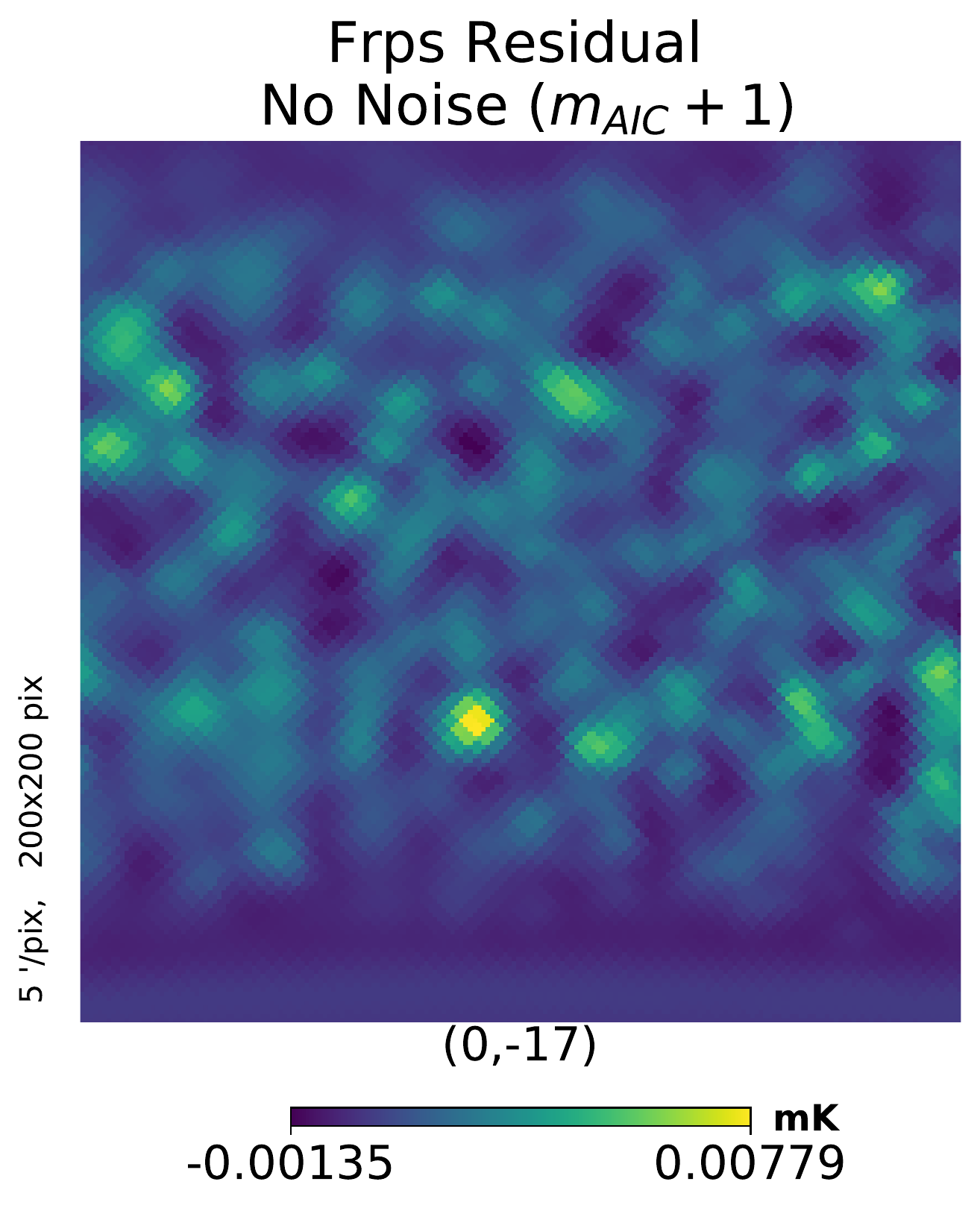}
\includegraphics[width=3.2cm]{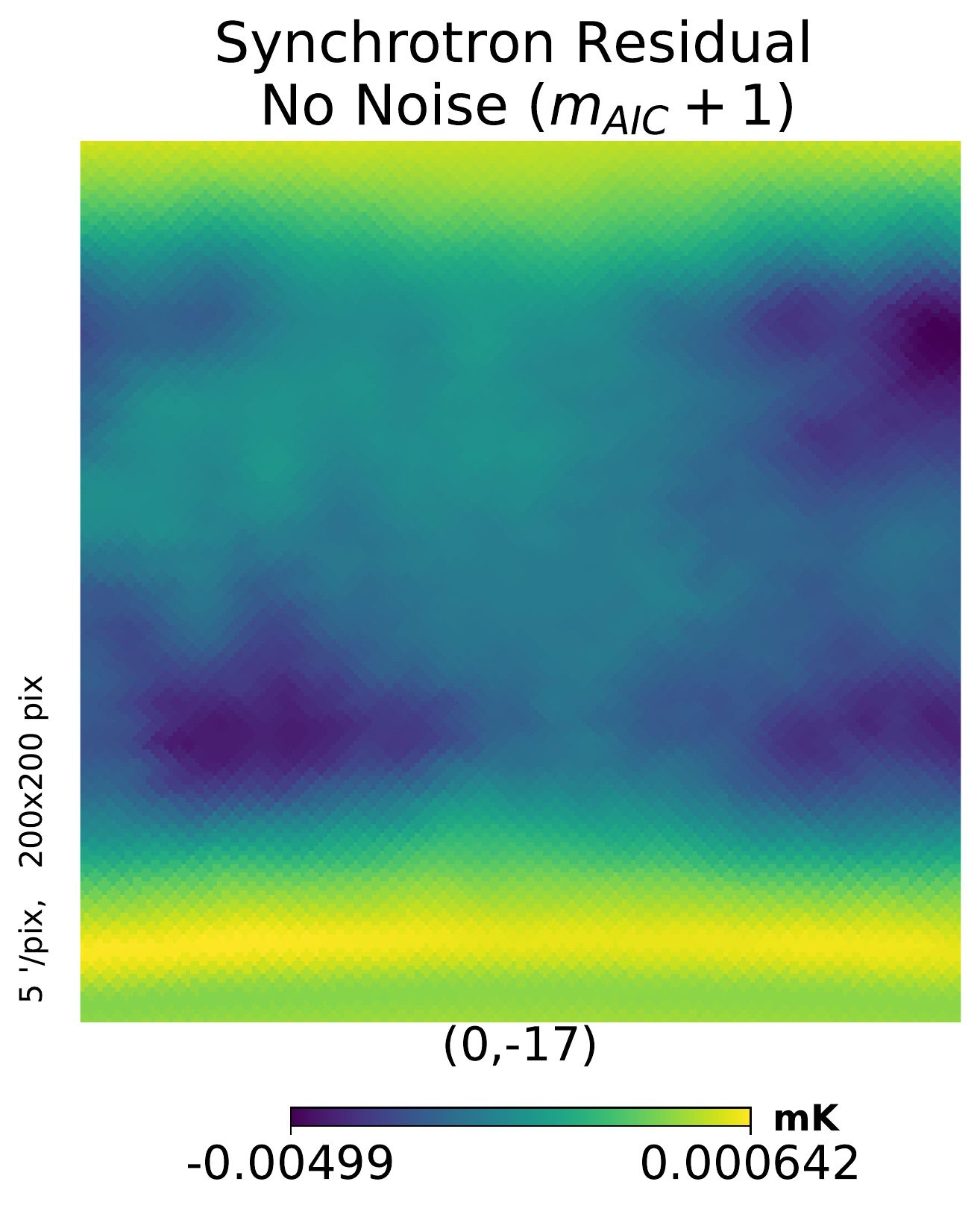} 
\includegraphics[width=3.2cm]{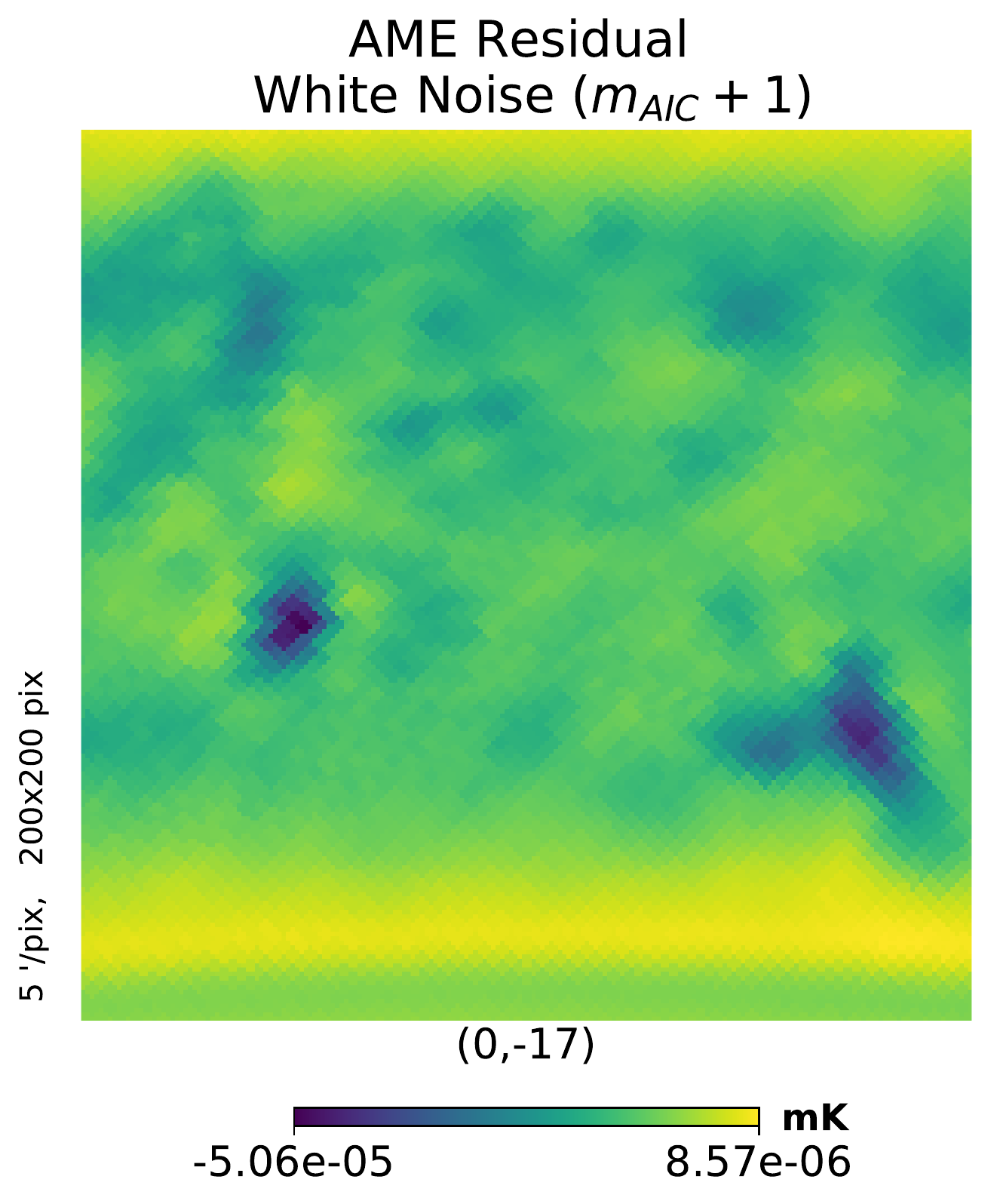}
\includegraphics[width=3.2cm]{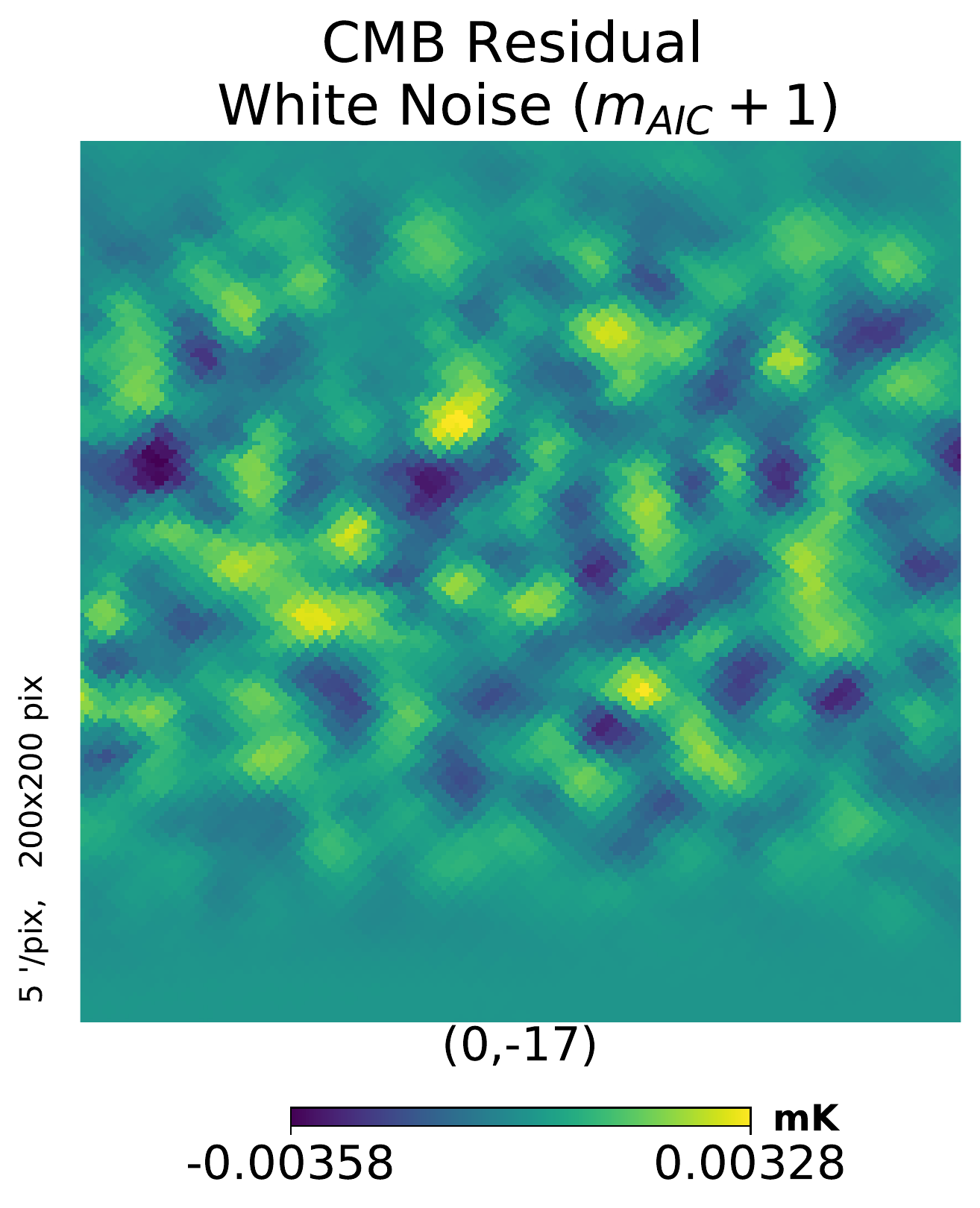}
\includegraphics[width=3.2cm]{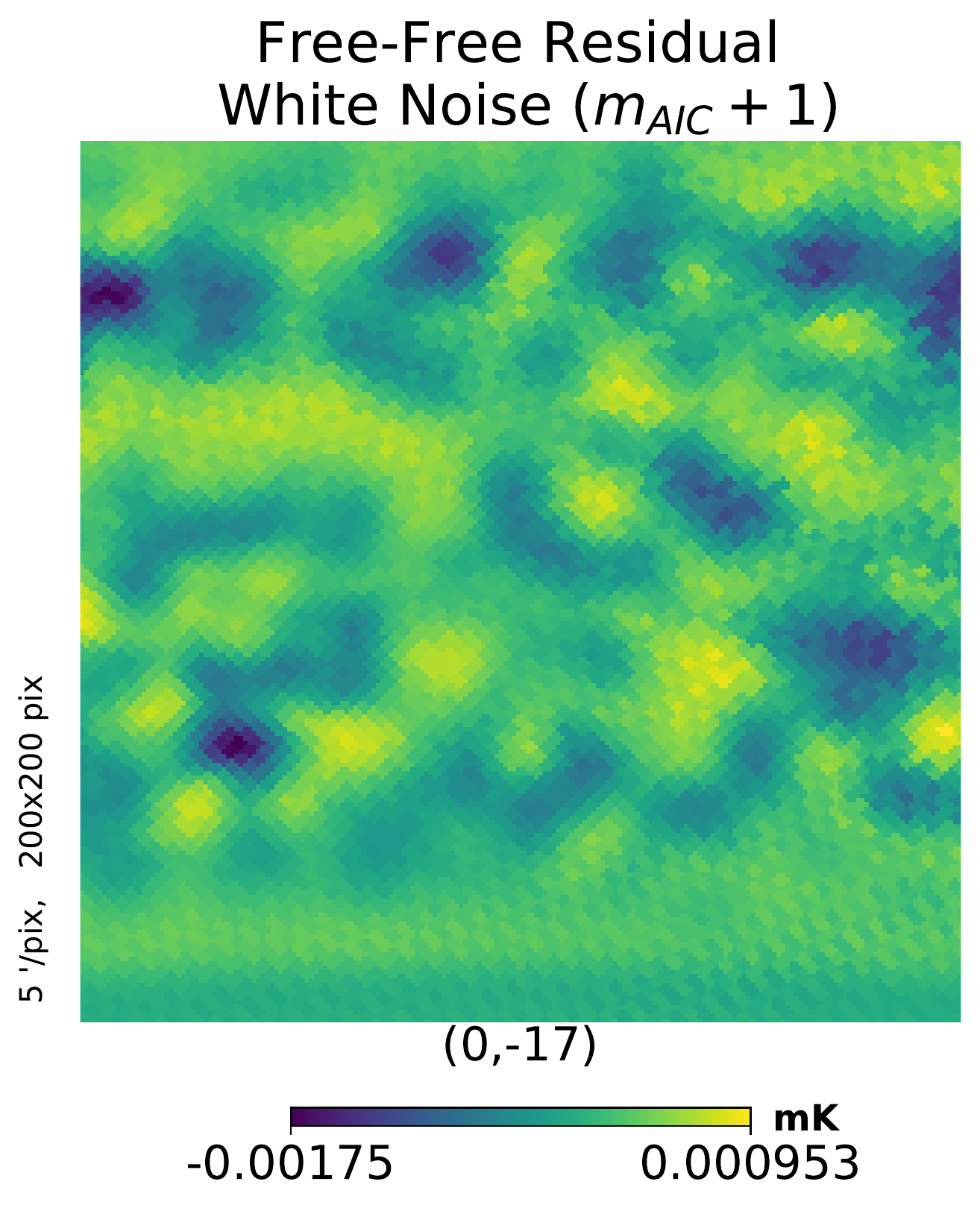}
\includegraphics[width=3.2cm]{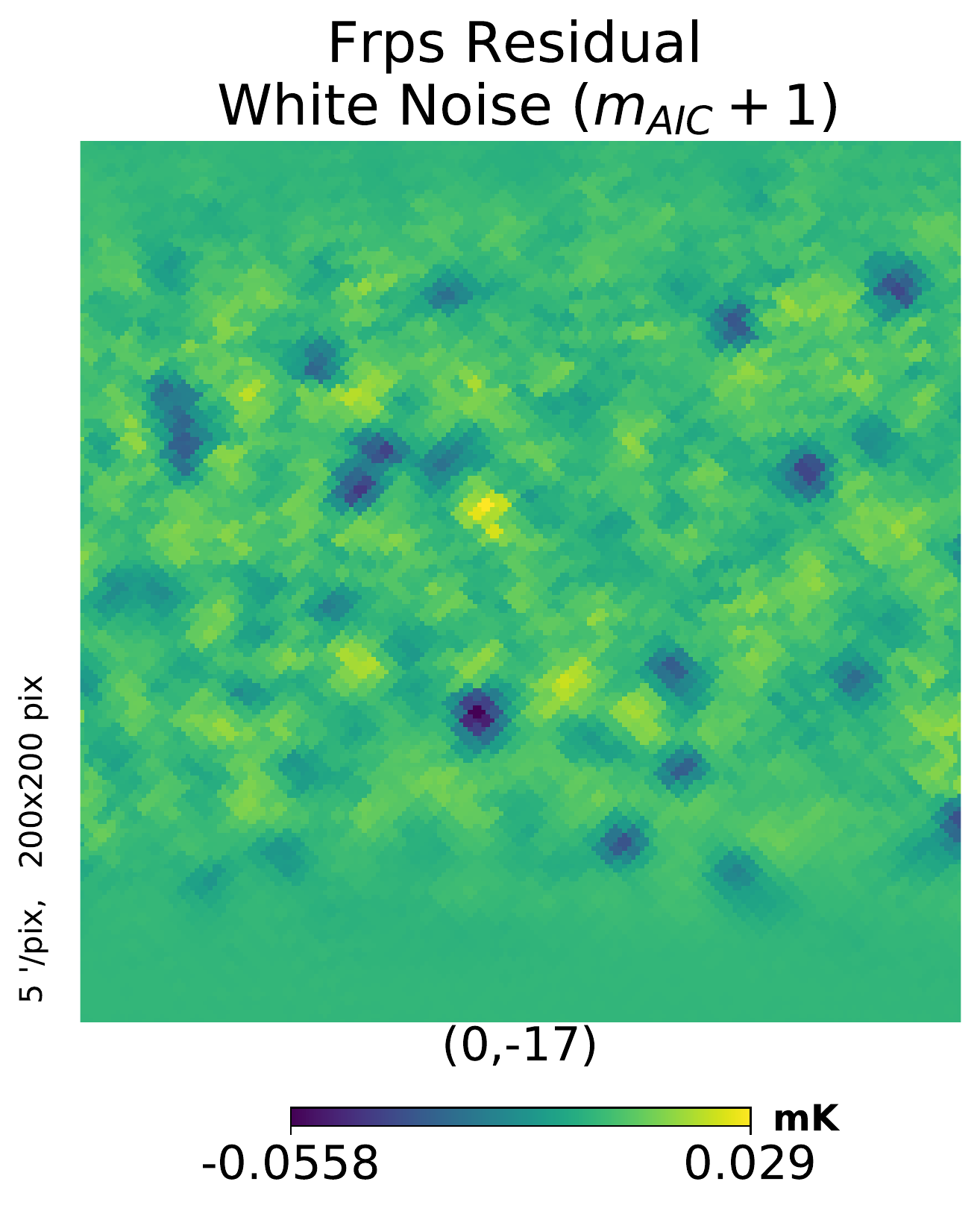}
\includegraphics[width=3.2cm]{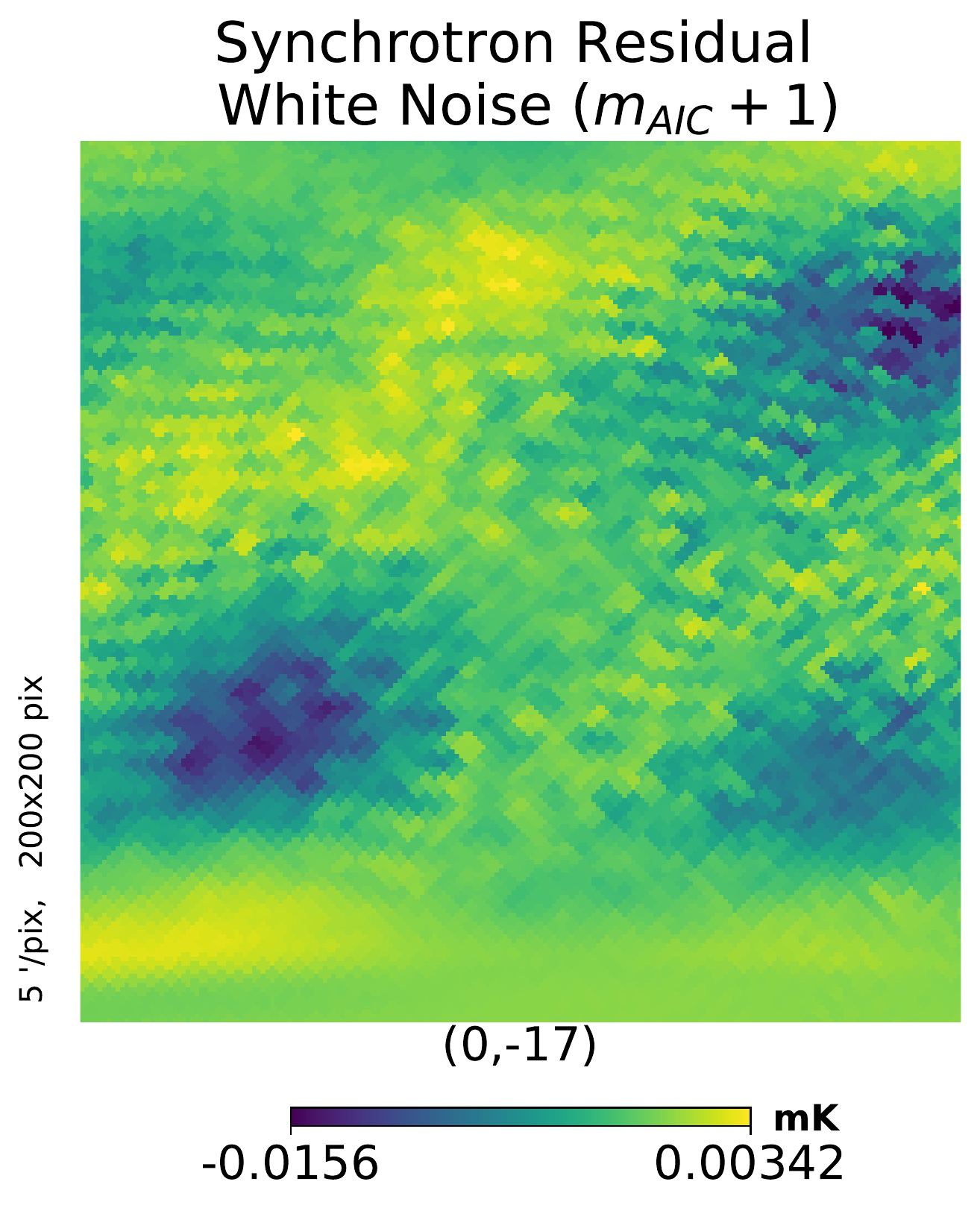} 

\caption{Gnomonic view of residual foreground components for dimension $m_\mathrm{AIC}-1$, $m_\mathrm{AIC}$ and $m_\mathrm{AIC}+1$, with and without noise in the simulation.}
\label{fig:gnom_ndims}
\end{figure*}

\begin{figure*}
\begin{centering}
\includegraphics[width=.33\textwidth]{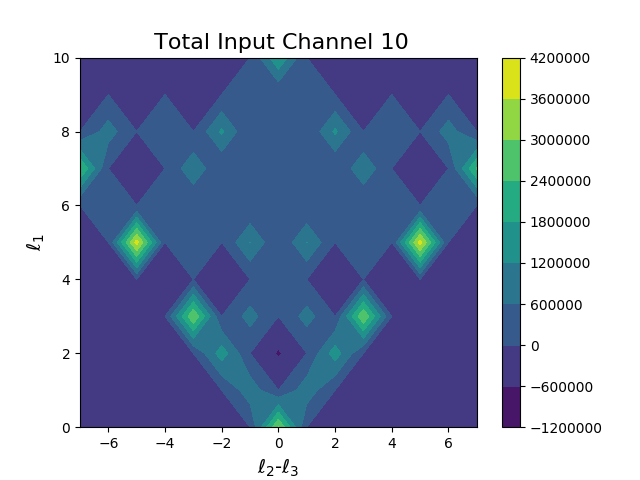}
\includegraphics[width=.33\textwidth]{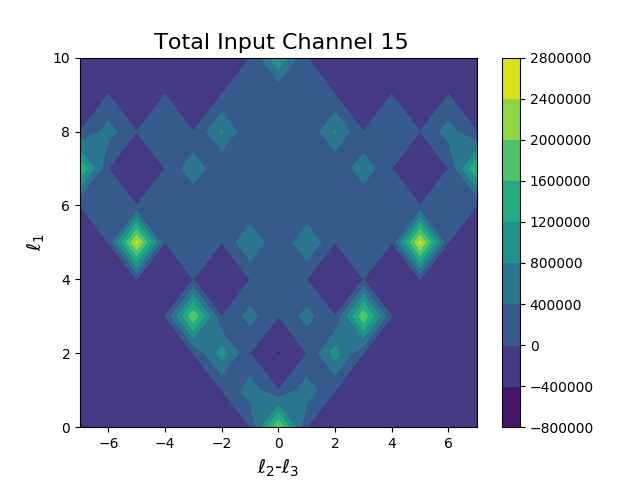}
\includegraphics[width=.33\textwidth]{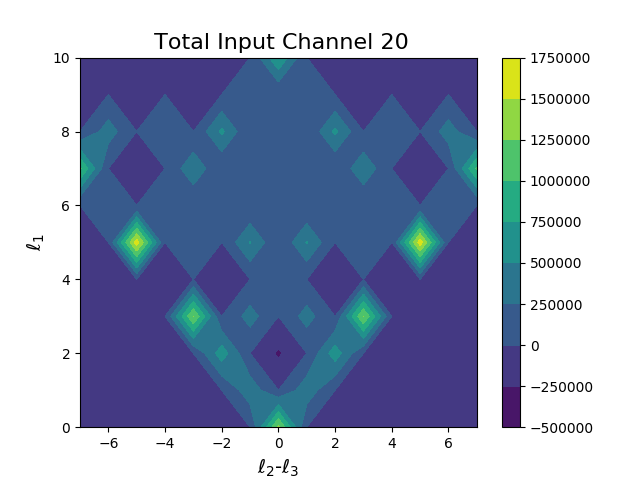}\\[\smallskipamount]
 
\includegraphics[width=.33\textwidth]{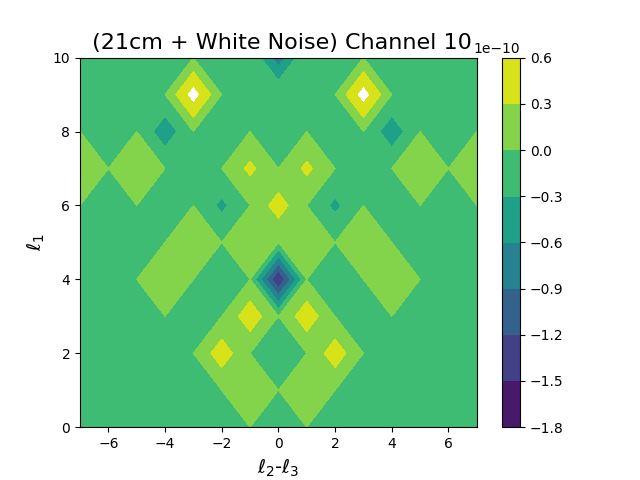}
\includegraphics[width=.33\textwidth]{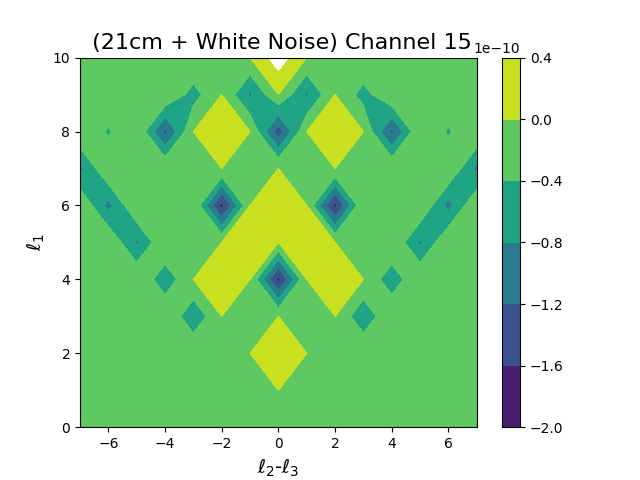}
\includegraphics[width=.33\textwidth]{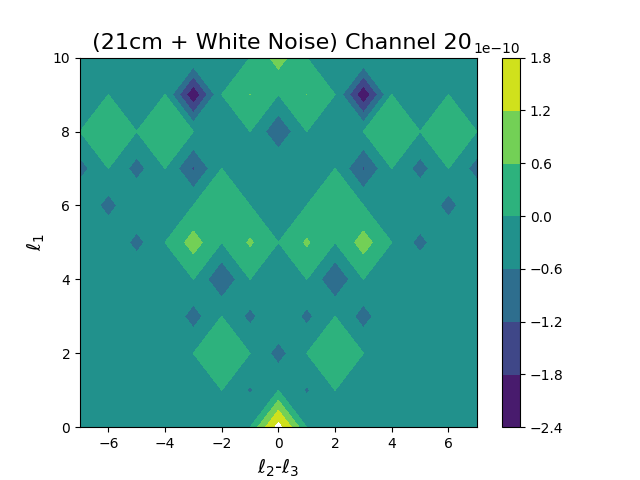}\\[\smallskipamount]

\includegraphics[width=.33\textwidth]{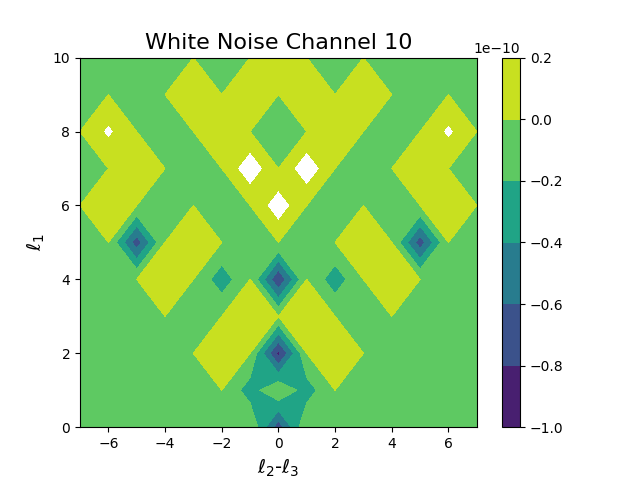}
\includegraphics[width=.33\textwidth]{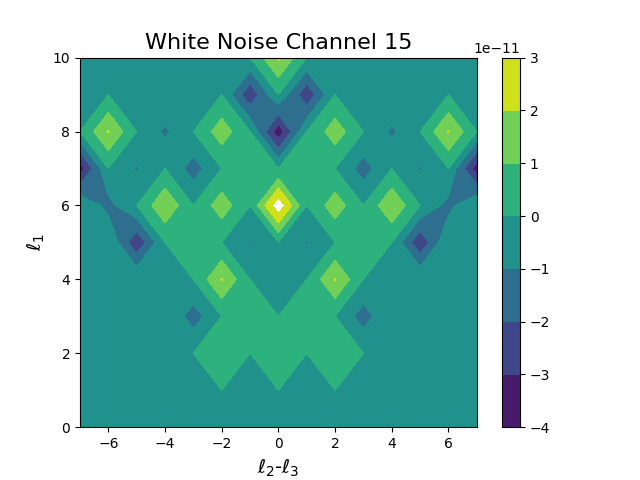}
\includegraphics[width=.33\textwidth]{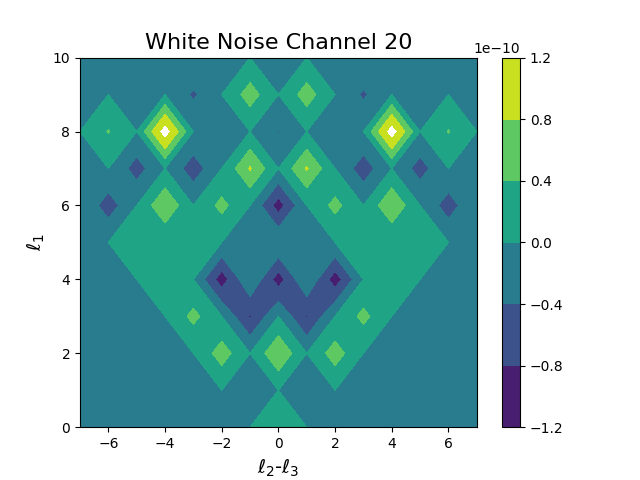}\\[\smallskipamount]

\caption{Contour charts for the  equisize configuration of the bispectrum in three different channels: 10 (first column), 15 (second column), and 20 (third column). Each line is related to each case analyzed: total foregrounds (first row), 21cm + white noise (second row), and white noise (third row).}\label{fig:equisizemaps}
    \end{centering}
\end{figure*}

In the case of a less conservative foreground subtraction than that suggested by the AIC method (i.e., a dimension of $m_\mathrm{AIC}-1$), we see in the top left panel of Fig.~\ref{fig:residuals} that there is significant residual contamination from free-free and synchrotron emission at \mbox{$\ell \lesssim 40$}. However, in the case of negligible noise, this configuration does manage to recover the majority of the power spectrum in the intermediate $\ell$ range $40< \ell < 200$. The overall residual foreground contamination is negligible compared to the 21cm signal over the larger range of multipoles $40\lesssim \ell \lesssim 500$, which is of interest for BAO measurements. 

In the case of $m_\mathrm{AIC} + 1$ we can see in Fig.~\ref{fig:residuals} that the residual foreground contributions, plotted in celestial coordinates, show little difference in comparison with $m_\mathrm{AIC}$; however, the loss of the 21cm signal in the reconstruction further increases in all channels, as highlighted in Fig.~\ref{fig:signal_bias}.

Figure~\ref{fig:gnom_ndims} shows maps of the residual foreground contributions to the reconstructed 21cm map from each individual component for the three different choices of the dimension of the foreground subspace, and for either noise-free or noisy simulations. The first two rows show larger residual foreground contamination for the $m_\mathrm{AIC}-1$ case compared to the more aggressive versions used in {\tt GNILC} in the middle ($m_\mathrm{AIC}$) and bottom ($m_\mathrm{AIC}+1$) rows. We can see from these results that there is a trade-off between the aggressiveness of the choice of $m$ where aggressive versions lose signal but have lower systematics, and the AIC technique implemented in {\tt GNILC} is doing a good job at determining this specific dimension.

\subsection{BINGO pipeline: Bispectrum estimation of foreground residual contamination}

\begin{figure*}
\begin{centering}

\includegraphics[width=.33\textwidth]{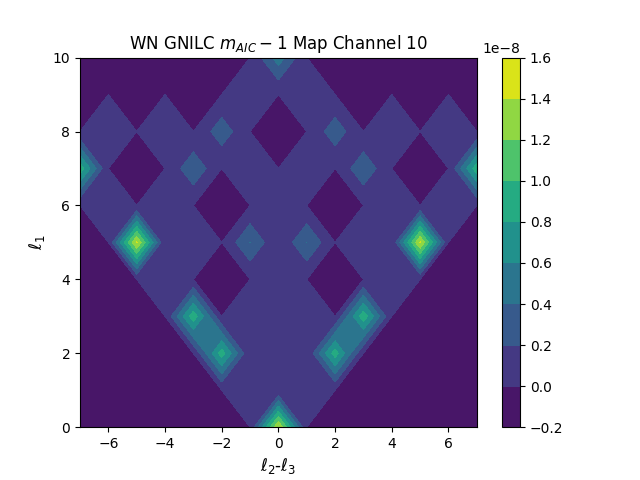}
\includegraphics[width=.33\textwidth]{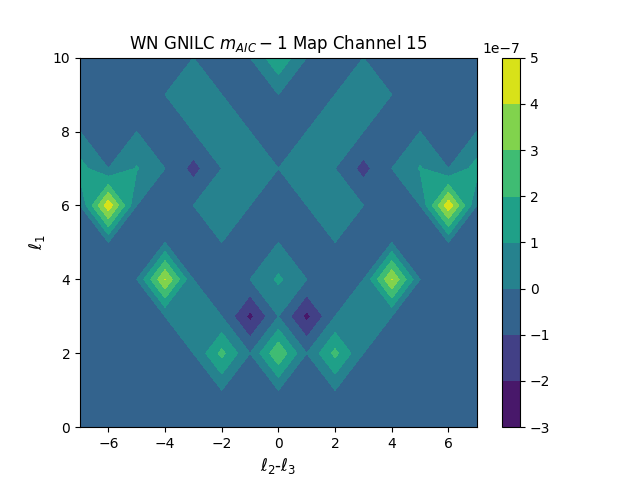}
\includegraphics[width=.33\textwidth]{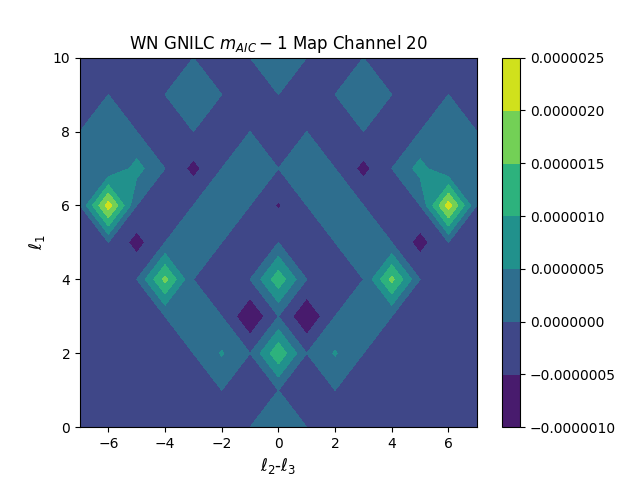}\\[\smallskipamount]

\includegraphics[width=.33\textwidth]{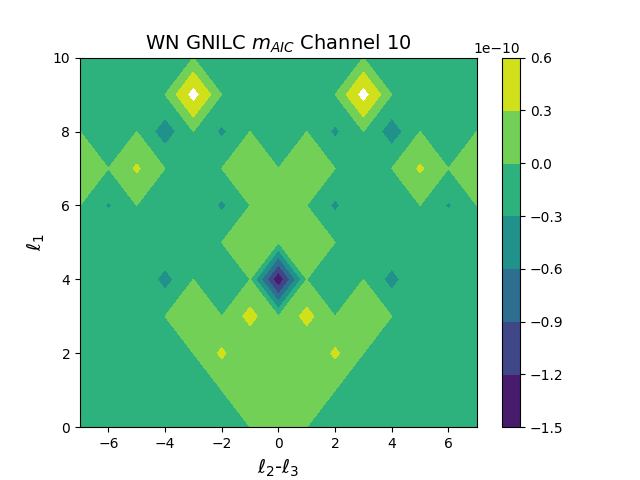}
\includegraphics[width=.33\textwidth]{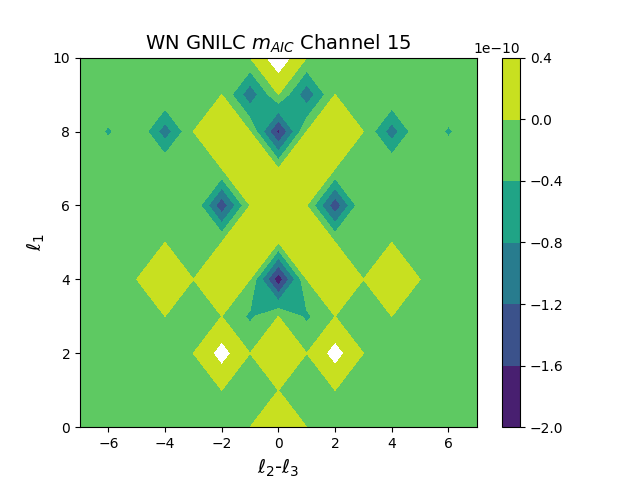}
\includegraphics[width=.33\textwidth]{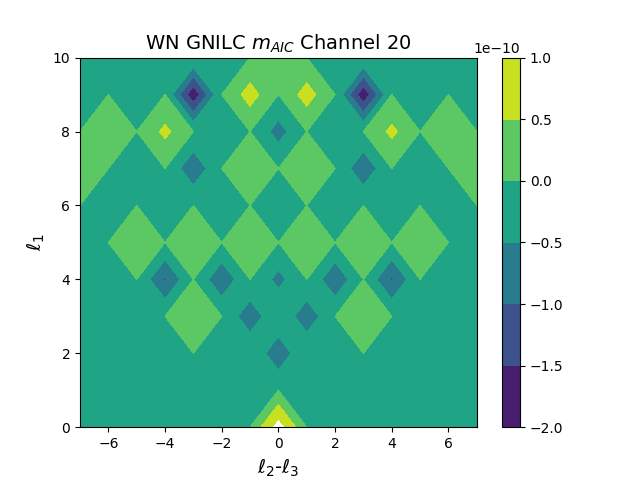}

\caption{Contour charts for the considered configurations of the bispectrum (where $\ell_1+\ell_2+\ell_3 = 30$) in three different channels: 10 (first column), 15 (second column), and 20 (third column). Each line is related to each case analyzed: {\tt GNILC} output using $m_\mathrm{AIC}-1$ (first row) and {\tt GNILC} output using $m_\mathrm{AIC}$ (second row). A very similar pattern arises from the results with $m_\mathrm{AIC}-1$ and the first row for the measurements with the input foregrounds in Fig.\ref{fig:equisizemaps}. These patterns are similar, albeit with a much smaller amplitude, which indicates a severe reduction of foreground levels, but not a complete removal, which is directly identified in our analysis.}
    \label{fig:equisizemaps2}
    \end{centering}
\end{figure*}

\begin{figure}

\begin{centering}
    \includegraphics[width=.45\textwidth]{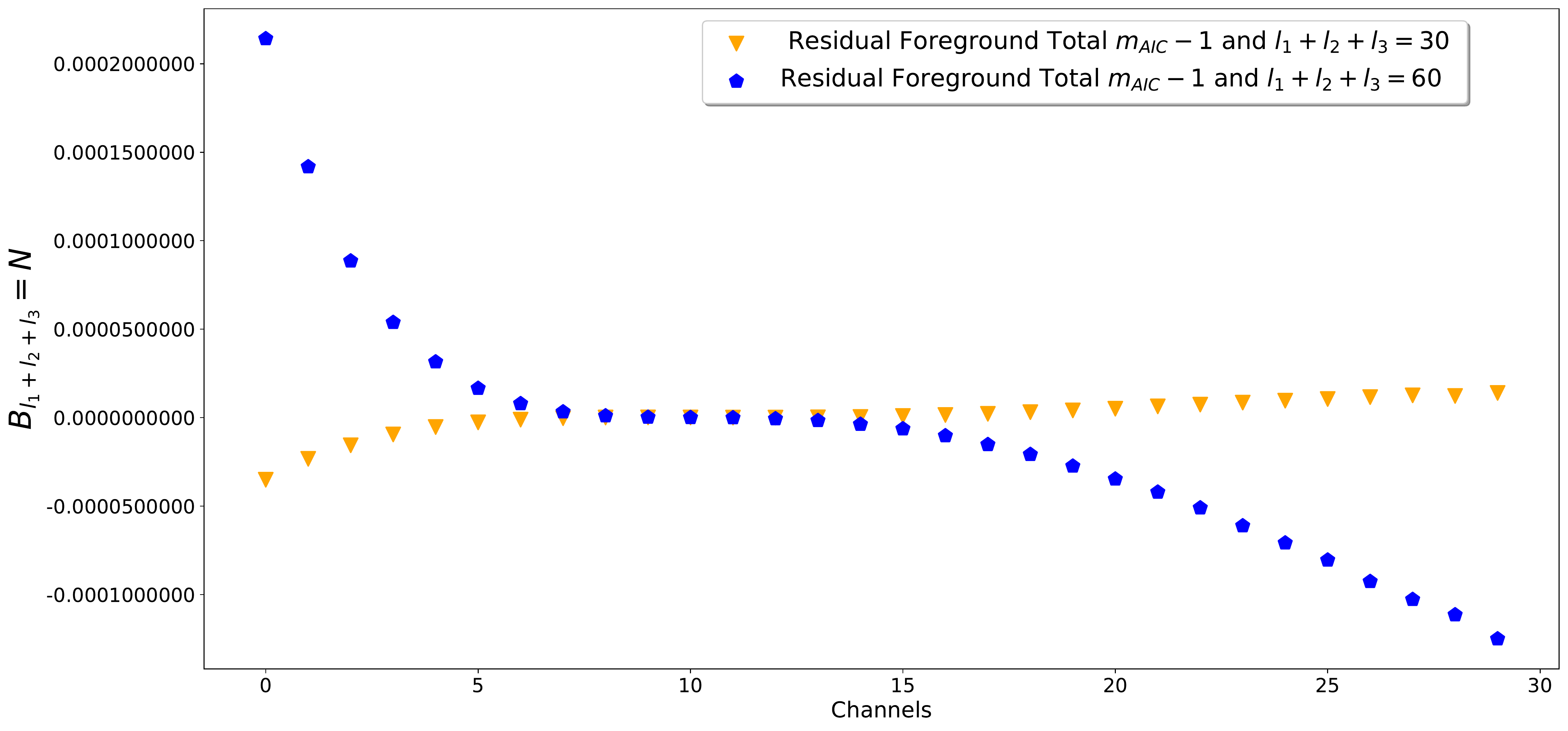}\hfill
    \includegraphics[width=.45\textwidth]{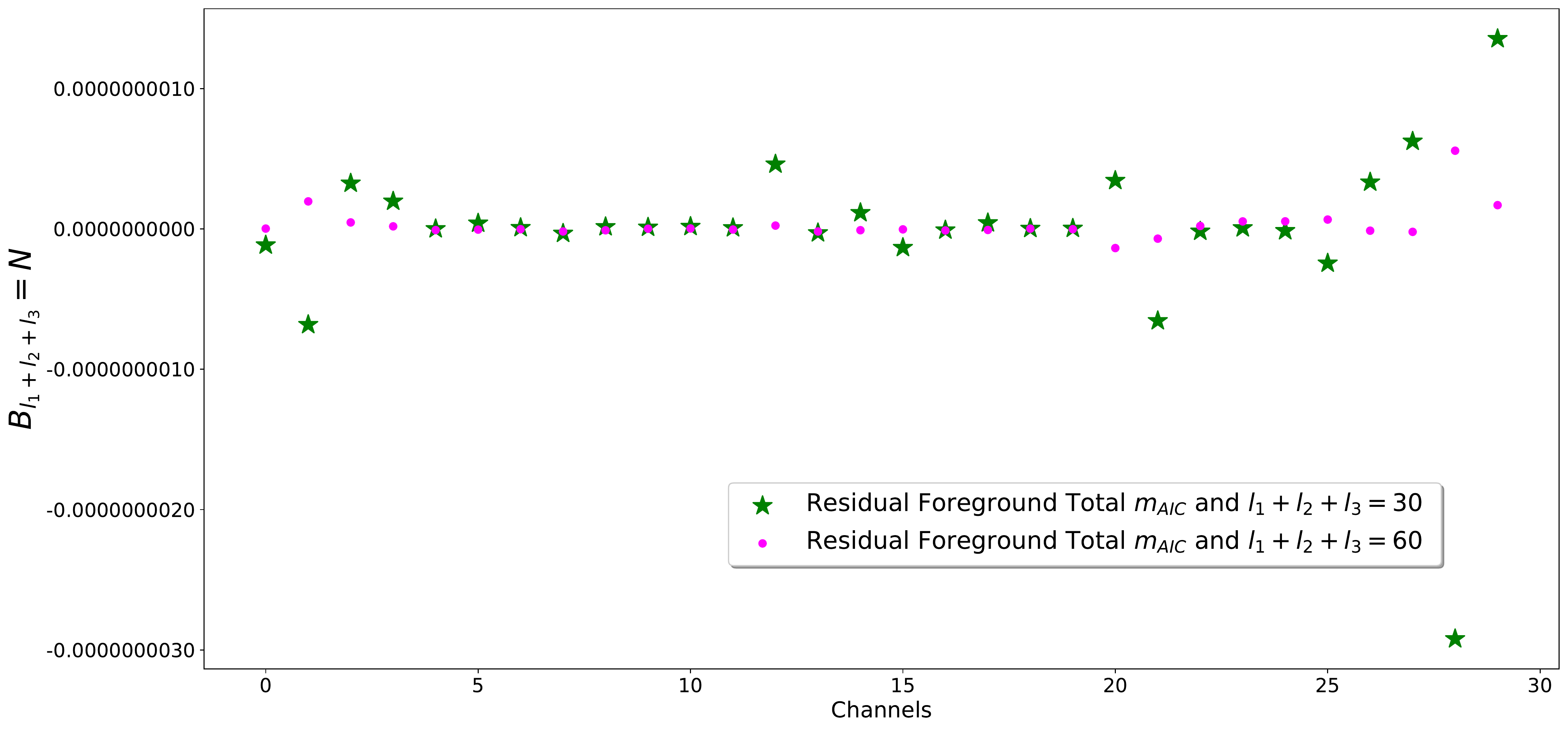}\hfill
  \caption{
  Residual values for $B_{\ell_1+\ell_2+\ell_3 = N}$ for $N=30$ and $N=60$, for the {\tt GNILC} reconstructed maps as a function of the channel number. We plot different values of $m_\mathrm{AIC}$ (top and bottom figures), and the residuals are a few orders of magnitude smaller than those presented in Fig. \ref{fig:equisizemaps2}, independently of which $m_\mathrm{AIC}$ is chosen for our bispectrum test for sub-optimal foreground extractions parameters.
  }\label{fig:belcompare}
\end{centering}
\end{figure}

We calculate the bispectrum of the reconstructed 21cm maps as well as the foreground maps and projections using Eqs. \ref{bisp15} and  \ref{bisp18}.
Since the bispectrum measures the non-Gaussian statistics intrinsic to the field, 
it is expected to return $B_{\ell}$ values that are compatible with zero for any Gaussian maps, including the noise input maps we have generated.

The reconstructed 21cm maps by {\tt GNILC} contain non-Gaussian information from both the intrinsic 21cm fluctuations, whose distribution is log-normal given that they were simulated with {\tt FLASK}, and the residual foreground contamination after component separation. The bispectrum module should therefore be able to detect non-Gaussian features in the input 21cm maps generated by {\tt FLASK} and in the recovered 21cm obtained by {\tt GNILC}, provided it has a sufficient signal-to-noise ratio.

There are several configurations we can select in order to estimate the non-Gaussian nature of a map via a bispectrum. In this analysis, as outlined in section \ref{sec:ang_bispectrum}, we select configurations where $\ell_1+\ell_2+\ell_3=N$, which we call  equisize configurations for the bispectrum \citep{CosmicString}. For the tests in this section we used a maximum value of $N = 30$ and $N = 60$ for the configurations.

In Fig.~\ref{fig:equisizemaps} we show the results for the $B_{\ell}$ values from the input of our simulations (21cm + noise + foregrounds) in the top row of the plots, from the 21cm maps of the {\tt FLASK} simulations plus the noise in the second row of the plot, and  for the simulated noise maps for the same simulation in the third row of the results. The plot shows the results for three different frequency bins in our simulation. We note that there is clearly a significant signal in the input foreground maps, which is natural and to be expected.

This signal in the bispectrum is around 16 orders of magnitude larger in the bispectrum than the signals obtained in the second and third rows. 
We expect the signal from the white noise configuration to have an expectation
value of zero as they contain no non-Gaussianities, whereas the signal in the second row should have a non-Gaussian signal in it, although   reasonably small given that this signal should come only from the log-normal transformation within {\tt FLASK}, which is meant to produce a slightly skewed one-point function. This means that, in essence, this will enhance some peaks of the density field according to the log-normal distribution and will have a reasonably small amount of non-Gaussianities.

\begin{figure*}
\begin{centering}

\includegraphics[width=.8\textwidth]{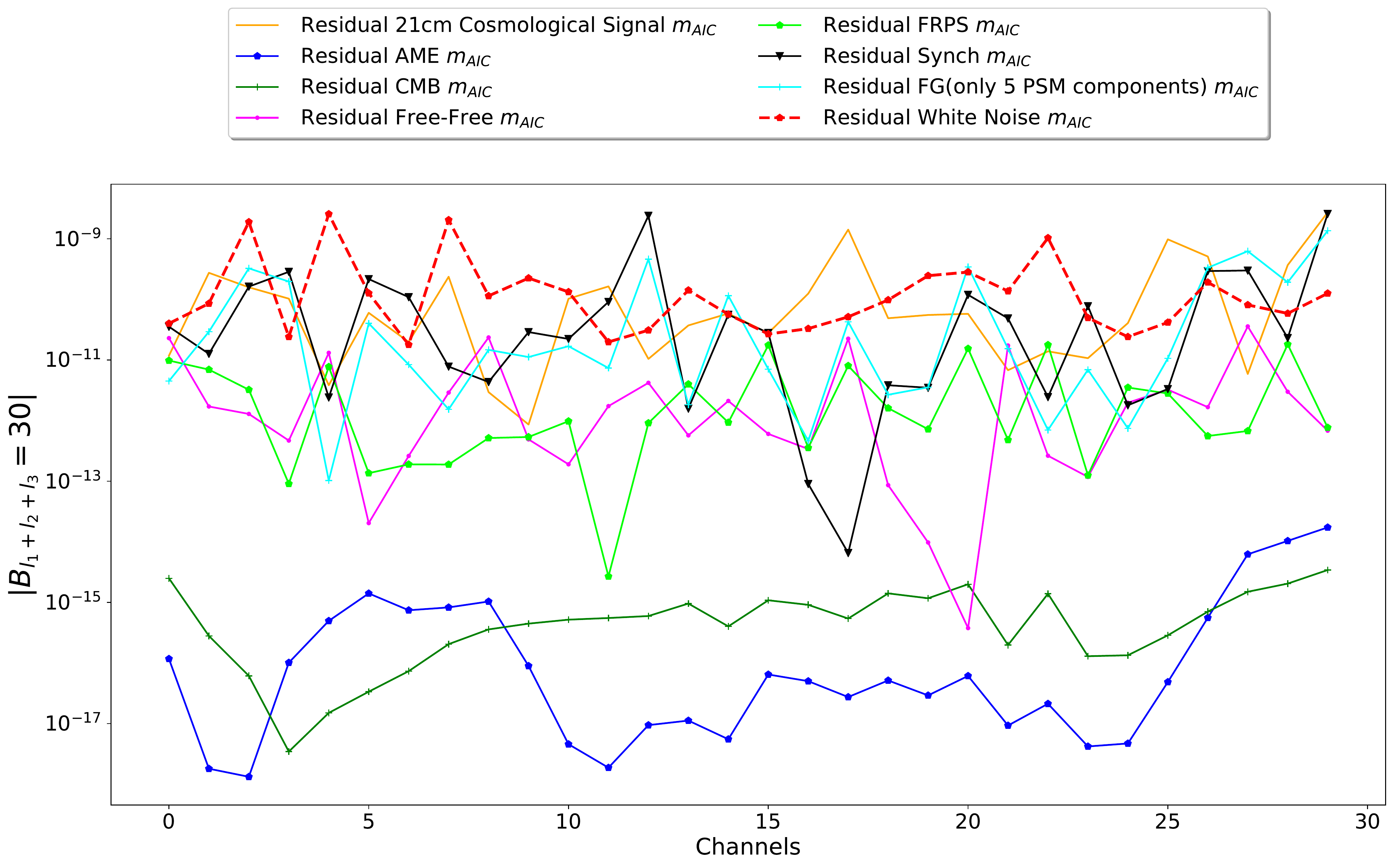}
    
\caption{
  $B_{\ell_1+\ell_2+\ell_3 = N}$ for $N=30$, for the {\tt GNILC} reconstructed maps as a function of the channel number. The results are plotted for different projected components from our foregrounds as well as the noise and 21cm input field, and show that the residuals can be compared to the original values expected within the 21cm field and noise realization. 
}
\label{fig:belcompare2}
\end{centering}
\end{figure*}

In Fig.~\ref{fig:equisizemaps2} we plot the same contours for the bispectrum that are  plotted in Fig.~\ref{fig:equisizemaps}, but for the {\tt GNILC} output maps obtained with a less aggressive value of $m_\mathrm{AIC}-1$, which we know is a configuration that is sub-optimal in terms of foreground subtraction from the previous sections, as well as the {\tt GNILC} output for $m_\mathrm{AIC}$, which is the preferred value chosen by the AIC algorithm. We can see that the bispectrum results for the value of $m_\mathrm{AIC}$ are comparable to the values obtained in the input signal. Although it is impossible from these plots alone to indicate if the {\tt GNILC} run with $m_\mathrm{AIC}$ has significant residuals in the bispectrum, we can certainly state that if there are any, they are at most of the same level of the noise inputs. 
However, we can clearly assert that the level of the signal in the first row, when the code is run with $m_\mathrm{AIC}-1$, has a measurable bispectrum above the bispectrum of the noise and the bispectrum of the 21cm signal. 

This indicates that we do have a detection of foreground residuals, and hence have an independent way to check if the {\tt GNILC} results are compatible with an efficient foreground subtraction. Although this check was made for {\tt GNILC}, we can use such methods to test and reliably check if there are any residual foregrounds in any such maps that are produced by other methods than {\tt GNILC}. We can also test optimal values for the ILC bias, which is also a parameter choice within {\tt GNILC} that we have fixed to be 0.01 in this work. 

In order to show this detection in a clearer way, we have also plotted Eq. \ref{bisp18} as a function of channel number in Fig. \ref{fig:belcompare}.
In this figure we can see a clear detection of the residuals, which are shown to be well measured by our bispectrum test for channels below 5 and above 15. 

Interestingly, the reconstruction for sub-optimal values of $m_\mathrm{AIC}$ are reasonable in the regions of channels from 5 to 10, which are exactly the channels where the suppression factor $S_\ell$ for the choice of $m_\mathrm{AIC}$ is close to the suppression factor for $m_\mathrm{AIC}-1$, whereas the suppression factor for the $m_\mathrm{AIC}$ case is closer to the $m_\mathrm{AIC}+1$ case above channel  15 and below channel  5. This clearly shows that there is an interplay between the dimension and aggressiveness of {\tt GNILC} as a function of the channel number, which is dictated by the data, and is encapsulated in the measurements of the bispectrum shown in this plot.


Finally, we plot the summed bispectrum again in Fig.\ref{fig:belcompare2}  increasing the scale so that
we can see all the foreground projections in our simulations. Given the smaller nature of each component, we plot the absolute value of $B_{\ell_1+\ell_2+\ell_3=N}$ and compare their magnitudes. 
We can see a channel-dependent structure in CMB (green curve) and AME (blue curve) foreground residuals, even though both are placed significantly below the noise level estimated by the bispectrum analysis (red dashed curve)


On the other hand, the level of the synchrotron residual (black curve) is equivalent to the white noise (red dashed curve) and 21cm (orange curve) projected components. Our tests indicate that {\tt GNILC} is able to remove foregrounds including any non-Gaussian residual all the way down to the noise level injected into the data. However, we cannot conclude that these  non-Gaussian residuals are removed below that noise level.


\section{Conclusions}\label{sec:conclusions}


In this paper we have presented an analysis of the GNILC method to remove foregrounds in a simulated scenario of BINGO Phase 1 observations. This analysis is based upon the simulations of the foregrounds and white noise, as expected from the electronics, and does not include the $1/f$ noise contribution. This paper discusses in more detail the tools that were  used in paper IV \citep{2020_sky_simulation}, complementing the time domain analysis described in that work. It is based upon the project, instrument, and optical descriptions presented in papers I to III \citep{2020_project,2020_instrument, 2020_optical_design}.


We used the simulated maps described in Sect. \ref{sec:simulations} combined with white noise realizations and foreground emission to produce a realistic sky, as should be observed by BINGO. The simulated maps were also masked to remove the Galactic plane contribution, so that the subsequent analysis was performed in a sky map in which we expect low contamination from our Galaxy.


Section \ref{sec:powerspectum} described the component separation performed by {\tt GNILC} on the simulated sky, as well as the additional steps to calibrate the residual bias left by the {\tt GNILC} filter. We conclude that the recovered \hi\, power spectrum is compatible with our input simulations within our noise levels and therefore should meet our scientific requirements.

The bispectrum module described in Sect. \ref{sec:bispectrum} was used to check for the presence of a non-Gaussian signal in the output \hi\, maps, which might indicate they still contain a significant level of foreground residuals. 


We also conclude that the bispectrum module is able to recognize if a non-Gaussian pattern is present in the output maps and that, for the BINGO Phase 1 configuration, we are able to reduce such residuals below the noise levels detected by the bispectrum verification. We also found that the residuals are clearly identified in the bispectrum analysis in cases 
where suboptimal foreground cleaning strategies are used in place of the nominal {\tt GNILC} method.
This can be a valuable tool for testing and verification of the quality of the foreground subtraction steps during the BINGO data analysis.

\begin{acknowledgements}
The BINGO project is supported by FAPESP grant 2014/07885-0. 
K.S.F.F. thanks S{\~a}o Paulo Research Foundation (FAPESP) for financial support through grant 2017/21570-0.
Support from CNPq is gratefully acknowledged (E.A.).
F.B.A. acknowledges the UKRI-FAPESP grant 2019/05687-0, and FAPESP and USP for Visiting Professor Fellowships where this work has been developed.
L.S. is supported by the National Key R\&D Program of China (2020YFC2201600).
E.J.M. acknowledges the support by CAPES.
M.R. acknowledges funding from the European Research Council (ERC) under the European Union’s Horizon 2020 research and innovation programme (grant agreement No~725456, CMBSPEC).
R.G.L. thanks  CAPES (process 88881.162206/2017-01) and the Alexander von Humboldt Foundation for the financial support.
C.P.N. thanks S{\~a}o Paulo Research Foundation (FAPESP) for financial support through grant 2019/06040-0.
A.R.Q., L.B., F.A.B. and M.V.S. acknowledge PRONEX/CNPq/FAPESQ-PB (Grant no. 165/2018). 
V.L. acknowledges the postdoctoral FAPESP grant 2018/02026-0. 
T.V. acknowledges CNPq Grant 308876/2014-8. 
C.A.W. acknowledges a CNPq grant 2014/313.597. 
J.Z. was supported by IBS under the project code, IBS-R018-D1.
 A.A.C. acknowledges financial support from the China Postdoctoral Science Foundation, grant number 2020M671611. 
B. W. and A.A.C. were also supported by the key project of NNSFC under grant 11835009.
M.P. acknowledges funding from a FAPESP Young Investigator fellowship, grant 2015/19936-1.
Some of the results in this paper have been derived using the {\tt HEALPix} package \citep{Gorski:2005}, and also {\tt Python} language, including the packages {\tt healpy}, {\tt astropy}, {\tt numpy},  and {\tt matplotlib}. 

\end{acknowledgements}

\bibliographystyle{aa}
\bibliography{projectpaperrefs}

\end{document}